\documentclass[12pt,preprint]{aastex} 
\usepackage{graphicx} 
\usepackage{epsfig}
\usepackage{xcolor}
\usepackage{ulem}
\usepackage{hyperref}
\input{colordvi}

\hypersetup{
    bookmarks=true,                 
    unicode=false,                  
    pdftoolbar=true,                
    pdfmenubar=true,                
    pdffitwindow=true,              
    pdfstartview={FitH},            
    pdftitle={CASH3}, 
    pdfauthor={hollek},           
    pdfsubject={Astronomy},         
    pdfcreator={dvipdf},            
    pdfproducer={dvipdf},           
    pdfkeywords={metal-poor stars}, 
    pdfnewwindow=true,              
    colorlinks=true,                
    linkcolor=red,                  
    citecolor=blue,                 
    filecolor=magenta,              
    urlcolor=cyan,                  
    breaklinks=true,
    linktocpage
}

\bibpunct{(}{)}{;}{a}{}{,} 
\usepackage{lscape}

\shorttitle{New Classification for CEMP s-process stars}  
\shortauthors{Hollek et al.} 
 
\begin{document} 

\title{The Chemical Abundances of Stars in the Halo (CASH)
  Project. III. A New Classification Scheme for Carbon-Enhanced
  Metal-poor Stars with S-process Element Enhancement\altaffilmark{1,2}}

\author{
Julie K. Hollek\altaffilmark{3}, 
Anna Frebel\altaffilmark{4}, 
Vinicius M. Placco\altaffilmark{5},
Amanda I. Karakas\altaffilmark{6},
Matthew Shetrone\altaffilmark{3,7},
Christopher Sneden\altaffilmark{3}, and
Norbert Christlieb\altaffilmark{8}}

\altaffiltext{1}{Based on observations gathered with the 6.5 meter
  Magellan Telescopes located at Las Campanas Observatory, Chile.}
\altaffiltext{2}{Based on observations obtained with the Hobby-Eberly
  Telescope, which is a joint project of the University of Texas at
  Austin, the Pennsylvania State University, Stanford University,
  Ludwig-Maximilians-Universit\"at M\"unchen, and
  Georg-August-Universit\"at G\"ottingen.}
\altaffiltext{3}{Department of Astronomy, University of Texas, Austin, TX 78712-0259, 
  USA; julie@astro.as.utexas.edu,chris@verdi.as.utexas.edu}

\altaffiltext{4}{Department of Physics \& Kavli Institute for
  Astrophysics and Space Research, Massachusetts Institute of
  Technology, Cambridge, MA 02139, USA; afrebel@mit.edu}

\altaffiltext{5}{Department of Physics and JINA Center for the Evolution of the Elements, 
University of Notre Dame, Notre Dame, IN 46556, USA; vplacco@nd.edu}

\altaffiltext{6}{Research School of Astronomy \& Astrophysics, Australian National University, 
  Canberra, Australia; amanda.karakas@anu.edu.au} 
\altaffiltext{7}{McDonald Observatory, University of Texas, Fort Davis, TX 79734, USA; 
  shetrone@astro.as.utexas.edu}
\altaffiltext{8}{Zentrum f\"ur Astronomie der
  Universit\"at Heidelberg, Landessternwarte, K\"onigstuhl 12, 69117
  Heidelberg, Germany; N.Christlieb@lsw.uni-heidelberg.de}

\begin{abstract} 
We present a detailed abundance analysis of 23 elements for a newly
discovered carbon-enhanced metal-poor (CEMP) star, HE~0414$-$0343,
from the Chemical Abundances of Stars in the Halo (CASH) Project. Its
spectroscopic stellar parameters are $T_{eff}=4863$\,K, $\log g=1.25$,
$\xi=2.20$\,km\,s$^{-1}$, and $\mbox{[Fe/H]}=-2.24$. Radial velocity
measurements covering seven years indicate HE~0414$-$0343 to be a
binary. HE~0414$-$0343 has $\mbox{[C/Fe]}=1.44$ and is strongly
enhanced in neutron-capture elements but its abundances cannot be
reproduced by a solar-type s-process pattern alone. Traditionally, it
could be classified as ``CEMP-r/s'' star. Based on abundance
comparisons with AGB star nucleosynthesis models, we suggest a new
physically-motivated origin and classification scheme for CEMP-s stars
and the still poorly-understood CEMP-r/s. The new scheme describes a
continuous transition between these two so-far distinctly treated
subgroups: CEMP-sA, CEMP-sB, and CEMP-sC. Possible causes for a
continuous transition include the number of thermal pulses the AGB
companion underwent, the effect of different AGB star masses on their
nucleosynthetic yields, and physics that is not well approximated
  in 1-D stellar models such as proton ingestion episodes and
  rotation.  Based on a set of detailed AGB models, we
suggest the abundance signature of HE~0414$-$0343 to have arisen from
a $>1.3$\,M$_{\odot}$ mass AGB star and a late-time mass transfer, that
transformed HE~0414$-$0343 into a CEMP-sC star. We also find the
[Y/Ba] ratio well parametrizes the classification and can thus be used
to easily classify any future such stars.

\end{abstract}
 
\keywords{Galaxy: halo -- methods: spectroscopy -- stars: abundances
  -- stars: atmospheres -- stars: Population II}
 
\section{Introduction}\label{introduction}
Metal-poor Population II (Pop II) stars were formed from gas that
contained the nucleosynthetic signatures of the first chemical
enrichment events in the Universe. Metal-poor stars preserve this information in their
atmospheres, which we observe today. By understanding the chemical abundance
patterns of metal-poor stars, we can probe the formation, initial mass
function, and fates of the first stars. Altogether, metal-poor
stars allow for a detailed reconstruction of the chemical enrichment
sources and processes operating in the early universe and leading to
the chemical evolution of the Milky Way: from core collapse supernovae
of the earliest, massive stars, to the later
contributions from the nucleosynthesis of lower-mass, evolved
asymptotic giant branch (AGB) stars, and even the Type Ia supernovae.
 
Many metal-poor stars have prominent molecular carbon features in
their spectra. The G-band feature near 4290\,{\AA}, the bandhead at
4313\,{\AA}, and the smaller band near 4323\,{\AA} are all CH
molecular features. These all become strong to the point of saturation
in the presence of large amounts of C. There are CN features across
the spectrum, including a prominent feature near 8005\,{\AA}. The
C$_{2}$ molecule is not often detected in metal-poor stars with
[C/Fe]\footnote{\mbox{[A/B]}$ = \log(N_{\rm A}/N_{\rm B}) -
  \log(N_{\rm A}/N_{\rm B})_\odot$ for N atoms of elements A, B, e.g.,
  $\mbox{[Fe/H]}=-2.0$ is 1/100 of solar Fe abundance.} ratios near
the solar ratio; however, in stars with large C abundances, the
bandheads near $\lambda$4735, 5165, and 5635 often become strong
enough from which to derive a [C/Fe] ratio.

\citet{bc05} define a carbon-enhanced metal-poor (CEMP) star to be any
metal-poor star with $\mbox{[C/Fe]} \geq 1$. \citet{aoki_cemp_2007}
presented a revised CEMP definition of $\mbox{[C/Fe]} > 0.7$ for stars with
log(L/L$_{\odot}$)$\leq$ 2.3 and $\mbox{[C/Fe]} >  3 - \log$(L/L$_{\odot}$)
for stars with $\log$(L/L$_{\odot}$)$\geq 2.3$. The CN cycle greatly
reduces the amount of C in the surface composition of a star over the
course of the later stages of stellar evolution on the giant
branch. Hence this definition allows for more evolved stars with lower
C abundances to be considered in the study of C in the early universe.

CEMP stars can be subdivided into distinct chemical
subgroups. \citet{masseroncemp2010} provides a comprehensive
description and study of the different types of CEMP stars, which we
will briefly outline here. CEMP-no stars are CEMP stars with normal
neutron-capture abundances (indicated by $\mbox{[Ba/Fe]}\le0$) and
otherwise typical abundances for metal-poor stars. The three most
iron-poor stars discovered are CEMP-no stars \citep{HE0107_Nature,
  HE1327Nature, kellernature}.  The majority of CEMP stars have
neutron-capture abundance enhancements. Among these, CS~22892$-$052 is
the only one discovered to date with a pure rapid neutron-capture (r-)
process abundance pattern \citep{sneden2000}. The largest subgroup of
CEMP stars is the CEMP-s stars \citep{masseroncemp2010}, which contain
enhancements in the slow neutron-capture (s-) process
elements. Finally, so-called CEMP-r/s stars are another CEMP group
with neutron-capture overabundances but their abundance distributions
do not display either a pure r- or pure s-process pattern
\citep{bisterzo09}; both processes have been suspected to have
contributed.

The designation of the CEMP-``r/s'' stars has undergone much
evolution. The term was introduced in \citet{bc05} as
``CEMP-r/s''. This term referred to stars with $\mbox{[C/Fe]} > $1.0
and 0$ < \mbox{[Ba/Eu]} < 0.5$. \citet{jonsell06} used the term
``r+s'' to refer to stars with $\mbox{[Ba/Fe]} > $1.0 and
$\mbox{[Ba/Eu]} > $0 or $\mbox{[Eu/Fe]} > $1 omitting the Ba criteria
altogether, making no mention of the C abundance in the definition
criteria, but noting that all of these r+s stars had significant C
enhancement. \citet{masseroncemp2010} essentially combined the two
previous definitions into the ``CEMP-rs'' designation for stars with
$\mbox{[C/Fe]} > $0.9 and $\mbox{[Eu/Fe]} > $1 and $\mbox{[Ba/Eu]} <
$0, or $\mbox{[C/Fe]} > $0.9 and $\mbox{[Ba/Fe]} > $2.1. The [Ba/Fe]
criterion allowed for stars with no Eu abundance to also be
classified. \citet{bisterzo12} even present additional subgroups.  The
\citet{jonsell06} designation was presented along with abundances for
HE~0338$-$3945, an r+s star near the main sequence turn-off with
[Fe/H] = $-$2.42. They presented nine possible scenarios for its
formation, including the suggestion that these stars themselves formed
from an r-enhanced gas cloud, although this scenario would be
difficult to confirm observationally. \citet{masseroncemp2010} aimed
to disentangle the contributions of AGB nucleosynthesis to the
abundance patterns in CEMP stars by investigating processes that occur
during the mass transfer from an AGB binary companion onto the
observed CEMP star. No satisfactory solution to explain the abundance
patterns of CEMP-r/s stars has been presented so far
\citep[see also discussion in][]{lugaro}.

The s-process occurs in the He-rich layers of evolved, low-mass
AGB stars 
\citep[e.g.,][]{gallino1998,karakas14}, and operates on timescales of
tens of thousands of years.  Seed nuclei acquire neutrons one at a
time and then $\beta$-decay as they climb the valley of stability on
the chart of the nuclides. Bismuth is the termination point of the
s-process. There are three stable peaks in the s-process pattern. They
are centered around Sr, Ba, and Pb and these elements are particularly
enhanced in s-process enriched metal-poor stars. This enrichment
occurs when a low-mass star receives s-process enhanced material from
a binary companion that underwent its AGB phase; observed today is the
low-mass recipient of the AGB material.  Many of these stars,
including CEMP-s and CEMP-r/s stars, have been monitored for radial
velocity variations and been shown be binary stars
\citep{lucatello2005,starkenburg14}.  
This principally confirms the mass-transfer scenario for these stars.

The Chemical Abundances of Stars in the Halo (CASH) project is a study
that aims to understand the chemical abundance trends and frequencies
of metal-poor halo stars as well as discover individual
astrophysically interesting stars, based on the chemical abundances
for $\sim 500$ stars from ``snapshot'' spectra observed using the
Hobby-Eberly Telescope (HET) at McDonald Observatory. The spectra have
moderate signal-to-noise ratios ($\sim$65) and resolution
(R$\sim$15,000). CEMP and s-process stars make up perhaps $\sim20$\%
of the population of metal-poor stars. The first result from this
project, \citet{cash1}, was the discovery of a CEMP-r/s giant star,
HK~II 17435-00532, with an enhanced Li abundance. The second paper,
\citet{CASH2}, presented the calibration of the automated stellar
parameter and abundance pipeline, Cashcode, using both the HET
snapshot spectra and higher-resolution, higher-S/N Magellan/MIKE
spectra of 16 new extremely metal-poor stars to test it, along with
the resultant comprehensive abundance analysis of the sample.

In this paper, we discuss HE~0414$-$0343, a CEMP star initially
identified in the Bright Metal-Poor Star (BMPS) sample of
\citet{frebel_bmps} and included in the CASH project for further
follow-up observations. This star was initially slated to be part of
the sample from \citet{CASH2}, but was singled out for special
attention in order to obtain an even higher-resolution spectrum to
better study this unique star. We discuss the observations and binary
status for HE~0414$-$0343 in Section~\ref{observations}. We present
our analysis methods used to determine the stellar parameters and
abundances in Section~\ref{analysis} and the results in
Section~\ref{abundances}. In Section~\ref{interpretation} we analyze
neutron-capture abundance ratios. We compare the abundances of
HE~0414$-$0343, as well as those of a literature sample to abundance
yields of AGB star models in Section~\ref{placcomodel} and propose a
new classification scheme based on our analysis of this star. In
Section~\ref{origins}, we use the classification to gain insight into
the origin of CEMP stars with neutron-capture element enhancement
associated with the s-process. We discuss the limitations of our
analysis in Section~\ref{caveats}, and summarize our results in
Section~\ref{conclusions}.

\section{Observations}\label{observations}
HE~0414$-$0343 has an R.A.  of 04\,h 17\,m 16.4\,s and declination of
$-$03$\degr$ 36$\arcmin$ 31$\farcs$0. Thus, it is accessible from both
the northern and southern hemispheres. Four separate spectra were
obtained for HE~0414$-$0343 between 2004 and 2011. A medium-resolution
spectrum was observed in September 2004 as part of the the Hamburg/ESO
Bright Metal-Poor Star Sample \citet{frebel_bmps}.  A high-resolution
spectrum was obtained with MIKE instrument \citep{mike} at the
Magellan-Clay Telescope at Las Campanas Observator in September 2006.
Using a 0\farcs7 slit with 2$\times$2 on-chip binning yielded a
resolution of $R\sim 35,000$ in the blue and 28,000 in the red.  MIKE
spectra have nearly full optical wavelength coverage from $\sim 3500$
- 9000\,{\AA}. In October 2008, HE~0414$-$0343 was observed as part of
the CASH project with the fiber-fed High Resolution Spectrograph
\citep{tull98} on the HET at McDonald Observatory.  All CASH spectra
were obtained with a 2$\farcs$0 fiber yielding $R\sim 15,000$. The
$2\times5$ on-chip CCD binning leads to 3.2 pixels per resolution
element. Two CCDs were used to record the red and blue portions of the
spectrum, spanning a wavelength range from 4200 - 7800\,{\AA}.

The highest resolution spectrum, taken in March 2011, was used for the
stellar parameter and chemical abundance determinations. This spectrum
was also obtained using MIKE on the Magellan-Clay Telescope, but taken
with a 0\farcs5 slit.  The spectral resolution is $R\sim 56,000$ at
4900\,{\AA} and $\sim 37,000$ at 5900\,{\AA} as measured from the ThAr
frames. Table~\ref{obs} lists the details of the observations for
HE~0414$-$0343. The high resolution spectra for HE~0414$-$0343 were
reduced using an echelle data reduction pipeline made for
MIKE\footnote{Available at
  http://obs.carnegiescience.edu/Code/python.}, initially described by
\citet{MIKEreduce}. We then utilized standard IRAF\footnote{IRAF is
  distributed by the National Optical Astronomy Observatories, which
  is operated by the Association of Universities for Research in
  Astronomy, Inc., under cooperative agreement with the National
  Science Foundation} routines to co-add and continuum normalize the
individual orders into a one-dimensional spectrum.

\begin{deluxetable}{lcccccc}
  \tablecolumns{7} 
  \tablewidth{0pc} 
  \tablecaption{Observations}
  \tablehead{\colhead{R}& \colhead{UT Date}& \colhead{UT Time} &\colhead{t$_{exp}$} 
  &\colhead{$S/N$}  &\colhead{Telescope} &\colhead{$v_{rad}$} \\
 \colhead{}&\colhead{}&\colhead{} &\colhead{s} &\colhead{at 5180\,{\AA}}& \colhead{}   
 & \colhead{km\,s$^{-1}$}  }
\startdata
2000   & 2004 September 20  & 18:13  & 120   & 50   &  SSO2.3m/DBS & $-$36 \\
35,000 & 2006 September 27  &  9:30  & 450   & 65   & Magellan-Clay/MIKE & $-$83.8  \\
15,000 & 2008 October 10    &  9:19  & 239   & 85   & Hobby-Eberly/HRS & 4.0 \\
56,000 & 2011 March 11      & 00:12  & 3000  & 120  & Magellan-Clay/MIKE& 11.3 \\
\enddata 
\label{obs}
\end{deluxetable} 

\subsection{Binary Status}\label{binarity}
Heliocentric radial velocities of HE~0414$-$0343 were measured based
on four independent observations taken over the course of seven years.
The results are given in Table~\ref{obs}. We note that Robospect
\citep{robospect}, an automated equivalent width measurement code that
can calculate radial velocity shifts through cross-correlation was
used for the $R\sim 35,000$ MIKE spectrum from 2006.  A
cross-correlation technique using the Mg b triplet was employed for
the vadial velocity determination of the HET snapshot spectrum
\citet{CASH2}. The radial velocity for the $R\sim 56,000$ MIKE
spectrum was determined by measuring the average velocity offset for a
set of 15 unblended lines in the red portion of the spectrum.

Typical radial velocity uncertainties for medium resolution spectra
are $\sim10$\,km\,s$^{-1}$, for snapshot spectra are
$\sim3$\,km\,s$^{-1}$, and for high-resolution spectra are
$\sim$1-2\,km\,s$^{-1}$. The radial velocity does vary
significantly over these seven years, indicating that the star has an
unseen binary companion. Establishing the binary status of
HE~0414$-$0343 aids in understanding the nature and mechanism of the
star's carbon-enhancement and the overabundances in neutron-capture
elements.

\section{Spectral Analysis}\label{analysis}

\subsection{Line Measurements}
The equivalent widths were measured with a customized ESO/Midas
program that automatically fits Gaussian profiles to each line.  The
user can adjust the fit to the continuum level by selecting line-free
continuum regions, if necessary. The linelist for these
stars is the same that was used in \citet{CASH2} for the MIKE spectra;
however, we rejected all Fe I and Fe II lines with wavelengths shorter
than 4450\,{\AA} due to severe blending with molecular C features. We
also omitted any strongly blended lines for the other elements whose
abundances were determined via equivalent width
measurements. Table~\ref{synthtab} lists the equivalent widths and
corresponding line abundances that are partially obtained through
spectrum synthesis (see also Section~\ref{abundances}).

\begin{deluxetable}{lccrrr}
  \tablecolumns{6} 
  \tablewidth{0pc} 
  \tablecaption{Equivalent Widths and Abundances}
  \tablehead{\colhead{Species}&\colhead{$\lambda$}& \colhead{$\chi$} &\colhead{log gf}& \colhead{W} &\colhead{$\log\epsilon$(X)}\\ 
 \colhead{}&\colhead{[\AA{}]}& \colhead{[eV]}&\colhead{} &\colhead{m\AA{}} &\colhead{}}
\startdata
C$_{2}$  & 5165     & \nodata &  \nodata & Synth & 7.63   \\
C$^{12}$/C$^{13}$& \nodata  & \nodata &  \nodata & Synth & 5\tablenotemark{a}      \\
Mg I &  4571.09  &  0.00  &  $-$5.69  &  98.7  &  5.94 \\
Mg I &  4702.99  &  4.33  &  $-$0.38  &  111.1  &  5.81 \\
Mg I &  5172.68  &  2.71  &  $-$0.45  &  240.5  &  5.75 \\
Mg I &  5528.40  &  4.34  &  $-$0.50  &  107.2  &  5.84 \\
Mg I &  5711.09  &  4.34  &  $-$1.72  &  24.7  &  5.76 \\
Ca I &  4455.89  &  1.90  &  $-$0.53  &  88.6  &  4.76 \\
Ca I &  5581.97  &  2.52  &  $-$0.56  &  34.0  &  4.51 \\
Ca I &  5588.76  &  2.52  &     0.21  &  80.2  &  4.49 \\
Ca I &  5590.12  &  2.52  &  $-$0.57  &  25.5  &  4.35 \\
Ca I &  5594.46  &  2.52  &     0.10  &  85.8  &  4.70 \\
Ca I &  5598.48  &  2.52  &  $-$0.09  &  68.6  &  4.60 \\
Ca I &  5601.28  &  2.53  &  $-$0.52  &  40.4  &  4.60 \\
Ca I &  5857.45  &  2.93  &     0.23  &  55.7  &  4.54 \\
Ca I &  6102.72  &  1.88  &  $-$0.79  &  63.5  &  4.46 \\
Ca I &  6122.22  &  1.89  &  $-$0.32  &  97.4  &  4.53 \\
Ca I &  6162.17  &  1.90  &  $-$0.09  &  102.1  &  4.39 \\
Ca I &  6439.07  &  2.52  &     0.47  &  102.0  &  4.55 \\
Ca I &  6449.81  &  2.52  &  $-$0.50  &  54.4  &  4.76 \\
Ca I &  6499.64  &  2.52  &  $-$0.82  &  35.2  &  4.76 \\
Sc II&  4415.54  &  0.59  &  $-$0.67  &  109.5  &  1.14 \\
Sc II&  5031.01  &  1.36  &  $-$0.40  &  71.9  &  1.02 \\
Sc II&  5031.01  &  1.36  &  $-$0.40  &  79.9  &  1.14 \\
Sc II&  5239.81  &  1.46  &  $-$0.77  &  68.0  &  1.43 \\
Sc II&  5526.78  &  1.77  &     0.02  &  70.0  &  1.01 \\
Sc II&  5641.00  &  1.50  &  $-$1.13  &  24.7  &  1.11 \\
Sc II&  5657.90  &  1.51  &  $-$0.60  &  51.2  &  1.04 \\
Sc II&  5658.36  &  1.50  &  $-$1.21  &  24.2  &  1.18 \\
Sc II&  5667.16  &  1.50  &  $-$1.31  &  28.5  &  1.37 \\
Sc II&  5684.21  &  1.51  &  $-$1.07  &  18.2  &  0.90 \\
Ti I &  3998.64  &  0.05  &     0.01  &  83.3  &  2.63 \\
Ti I &  4518.02  &  0.83  &  $-$0.27  &  42.6  &  3.06 \\
Ti I &  4533.24  &  0.85  &     0.53  &  67.0  &  2.66 \\
Ti I &  4534.78  &  0.84  &     0.34  &  58.0  &  2.70 \\
Ti I &  4535.56  &  0.83  &     0.12  &  55.8  &  2.87 \\
Ti I &  4548.76  &  0.83  &  $-$0.30  &  35.1  &  2.95 \\
Ti I &  4555.49  &  0.85  &  $-$0.43  &  35.2  &  3.11 \\
Ti I &  4656.47  &  0.00  &  $-$1.29  &  31.9  &  2.91 \\
Ti I &  4681.91  &  0.05  &  $-$1.01  &  63.8  &  3.20 \\
Ti I &  4840.87  &  0.90  &  $-$0.45  &  22.5  &  2.91 \\
Ti I &  4840.87  &  0.90  &  $-$0.45  &  25.1  &  2.97 \\
Ti I &  4981.73  &  0.84  &  0.56  &  85.7  &  2.88 \\
Ti I &  4981.73  &  0.84  &  0.56  &  90.8  &  2.98 \\
Ti I &  4991.07  &  0.84  &  0.44  &  87.1  &  3.02 \\
Ti I &  4991.07  &  0.84  &  0.44  &  96.5  &  3.20 \\
Ti I &  5007.20  &  0.82  &  0.17  &  85.6  &  3.25 \\
Ti I &  5007.20  &  0.82  &  0.17  &  88.2  &  3.29 \\
Ti I &  5016.16  &  0.85  &  $-$0.52  &  26.8  &  3.00 \\
Ti I &  5016.16  &  0.85  &  $-$0.52  &  28.5  &  3.04 \\
Ti I &  5020.02  &  0.84  &  $-$0.36  &  28.0  &  2.86 \\
Ti I &  5024.84  &  0.82  &  $-$0.55  &  26.8  &  3.00 \\
Ti I &  5024.84  &  0.82  &  $-$0.55  &  28.8  &  3.04 \\
Ti I &  5035.90  &  1.46  &  0.26  &  28.1  &  2.95 \\
Ti I &  5035.90  &  1.46  &  0.26  &  35.9  &  3.10 \\
Ti I &  5036.46  &  1.44  &  0.19  &  19.0  &  2.79 \\
Ti I &  5036.46  &  1.44  &  0.19  &  23.5  &  2.90 \\
Ti I &  5039.96  &  0.02  &  $-$1.13  &  55.6  &  3.12 \\
Ti I &  5039.96  &  0.02  &  $-$1.13  &  57.3  &  3.15 \\
Ti I &  5064.65  &  0.05  &  $-$0.94  &  71.1  &  3.20 \\
Ti I &  5210.39  &  0.05  &  $-$0.83  &  65.3  &  2.99 \\
Ti II&  4418.33  &  1.24  &  $-$1.97  &  84.1  &  3.12 \\
Ti II&  4441.73  &  1.18  &  $-$2.41  &  85.3  &  3.51 \\
Ti II&  4464.44  &  1.16  &  $-$1.81  &  97.1  &  3.10 \\
Ti II&  4470.85  &  1.17  &  $-$2.02  &  105.6  &  3.49 \\
Ti II&  4488.34  &  3.12  &  $-$0.82  &  21.2  &  3.03 \\
Ti II&  4529.48  &  1.57  &  $-$2.03  &  77.7  &  3.43 \\
Ti II&  4563.77  &  1.22  &  $-$0.96  &  138.8  &  3.17 \\
Ti II&  4583.40  &  1.16  &  $-$2.92  &  27.2  &  3.01 \\
Ti II&  4589.91  &  1.24  &  $-$1.79  &  90.0  &  3.01 \\
Ti II&  4636.32  &  1.16  &  $-$3.02  &  29.5  &  3.16 \\
Ti II&  4657.20  &  1.24  &  $-$2.24  &  58.6  &  2.94 \\
Ti II&  4708.66  &  1.24  &  $-$2.34  &  67.8  &  3.17 \\
Ti II&  4779.97  &  2.05  &  $-$1.37  &  62.7  &  3.07 \\
Ti II&  4798.53  &  1.08  &  $-$2.68  &  61.0  &  3.21 \\
Ti II&  4805.08  &  2.06  &  $-$1.10  &  84.7  &  3.17 \\
Ti II&  4805.08  &  2.06  &  $-$1.10  &  89.8  &  3.25 \\
Ti II&  4865.61  &  1.12  &  $-$2.81  &  29.8  &  2.89 \\
Ti II&  4865.61  &  1.12  &  $-$2.81  &  45.7  &  3.15 \\
Ti II&  4911.17  &  3.12  &  $-$0.34  &  43.8  &  2.97 \\
Ti II&  5185.90  &  1.89  &  $-$1.49  &  62.7  &  2.97 \\
Ti II&  5226.53  &  1.57  &  $-$1.26  &  124.2  &  3.41 \\
Ti II&  5336.78  &  1.58  &  $-$1.59  &  96.0  &  3.21 \\
Ti II&  5381.02  &  1.57  &  $-$1.92  &  64.3  &  3.03 \\
Ti II&  5418.76  &  1.58  &  $-$2.00  &  46.7  &  2.86 \\
Cr I &  4545.95  &  0.94  &  $-$1.37  &  31.4  &  3.27 \\
Cr I &  4600.75  &  1.00  &  $-$1.26  &  40.1  &  3.38 \\
Cr I &  4626.18  &  0.97  &  $-$1.32  &  32.6  &  3.27 \\
Cr I &  4646.15  &  1.03  &  $-$0.74  &  69.1  &  3.36 \\
Cr I &  4652.15  &  1.00  &  $-$1.03  &  48.3  &  3.28 \\
Cr I &  5206.04  &  0.94  &     0.02  & 113.1  &  3.24 \\
Cr I &  5247.56  &  0.96  &  $-$1.64  &  18.8  &  3.24 \\
Cr I &  5296.69  &  0.98  &  $-$1.36  &  27.5  &  3.18 \\
Cr I &  5348.31  &  1.00  &  $-$1.21  &  35.9  &  3.21 \\
Cr I &  5409.77  &  1.03  &  $-$0.67  &  68.9  &  3.22 \\
Mn I &  4754.04  &  2.28  &  $-$0.09  & Synth  &  2.71 \\
Mn I &  4783.52  &  2.32  &     0.14  & Synth  &  2.79 \\
Fe I &  4443.19  &  2.86  &  $-$1.04  &  71.1  &  5.32 \\
Fe I &  4466.55  &  2.83  &  $-$0.60  &  99.2  &  5.40 \\
Fe I &  4476.01  &  2.85  &  $-$0.82  &  88.6  &  5.42 \\
Fe I &  4484.22  &  3.60  &  $-$0.86  &  36.1  &  5.40 \\
Fe I &  4592.65  &  1.56  &  $-$2.46  &  79.8  &  5.37 \\
Fe I &  4632.91  &  1.61  &  $-$2.91  &  51.8  &  5.41 \\
Fe I &  4643.46  &  3.64  &  $-$1.15  &  21.1  &  5.41 \\
Fe I &  4678.84  &  3.60  &  $-$0.83  &  39.3  &  5.41 \\
Fe I &  4859.74  &  2.88  &  $-$0.76  &  92.7  &  5.41 \\
Fe I &  4859.74  &  2.88  &  $-$0.76  &  92.7  &  5.41 \\
Fe I &  4903.31  &  2.88  &  $-$0.93  &  60.4  &  5.01 \\
Fe I &  4903.31  &  2.88  &  $-$0.93  &  67.9  &  5.13 \\
Fe I &  4918.99  &  2.85  &  $-$0.34  &  99.3  &  5.08 \\
Fe I &  4918.99  &  2.85  &  $-$0.34  &  99.3  &  5.08 \\
Fe I &  4924.77  &  2.28  &  $-$2.11  &  39.9  &  5.17 \\
Fe I &  4924.77  &  2.28  &  $-$2.11  &  47.1  &  5.29 \\
Fe I &  4938.81  &  2.88  &  $-$1.08  &  58.5  &  5.13 \\
Fe I &  4938.81  &  2.88  &  $-$1.08  &  63.6  &  5.21 \\
Fe I &  4966.08  &  3.33  &  $-$0.87  &  41.0  &  5.15 \\
Fe I &  4966.08  &  3.33  &  $-$0.87  &  46.7  &  5.25 \\
Fe I &  4973.10  &  3.96  &  $-$0.95  &  17.6  &  5.45 \\
Fe I &  4994.13  &  0.92  &  $-$2.97  &  97.9  &  5.39 \\
Fe I &  4994.13  &  0.92  &  $-$2.97  &  98.8  &  5.42 \\
Fe I &  5006.11  &  2.83  &  $-$0.61  &  87.1  &  5.07 \\
Fe I &  5006.11  &  2.83  &  $-$0.61  &  88.7  &  5.11 \\
Fe I &  5014.94  &  3.94  &  $-$0.30  &  44.1  &  5.33 \\
Fe I &  5014.94  &  3.94  &  $-$0.30  &  44.9  &  5.34 \\
Fe I &  5041.07  &  0.96  &  $-$3.09  &  96.9  &  5.54 \\
Fe I &  5041.75  &  1.49  &  $-$2.20  &  98.4  &  5.30 \\
Fe I &  5049.82  &  2.28  &  $-$1.35  &  89.4  &  5.22 \\
Fe I &  5049.82  &  2.28  &  $-$1.35  &  93.6  &  5.30 \\
Fe I &  5051.63  &  0.92  &  $-$2.76  &  99.8  &  5.21 \\
Fe I &  5051.63  &  0.92  &  $-$2.76  &  99.8  &  5.21 \\
Fe I &  5068.76  &  2.94  &  $-$1.04  &  77.0  &  5.45 \\
Fe I &  5074.74  &  4.22  &  $-$0.20  &  35.2  &  5.39 \\
Fe I &  5079.22  &  2.20  &  $-$2.10  &  52.2  &  5.25 \\
Fe I &  5127.36  &  0.92  &  $-$3.25  &  83.4  &  5.39 \\
Fe I &  5141.73  &  2.42  &  $-$2.24  &  30.2  &  5.27 \\
Fe I &  5142.92  &  0.96  &  $-$3.08  &  99.2  &  5.55 \\
Fe I &  5192.34  &  3.00  &  $-$0.42  &  95.5  &  5.22 \\
Fe I &  5198.71  &  2.22  &  $-$2.09  &  50.8  &  5.24 \\
Fe I &  5202.33  &  2.18  &  $-$1.87  &  72.5  &  5.31 \\
Fe I &  5216.27  &  1.61  &  $-$2.08  &  85.9  &  5.07 \\
Fe I &  5217.39  &  3.21  &  $-$1.16  &  37.0  &  5.22 \\
Fe I &  5242.49  &  3.63  &  $-$0.97  &  20.4  &  5.17 \\
Fe I &  5263.30  &  3.27  &  $-$0.88  &  49.4  &  5.21 \\
Fe I &  5266.55  &  3.00  &  $-$0.39  &  87.5  &  5.04 \\
Fe I &  5281.79  &  3.04  &  $-$0.83  &  62.6  &  5.11 \\
Fe I &  5283.62  &  3.24  &  $-$0.52  &  77.3  &  5.27 \\
Fe I &  5302.30  &  3.28  &  $-$0.72  &  64.1  &  5.30 \\
Fe I &  5307.36  &  1.61  &  $-$2.91  &  46.1  &  5.26 \\
Fe I &  5324.17  &  3.21  &  $-$0.10  &  93.5  &  5.09 \\
Fe I &  5332.90  &  1.55  &  $-$2.78  &  56.4  &  5.22 \\
Fe I &  5339.93  &  3.27  &  $-$0.72  &  59.0  &  5.20 \\
Fe I &  5364.87  &  4.45  &     0.23  &  34.6  &  5.20 \\
Fe I &  5365.40  &  3.56  &  $-$1.02  &  19.7  &  5.11 \\
Fe I &  5367.46  &  4.42  &     0.44  &  39.2  &  5.04 \\
Fe I &  5369.96  &  4.37  &     0.54  &  53.0  &  5.11 \\
Fe I &  5389.47  &  4.42  &  $-$0.41  &  13.1  &  5.27 \\
Fe I &  5410.91  &  4.47  &     0.40  &  36.5  &  5.08 \\
Fe I &  5415.19  &  4.39  &     0.64  &  52.6  &  5.02 \\
Fe I &  5424.06  &  4.32  &     0.52  &  74.7  &  5.43 \\
Fe I &  5569.61  &  3.42  &  $-$0.54  &  66.3  &  5.29 \\
Fe I &  5572.84  &  3.40  &  $-$0.28  &  71.7  &  5.10 \\
Fe I &  5576.08  &  3.43  &  $-$1.00  &  44.6  &  5.43 \\
Fe I &  5586.75  &  3.37  &  $-$0.14  &  78.4  &  5.03 \\
Fe I &  5615.64  &  3.33  &     0.05  &  98.5  &  5.15 \\
Fe I &  5624.54  &  3.42  &  $-$0.76  &  51.4  &  5.28 \\
Fe I &  5658.81  &  3.40  &  $-$0.79  &  46.7  &  5.21 \\
Fe I &  5662.51  &  4.18  &  $-$0.57  &  15.4  &  5.23 \\
Fe I &  5701.54  &  2.56  &  $-$2.14  &  26.1  &  5.22 \\
Fe I &  6065.48  &  2.61  &  $-$1.41  &  71.0  &  5.26 \\
Fe I &  6136.61  &  2.45  &  $-$1.41  &  82.8  &  5.26 \\
Fe I &  6137.69  &  2.59  &  $-$1.35  &  74.8  &  5.24 \\
Fe I &  6191.55  &  2.43  &  $-$1.42  &  84.4  &  5.27 \\
Fe I &  6200.31  &  2.61  &  $-$2.44  &  20.2  &  5.41 \\
Fe I &  6213.42  &  2.22  &  $-$2.48  &  44.6  &  5.46 \\
Fe I &  6230.72  &  2.56  &  $-$1.28  &  84.0  &  5.28 \\
Fe I &  6252.55  &  2.40  &  $-$1.69  &  68.5  &  5.25 \\
Fe I &  6254.25  &  2.28  &  $-$2.44  &  47.3  &  5.53 \\
Fe I &  6322.68  &  2.59  &  $-$2.47  &  20.7  &  5.43 \\
Fe I &  6335.33  &  2.20  &  $-$2.18  &  48.1  &  5.19 \\
Fe I &  6393.60  &  2.43  &  $-$1.58  &  84.4  &  5.42 \\
Fe I &  6400.00  &  3.60  &  $-$0.29  &  65.5  &  5.20 \\
Fe I &  6411.64  &  3.65  &  $-$0.59  &  47.9  &  5.28 \\
Fe I &  6421.35  &  2.28  &  $-$2.01  &  64.1  &  5.35 \\
Fe I &  6430.84  &  2.18  &  $-$1.95  &  87.8  &  5.54 \\
Fe I &  6592.91  &  2.73  &  $-$1.47  &  52.2  &  5.15 \\
Fe II&  4489.18  &  2.83  &  $-$2.97  &  47.7  &  5.28 \\
Fe II&  4508.28  &  2.86  &  $-$2.58  &  68.5  &  5.26 \\
Fe II&  4515.34  &  2.84  &  $-$2.60  &  75.8  &  5.38 \\
Fe II&  4520.22  &  2.81  &  $-$2.60  &  64.1  &  5.15 \\
Fe II&  4583.84  &  2.81  &  $-$1.93  &  95.3  &  5.05 \\
Fe II&  4620.52  &  2.83  &  $-$3.21  &  36.3  &  5.32 \\
Fe II&  4731.43  &  2.89  &  $-$3.36  &  29.4  &  5.40 \\
Fe II&  4993.35  &  2.81  &  $-$3.67  &  13.6  &  5.19 \\
Fe II&  4993.35  &  2.81  &  $-$3.67  &  17.9  &  5.33 \\
Fe II&  5197.58  &  3.23  &  $-$2.22  &  62.7  &  5.18 \\
Fe II&  5234.63  &  3.22  &  $-$2.18  &  61.9  &  5.11 \\
Fe II&  5276.00  &  3.20  &  $-$2.01  &  70.7  &  5.06 \\
Fe II&  5284.08  &  2.89  &  $-$3.19  &  30.2  &  5.21 \\
Fe II&  5325.55  &  3.22  &  $-$3.16  &  16.1  &  5.22 \\
Fe II&  5534.83  &  3.25  &  $-$2.93  &  27.1  &  5.29 \\
Fe II&  6247.54  &  3.89  &  $-$2.51  &  20.5  &  5.42 \\
Fe II&  6432.68  &  2.89  &  $-$3.71  &  22.2  &  5.51 \\
Fe II&  6456.38  &  3.90  &  $-$2.08  &  37.2  &  5.34 \\
Ni I &  4605.00  &  3.48  &  $-$0.29  &  22.5  &  4.10 \\
Ni I &  4648.65  &  3.42  &  $-$0.16  &  23.7  &  3.92 \\
Ni I &  4855.41  &  3.54  &  0.00  &  20.7  &  3.81 \\
Ni I &  4904.41  &  3.54  &  $-$0.17  &  12.2  &  3.71 \\
Ni I &  4904.41  &  3.54  &  $-$0.17  &  20.3  &  3.97 \\
Ni I &  4980.16  &  3.61  &  $-$0.11  &  22.6  &  4.05 \\
Ni I &  4980.16  &  3.61  &  $-$0.11  &  25.0  &  4.10 \\
Ni I &  5035.37  &  3.63  &  0.29  &  22.7  &  3.67 \\
Ni I &  5035.37  &  3.63  &  0.29  &  34.7  &  3.92 \\
Ni I &  5080.52  &  3.65  &  0.13  &  34.2  &  4.09 \\
Ni I &  5081.11  &  3.85  &  0.30  &  15.6  &  3.71 \\
Ni I &  5084.08  &  3.68  &  0.03  &  16.1  &  3.80 \\
Ni I &  5754.67  &  1.94  &  $-$2.33  &  25.2  &  4.37 \\
Ni I &  6643.64  &  1.68  &  $-$2.30  &  33.4  &  4.14 \\
Ni I &  6767.77  &  1.83  &  $-$2.17  &  34.6  &  4.20 \\
Zn I  & 4810.52  &  4.08  &  $-$0.13  & Synth  &  2.38 \\
Sr II   & 3464.45  &   3.04  &    0.49  & Synth &  1.11   \\
Y II    & 4883.68  &   1.08  &    0.07  & Synth & 0.13 \\
Y II    & 5087.42  &   1.08  & $-$0.17  & Synth & 0.21 \\
Zr II & 4208.99 & 0.71 &$-$0.46 & Synth & 0.71 \\
Zr II & 4613.95 & 0.97 &$-$1.54 & Synth & 0.87 \\
Zr II & 5112.28 & 1.66 &$-$0.59 & Synth & 0.99 \\
Ba II & 5853.69 & 0.60 &$-$0.91 & Synth & 1.80 \\
Ba II & 6141.73 & 0.70 &$-$0.08 & Synth & 1.77 \\
Ba II & 6496.91 & 0.60 &$-$0.38 & Synth & 1.87 \\
La II & 4740.28 & 0.13 &$-$0.94 & Synth & 0.45 \\
La II & 4748.73 & 0.93 &$-$0.54 & Synth & 0.42 \\
La II & 4824.05 & 0.65 &$-$1.19 & Synth & 0.64 \\
La II & 6262.29 & 0.40 &$-$1.24 & Synth & 0.40 \\
La II & 6390.48 & 0.32 &$-$1.45 & Synth & 0.41 \\
Ce II & 4739.51 & 1.25 &$-$0.53 & Synth & 0.67 \\
Ce II & 4739.52 & 0.53 &$-$1.02 & Synth & 0.69 \\
Ce II & 4747.26 & 0.90 &$-$1.79 & Synth & 0.89 \\
Ce II & 4882.46 & 1.53 &0.19 & Synth & 0.70 \\
Pr II & 4744.91 & 0.20 &$-$1.14 & Synth & $-$0.17 \\
Nd II & 5310.04 & 1.14 &$-$0.98 & Synth & 0.83 \\
Nd II & 5311.45 & 0.98 &$-$0.42 & Synth & 0.87 \\
Nd II & 5319.81 & 0.55 &$-$0.14 & Synth & 0.73 \\
Sm II & 4318.93 & 0.28 &$-$0.25 & Synth & 0.04 \\
Sm II & 4434.32 & 0.38 &$-$0.07 & Synth & 0.17 \\
Sm II & 4519.63 & 0.54 &$-$0.35 & Synth & 0.21 \\
Eu II & 6645.06 & 1.38 &0.20 & Synth & $-$0.49 \\
Dy II & 4449.70 & 0.00 &$-$1.03 & Synth & 0.45 \\
Er I & 3682.70 & 0.89 &$-$0.38 & Synth & 0.24 \\
Yb II & 3694.19 & 0.00 &$-$0.82 & Synth & 0.03 \\
Pb I & 3683.46 & 0.97 &$-$0.46 & Synth & 1.97 \\
Pb I & 4057.81 & 1.32 &$-$0.22 & Synth & 2.11 \\
\enddata
\tablenotetext{a}{The C$^{12}$/C$^{13}$ value is a ratio and not a $\log\epsilon$(X) value.}
\label{synthtab}
\end{deluxetable}

\subsection{Stellar Parameters}\label{stellarparams}
Stellar parameters and elemental abundances derived from equivalent
widths were determined using the spectroscopic stellar parameter and
abundance analysis pipeline, Cashcode. Cashcode is written around the
LTE line analysis and spectral synthesis code, MOOG \citep{moog}. It
employs the latest version of MOOG \citep{sobeckmoog}, which properly
treats Rayleigh scattering, an opacity source that is important in
cool giants like HE~0414$-$0343. We used a Kurucz stellar atmosphere
with $\alpha$-enhancement \citep{CastelliKurucz}. Cashcode iterates to
determine the set of stellar parameters which yield a flat relation
between the line abundances and excitation potential, a flat relation
between the line abundances and reduced equivalent width values, and
to ensure that the Fe I and Fe II abundance values are consistent with
each other. See \citet{CASH2} for a detailed description of the
stellar parameter determination technique.

\begin{figure}[!th]
\begin{center}
\includegraphics[width=0.5\textwidth]{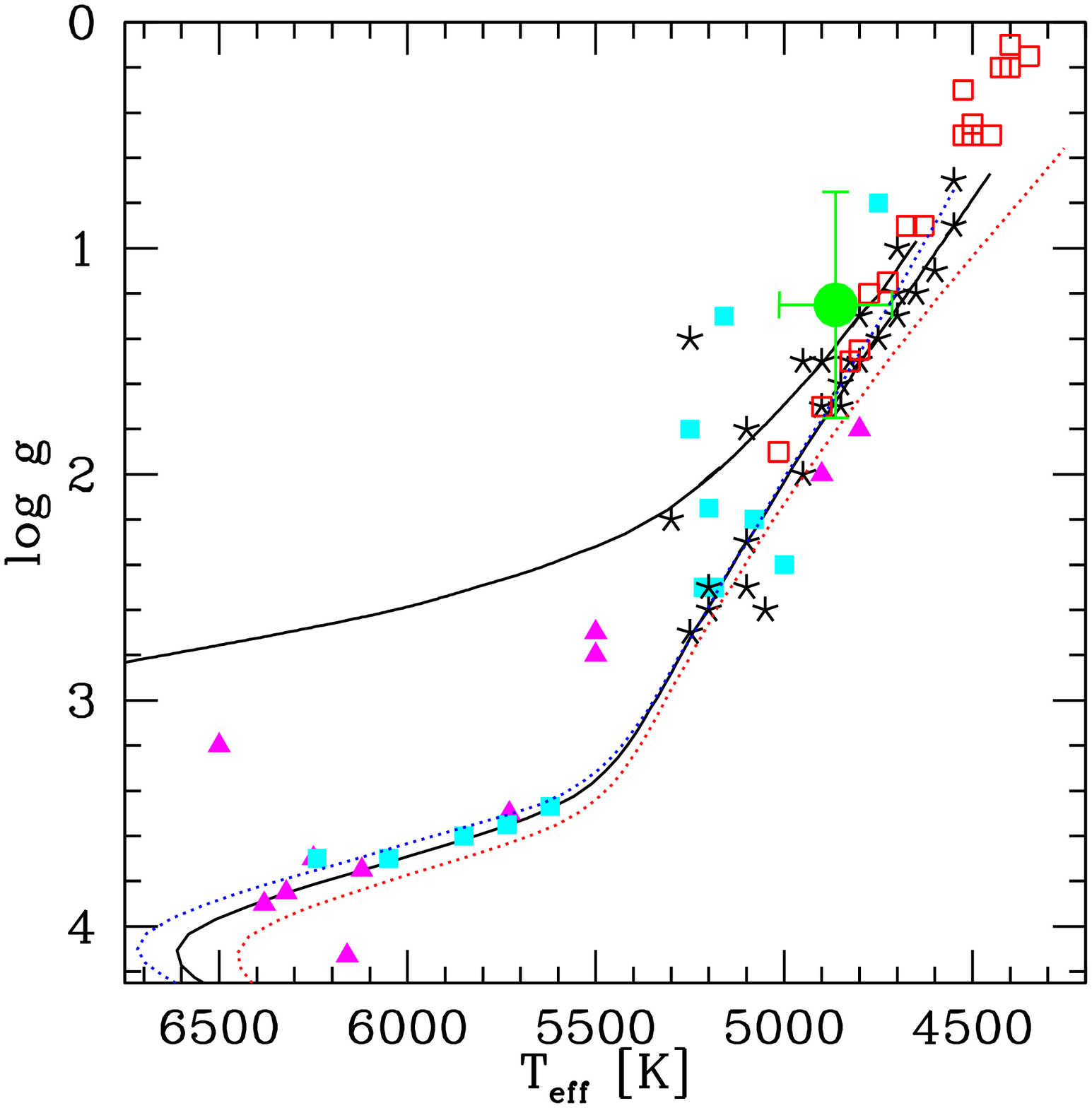}
\caption{HR diagram with HE~0414$-$0343 (green filled circle) plotted
  against the \citet{CASH2} sample (red open squares), the
  \citet{cayrel2004} sample (black stars), and selected s-process
  (cyan filled squares) and r/s-process (magenta filled triangles)
  enhanced stars as collated by \citet{placco2013}.  Overplotted are
  the Yale-Yonsei isochrones \citep{Y2_iso,green} for 12 Gyr, at
  $\mbox{[Fe/H]} =-2.0$ (red line), $-2.5$ (black line), and $-3.0$ (blue
  line, as well as a horizontal-branch mass track from
  \citet{CassisiHB}.\label{hrd}}
 \end{center}
\end{figure}

We determined the spectroscopic stellar parameters using equivalent
width measurements of 88 Fe I and 18 Fe II lines resulting in
T$_{eff}=4660$\,K, $\log g = 0.75$, $\xi = 2.05$\,km\,s$^{-1}$ and
$\mbox{[Fe/H]} = -2.38$. The resonance lines of Fe I were excluded in
this analysis, as they are strong enough that they are often near
the flat portion of the curve of growth. It should be noted
  that photometric temperatures are difficult to determine in CEMP
  stars because the molecular carbon bands interfere with the
  different photometric band passes in varying degree, thus making the
  photometric temperatures unreliable. Spectroscopic temperatures are
  often several hundred degrees cooler than photometrically-derived
  values. We thus adjusted the stellar parameters to make them more
  closely reflect photometric stellar parameters, following the
  procedure outlined in \citet{frebelcalib}. These values are
  T$_{eff} = 4863$\,K, $\log g = 1.25$, $\xi = 2.20$\,km\,s$^{-1}$ and
  $\mbox{[Fe/H]} = -2.24$, which we adopt. In Figure~\ref{hrd}, we
  show the derived effective temperature and surface gravity for
  HE~0414$-$0343 plotted together with 12\,Gyr Yale-Yonsei isochrones
  \citep{Y2_iso,green} for $\mbox{[Fe/H]} =-2.0$, $-2.5$, and $-3.0$
  as well as a \citet{CassisiHB} horizontal branch track.

We determined the random uncertainty in the surface gravity by
allowing the Fe~I and Fe~II values to vary until they no longer agree
within the uncertainty of Fe I, which is 0.12\,dex.  Since
HE~0414$-$0343 is on the giant branch, uncertainties in effective
temperature at the $\sim150$ K level lead to changes in the surface
gravity of $\sim0.5$\,dex.  We conservatively adopt this as our
$\sigma_{log g}$ uncertainty. We determined the standard error of the
mean Fe I abundance to be $\sim0.01$\,dex; however, we adopt the
scatter of the individual Fe line abundances as our final [Fe/H]
uncertainty ($\sim 0.12$\,dex) as the standard error is quite low and
does not account for uncertainties in the continuum placement for each
measured line, which is especially difficult in a CEMP star.

\section{Chemical Abundance Analysis}\label{abundances}
The equivalent widths were used to determine abundances for seven
different elements as well as the stellar parameters.  Spectral
syntheses of blended lines or lines with hyper-fine structure were
performed manually, given the often severe blending due to the
C-enhancement in the star. Table~\ref{eqwabunds} lists the
abundances. Solar abundances of \citet{asplund09} were used to
calculate [X/H] and [X/Fe] values. Further details on the elemental
abundances are given below.

Table~\ref{err} lists our abundance uncertainties. We determined the
systematic uncertainties by varying the stellar parameters of
effective temperature, log g, and microturbulence in the model
atmosphere used in proportion to the uncertainty of each parameter.
The abundances were then recalculated with the new model atmospheres
either by averaging the individual line abundances determined from
equivalent width or by re-fitting a synthetic spectrum. The random
uncertainty for each abundance determined via equivalent width was
taken as the standard deviation of the individual line abundances. We
use the standard deviation rather than the standard error because it
better reflects that our abundances are hampered by the presence of
molecular C. For the abundances derived via spectral synthesis, we
used the original model atmosphere and varied the abundance of the
synthetic spectrum until the fit no longer matched the input
spectrum. In the cases where there were fewer than 5 measurements, we
used a special treatment for low number statistics. We adopt a minimum
standard error of 0.12\,dex and use this for all measurements with
formally calculated smaller values. For elements with just one
  available line, we conservatively assigned an 0.3\,dex uncertainty.
The systematic uncertainties based on the stellar parameters and the
random uncertainties were then added in quadrature to determine the
total error value.

\begin{deluxetable}{lrrrrrl}
\tabletypesize{\scriptsize}
  \tablecolumns{7}
  \tablewidth{0pc}
  \tablecaption{Elemental Abundances of HE~0414$-$0343\label{eqwabunds}}
  \tablehead{\colhead{Element} & \colhead{$\log\epsilon$(X)} & \colhead{$\sigma$} 
  & \colhead{[X/Fe]} & \colhead{n} & \colhead{$\log\epsilon$(X)$_{\odot}$} &\colhead{Method}}
\startdata
C (C$_{2}$) &    7.63 &  0.30 &    1.44 &   1 & 8.43 &   Synth \\
Mg I        &    5.82 &  0.12 &    0.46 &   5 & 7.60 &   EW    \\       
Ca I        &    4.57 &  0.13 &    0.47 &  14 & 6.34 &   EW    \\      
Sc II       &    1.14 &  0.12 &    0.23 &  10 & 3.15 &   EW    \\      
Ti I        &    2.99 &  0.17 &    0.28 &  30 & 4.95 &   EW    \\      
Ti II       &    3.14 &  0.18 &    0.43 &  24 & 4.95 &   EW    \\      
Cr I        &    3.26 &  0.12 & $-$0.14 &  10 & 5.64 &   EW    \\      
Mn I        &    2.75 &  0.12 & $-$0.44 &   2 & 5.43 &   Synth \\      
Fe I        &    5.26 &  0.12 & \nodata &  88 & 7.50 &   EW    \\
Fe II       &    5.26 &  0.13 & \nodata &  18 & 7.50 &   EW    \\
Ni I        &    3.97 &  0.20 & $-$0.01 &  15 & 6.22 &   EW    \\
Zn I        &    2.38 &  0.20 &    0.06 &   1 & 4.56 &   Synth \\  
Sr II       &    1.11 &  0.30 &    0.48 &   1 & 2.87 &   Synth \\
Y  II       &    0.17 &  0.12 &    0.20 &   2 & 2.21 &   Synth \\ 
Zr II       &    0.85 &  0.12 &    0.51 &   3 & 2.58 &   Synth \\
Ba II       &    1.81 &  0.12 &    1.87 &   3 & 2.18 &   Synth \\
La II       &    0.46 &  0.12 &    1.60 &   5 & 1.10 &   Synth \\ 
Ce II       &    0.74 &  0.12 &    1.40 &   3 & 1.58 &   Synth \\    
Pr II       & $-$0.17 &  0.30 &    1.35 &   1 & 0.72 &   Synth \\
Nd II       &    0.81 &  0.12 &    1.63 &   3 & 1.42 &   Synth \\
Sm II       &    0.14 &  0.12 &    1.42 &   3 & 0.96 &   Synth \\     
Eu II       & $-$0.49 &  0.30 &    1.23 &   1 & 0.52 &   Synth \\  
Dy II       &    0.45 &  0.30 &    1.59 &   1 & 1.10 &   Synth \\
Er I        &    0.24 &  0.30 &    1.56 &   1 & 0.92 &   Synth \\
Yb II       &    0.03 &  0.30 &    1.43 &   1 & 0.84 &   Synth \\ 
Pb I        &    2.04 &  0.12 &    2.53 &   2 & 1.75 &   Synth \\ 
\enddata
\end{deluxetable}

\begin{deluxetable}{lrrrrr}
\tabletypesize{\scriptsize}
\tablecolumns{3}
\tablewidth{0pc}
\tablecaption{\label{err} Abundance Uncertainties}
\tablehead{\colhead{Element}&\colhead{Random}&
\colhead{$\Delta$\mbox{T$_{\rm eff}$}}&\colhead{$\Delta\log g$}&
\colhead{$\Delta v_{micr}$}&\colhead{Total\tablenotemark{a}}\\
\colhead{}&\colhead{Unc.}&\colhead{+150\,K}&
\colhead{$+$0.5\,dex}&\colhead{+0.3\,km\,s$^{-1}$}&\colhead{Unc.}}
\startdata
C (C$_{2}$) & 0.30 & 0.25  & $-$0.10  &    0.00  & 0.40 \\
Mg I        & 0.12 & 0.12  & $-$0.12  & $-$0.09  & 0.23 \\
Ca I        & 0.13 & 0.12  & $-$0.05  & $-$0.07  & 0.20 \\
Sc II       & 0.12 & 0.06  &    0.16  & $-$0.05  & 0.21 \\
Ti I        & 0.17 & 0.21  & $-$0.07  & $-$0.05  & 0.28 \\
Ti II       & 0.18 & 0.05  &    0.15  & $-$0.09  & 0.26 \\
Cr I        & 0.12 & 0.19  & $-$0.07  & $-$0.05  & 0.24 \\
Mn I        & 0.12 & 0.16  & $-$0.06  & $-$0.04  & 0.21 \\
Fe I        & 0.12 & 0.17  & $-$0.06  & $-$0.08  & 0.23 \\
Fe II       & 0.13 & 0.00  &    0.17  & $-$0.04  & 0.22 \\
Ni I        & 0.20 & 0.14  & $-$0.04  & $-$0.02  & 0.25 \\
Ba II       & 0.12 & 0.10  &    0.10  & $-$0.25  & 0.31 \\
\enddata
\tablenotetext{a}{Obtained by adding all uncertainties in quadrature.}
\end{deluxetable}

\subsection{Carbon, Nitrogen, and Oxygen}
\begin{figure}[!t]
\includegraphics[width=0.5\textwidth]{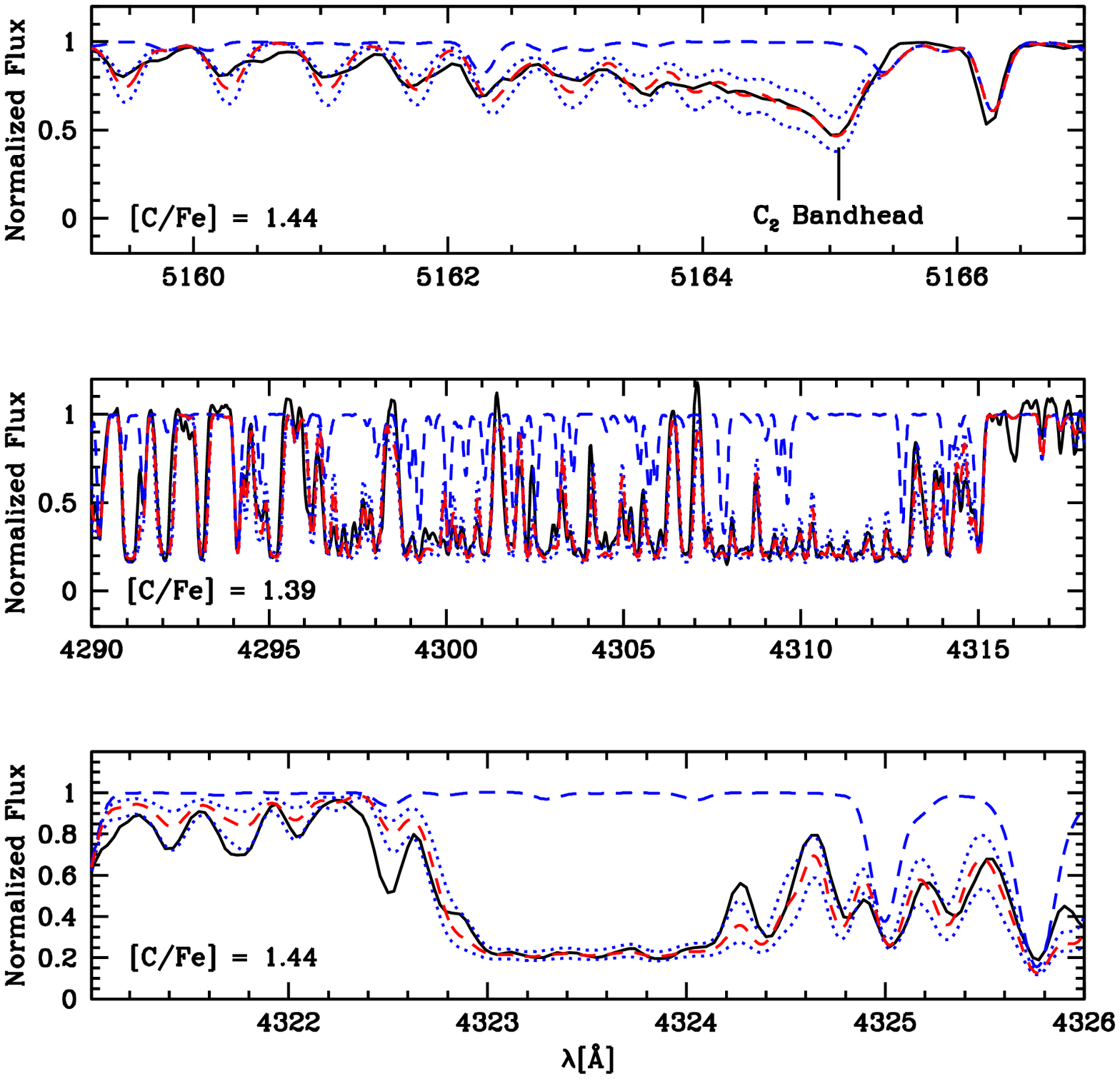}\\
\includegraphics[width=0.48\textwidth,clip=true, bbllx=20, bblly=345,
  bburx=565, bbury=740]{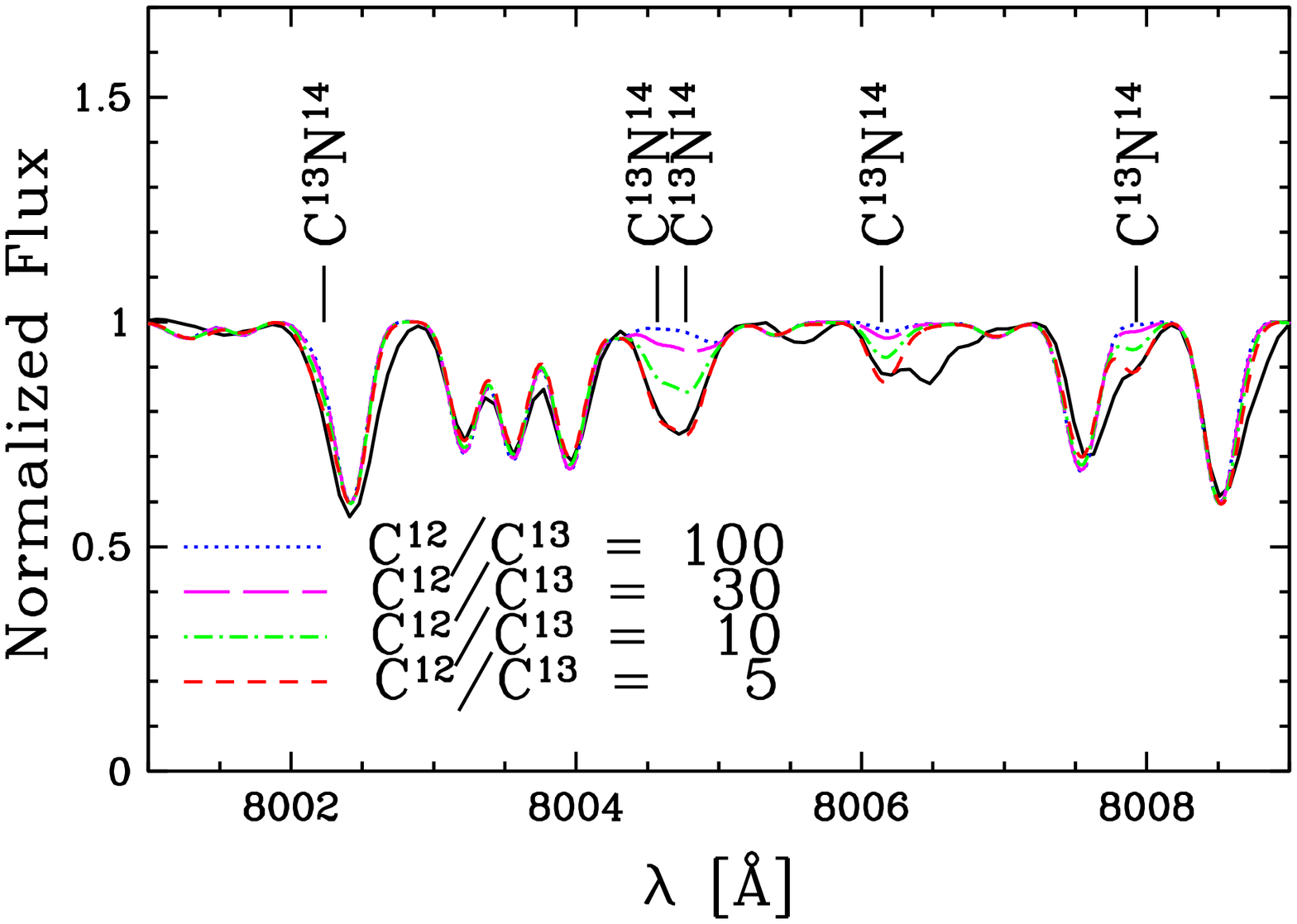}
\caption{\label{he0414_c}Top: $\lambda$5165 C$_{2}$ feature
  (solid black line), along with the best fit
  synthetic abundance (red dashed line) with $\mbox{[C/Fe]}= 1.44$, C
  abundances changed a factor of two above and below the best fit
  abundance (blue dotted line), and a synthetic spectrum for which no
  C is present (blue dashed line). Two middle panels: CH G-band and
  the 4323\,{\AA} CH feature, yielding $\mbox{[C/Fe]} = 1.39$ and
  1.44, respectively, confirming the $\mbox{[C/Fe]}= 1.44$ abundance
  ratio adopted from the $\lambda$5165 C$_{2}$ feature. Bottom:
  $\lambda$8005 CN feature in HE~0414$-$0343 (solid black line) from
  which the $^{12}$C/$^{13}$C ratio was derived, along with synthetic
  spectra of varying $^{12}$C/$^{13}$C ratios where $^{12}$C/$^{13}$C$
  = 5$ (red dashed line), $^{12}$C/$^{13}$C $= 10$ (green
  dotted-dashed line), $^{12}$C/$^{13}$C $= 30$ (magenta long dashed
  line), and $^{12}$C/$^{13}$C $= 100$ (blue dotted line).  }
\end{figure}

\begin{figure*}[!t]
\includegraphics[width=0.9\textwidth]{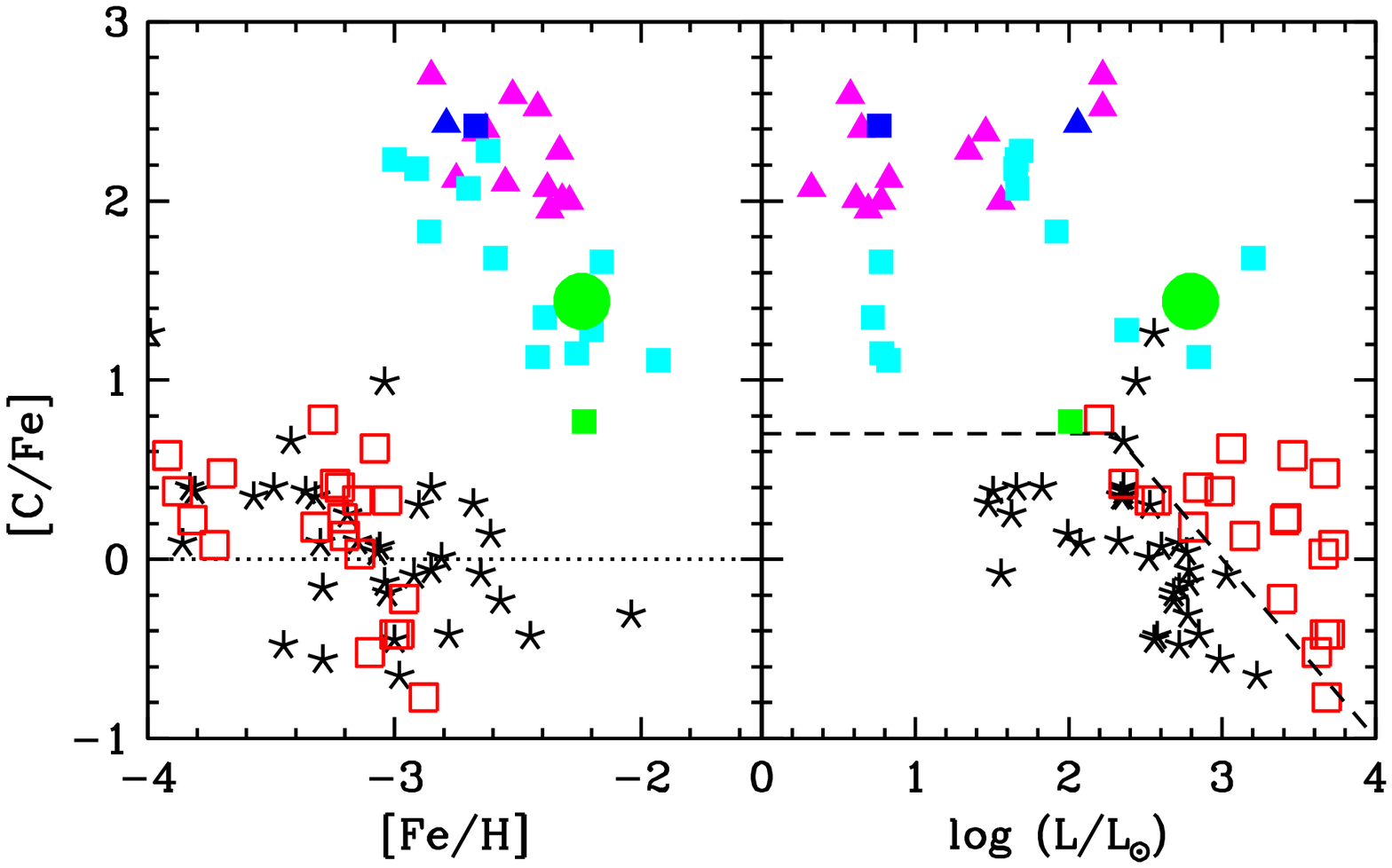}
\caption{Left: [C/Fe] plotted against [Fe/H], where the green filled
  circle represents HE~0414$-$0343, the red open squares are stars
  from \citet{CASH2}, and the black stars represent \citet{cayrel2004}
  data. The magenta triangles correspond to CEMP-r/s stars, the cyan
  squares represent CEMP-s stars, and the blue stars are those from
  \citet{placco2013} with the triangle and the square representing the
  CEMP-r/s and CEMP-s stars respectively. Right: [C/Fe] abundance
  against luminosity, with the black dashed line representing the CEMP
  cut-off as prescribed by \citet{aoki_cemp_2007}. \label{clum}}
\end{figure*}

There are several strong molecular C features in the spectrum of
HE~0414$-$0343, as seen in Figure~\ref{he0414_c}. In fact, the CH
features at 4313\,{\AA} (the G-band) and another smaller feature at
4323\,{\AA} are essentially saturated. The bandhead of the
$\lambda$5165 C$_{2}$ feature is not saturated and thus was used to
determine the C abundance via spectral synthesis. We find
$\mbox{[C/Fe]} =1.44$. However, attempting to measure the G-band and
$\lambda$4323 features yield $\mbox{[C/Fe]} =1.39$ and 1.44,
respectively. These are consistent with the adopted abundance
ratio. The CH and CN linelists (B. Plez 2006, private communication)
are described in \citet{he1523}, with further description of the CN
linelist available in \citet{Hilletal:2002}. The linelist used to
determine the adopted C abundance from the C$_{2}$ feature is based on
the \citet{kurucz_lines} linelist. In Figure~\ref{he0414_c} we show
the best fit abundances derived from the $\lambda$5165 C$_{2}$
feature, as well as those from the CH G-band and the 4323\,{\AA}
features. The large [C/Fe] ratio of HE~0414$-$0343 categorizes it as a
CEMP star using both the \citet{bc05} and \citet{aoki_cemp_2007}
definitions, as demonstrated in both panels of Figure~\ref{clum}.

In order to obtain an accurate C abundance, we determined the
$^{12}$C/$^{13}$C ratio from the $^{12}$CN and $^{13}$CN features near
$\lambda$8005 seen in Figure~\ref{he0414_c}. For analysis of the CN
features, the N abundance was used as a free parameter. HE~0414$-$0343
is an evolved red giant star so the $^{12}$C/$^{13}$C ratio should be
low due to the mixing of CN-cycled material into its atmosphere with
much of the $^{12}$C converted to $^{13}$C. Indeed, we find
$^{12}$C/$^{13}$C = 5, which was adopted uniformly throughout
subsequent spectrum syntheses that required the C abundance.

We also measured the CN bandhead near 4215\,{\AA} and the CH feature
near 4237\,{\AA}, to confirm our $^{12}$C/$^{13}$C result. From both
features, we derive a ratio of $\sim$5-10. Using a new C$_{2}$
linelist from \citet{c2ll} and \citet{c2iso}, we determined a [C/Fe]
ratio of $\sim 1.4$ using several features across the spectrum, including
one near the $\lambda$4736 C$_{2}$ bandhead, which confirms our C
abundance. 

The N abundance can be determined from diatomic CN and monatomic
NH. While it is desirable to determine the N abundance independently,
we were unable to derive an abundance from the $\lambda$3360 NH
molecular feature given its blue wavelength and the corresponding low
S/N ratio. The N abundance in the CN molecule was treated as a free
parameter and also yielded no useful N abundance given the overwhelming
amount of carbon in these features.

The O abundance is difficult to measure in metal-poor stars due to the
paucity of lines. The O features principally available in
HE~0414$-$0343 are the [O I] forbidden line at 6300\,{\AA} and the O
triplet. The forbidden line is weak and is difficult to discern from
molecular C in our spectrum.  We measured the equivalent widths of the
three lines of the O triplet near 7772\,{\AA} in this star; however,
no reliable abundance could be determined since these lines all give
varying abundances.

\subsection{Light elements: Z $\leq$ 30}
\begin{figure*}[!th]
\includegraphics[width=1\textwidth]{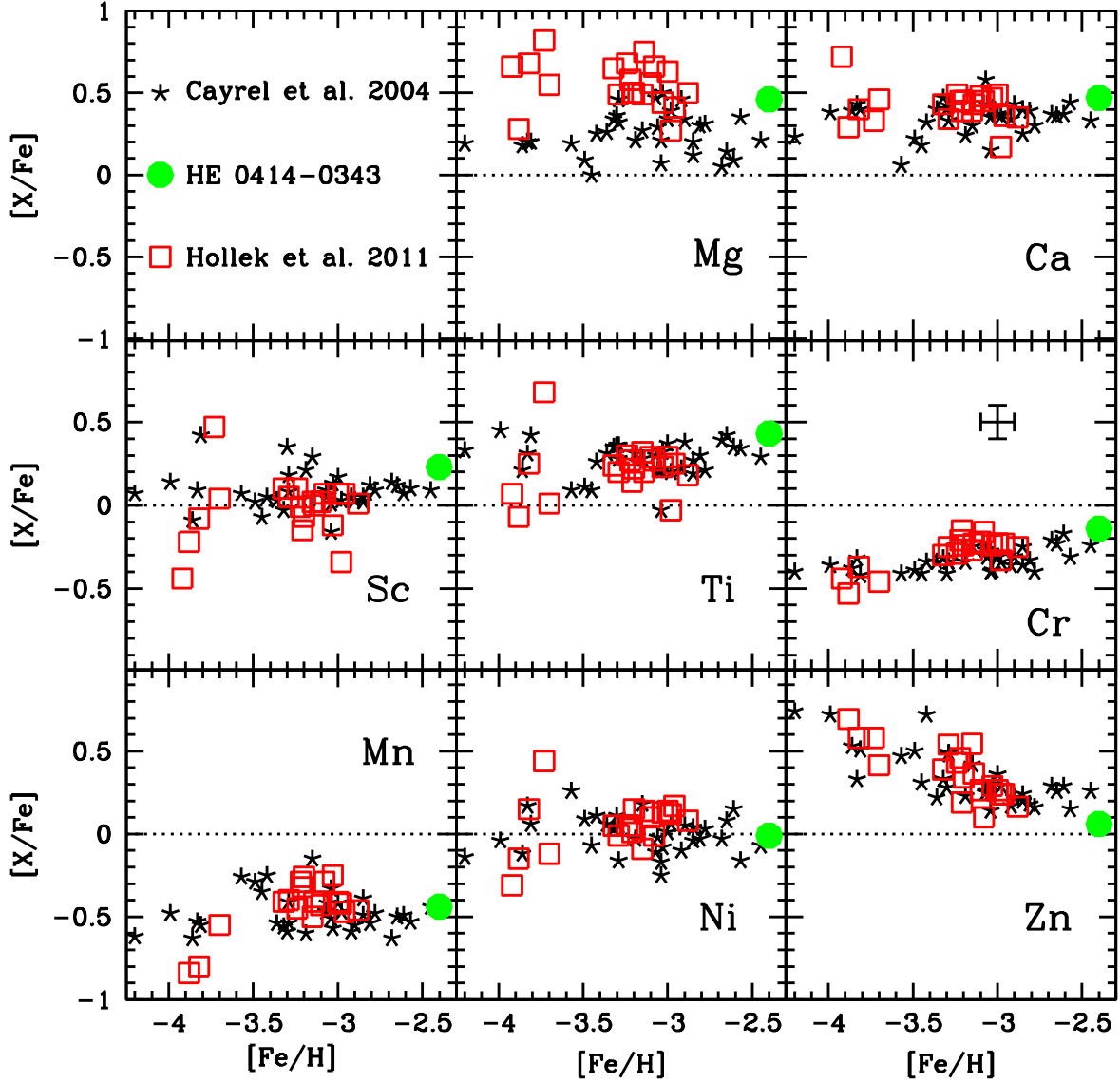}
\caption{[X/Fe] abundance ratios plotted against [Fe/H] for elements
  up to Zn measured in the spectrum of HE~0414$-$0343 (green filled
  circle) compared with the \citet{cayrel2004} (black stars) and the
  \citet{CASH2} abundances (red open squares). The black dotted line
  represents the solar abundance ratio. The Zn and Mn abundances were
  determined via spectral synthesis while the rest were determined
  from equivalent width measurements. In the Cr panel, we include a
  typical error bar.
\label{lea}}
\end{figure*}

Figure~\ref{lea} shows the abundances of HE~0414$-$0343 together with
those of the \citet{cayrel2004} and \citet{CASH2} studies. We now
discuss individual elements and abundance results.

We do not detect the $\lambda$6707 \ion{Li}{1} doublet in our spectra.
The Li abundance in evolved stars is expected to be low. During the
first and second dredge-ups, the Li surface abundance is greatly
diluted, as Li-poor material is brought to the surface.  Our
non-detection of Li in HE~0414-0343 is consistent with this standard
scenario. We derived a 3$\sigma$ upper limit of A(Li) = $\log
\epsilon$(Li)\footnote{ $\log\epsilon(X) = \log(N_{\rm X}/N_{\rm
    H})+12$} = 1.08.  We also do not derive an abundance from the Na~D
features near 5890\,{\AA}, as they are heavily blended with
interstellar Na. We did not calculate the Al or Si abundance because
the corresponding lines are located exclusively in the blue portion of
the spectrum between 3900 and 4105\,{\AA} and are heavily blended with
CH lines.  Additionally, the $\lambda$4102 Si line is located in the
pseudocontinuum of the nearby H$\delta$ line. The Co I lines available
are $\lambda$3502, $\lambda$3995, and $\lambda$4020, all of which are
well within a C-rich region and give spurious results. Thus, we do not
present an abundance for Co.

With the exceptions of C, Mn, and Zn, all Z$\leq$30 elemental
abundances were derived from equivalent width measurements.  The
abundances derived for the light elements are all consistent with what
is expected from the typical metal-poor halo star.  We find
enhancement in the $\alpha$-elements of Mg, Ca, and Ti, with
$\mbox{[$\alpha$/Fe]} = 0.49$. For the purposes of synthesis,
plotting, and determining the [$\alpha$/Fe] ratio, we adopt the Ti II
abundance as the Ti abundance, as Ti I and Ti II differ by 0.16\,dex.
Using only those Ti I lines with newly-determined gf values from
\citet{lawlerti1} and Ti II lines from \citet{lawlerti2}, the
abundance discrepancy shrinks to 0.12\,dex. This agreement supports our
Fe-derived $\log g$ value.  We find depletion in the Fe-peak elements
of Cr, Mn, and Ni and enhancement of Sc and Zn, all of which is
consistent with the \citet{CASH2} and \citet{cayrel2004} studies.  For
Mn and Zn, we derived abundances from synthetic spectrum
computations. We obtain $\mbox{[Mn/Fe]} = -0.48$ from the
$\lambda$4754 line and $\mbox{[Zn/Fe]} = 0.06$ from the $\lambda$4810
line.

\subsection{Neutron-Capture Elements}

All abundances for neutron-capture elements discussed in this section
were determined with spectrum synthesis due to blending with
other species or hyperfine structure. We discuss each element in
detail below.

The Sr abundance was obtained from the $\lambda$3464 Sr II line,
yielding $\mbox{[Sr/Fe]} = 0.48$, though it is in a region of low S/N,
this line has the cleanest spectral region. The typical Sr lines used
in abundance analyses of metal-poor stars, $\lambda$4215 and
$\lambda$4077, both suffer from extensive blending with molecular C
features. In fact, the $\lambda$4215 line is blended with so much CN
that we were able to use that feature to measure the $^{12}$C/$^{13}$C
ratio, but could not determine a Sr abundance.

The Y abundance was determined from the $\lambda$4883 line, which is
blended with CN accounted for in the linelist, and the unblended
$\lambda$5087 line. We derived $\mbox{[Y/Fe]} = 0.16$ and 0.24 from
these lines, respectively and adopted the average abundance,
$\mbox{[Y/Fe]} = 0.20$.

The Zr abundance is based on the $\lambda$4208, $\lambda$4613, and
$\lambda$5112 lines. Though the $\lambda$4208 feature resides within
the same CN bandhead as the Sr\,II $\lambda$4215 line, it is strong
and unblended enough to allow an abundance measurement. We adopt the
mean value of these three lines, $\mbox{[Zr/Fe]} = 0.52$.

\begin{figure*}[!th]
\includegraphics[width=0.9\textwidth]{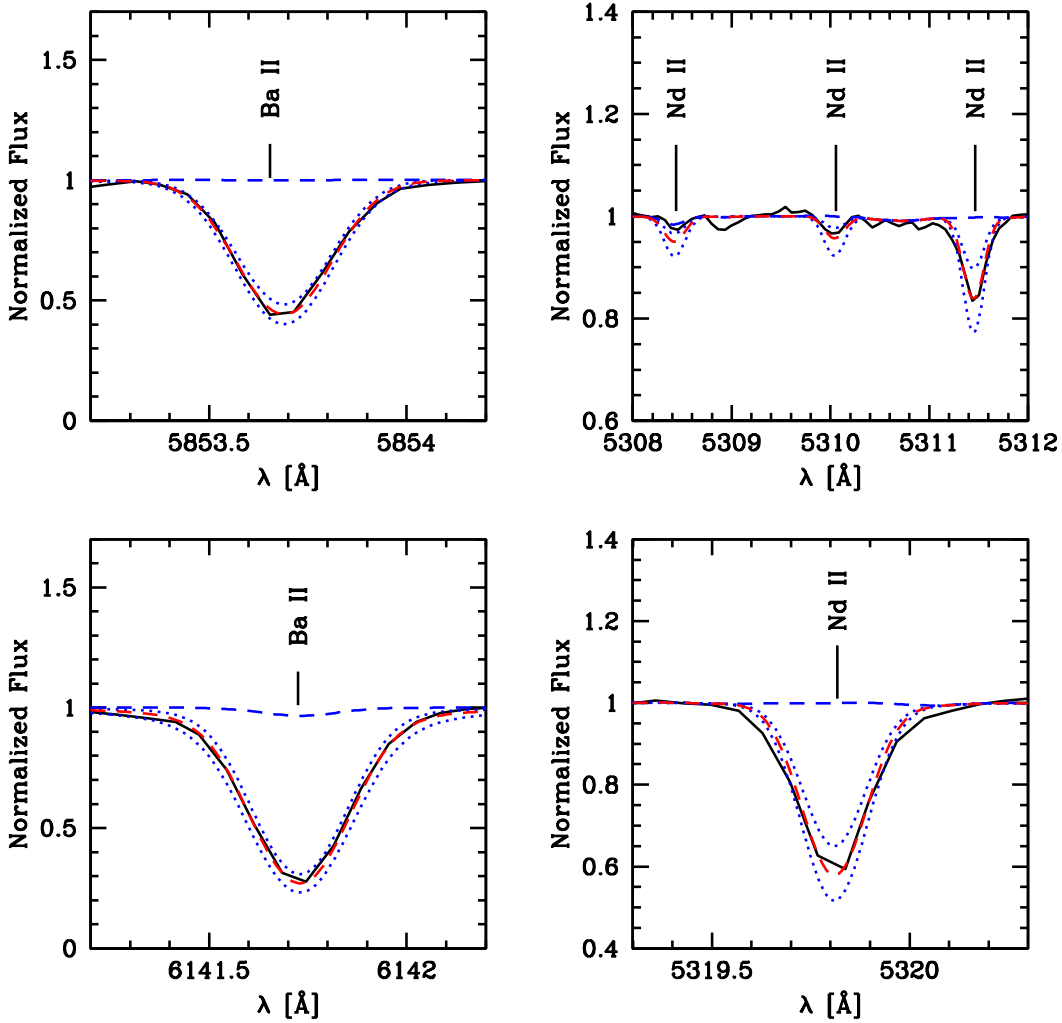}
\caption{Ba features (left panels) in the HE~0414$-$0343 spectrum
  (solid black line), along with the best fit abundance (red dashed
  line), Ba abundances changed a factor of two above and below the
  best fit abundance (blue dotted line), and a synthetic spectrum for
  which no Ba is present (blue dashed line). The top left panel shows
  the $\lambda$4554 line, the middle left panel shows the
  $\lambda$5853 lines, and the bottom left panel shows the
  $\lambda$6141 line. The Nd features (right panels) are plotted in
  the same scheme as Ba.  The top right panel shows the $\lambda$5310
  and $\lambda$5311 lines, the middle right panel shows the
  $\lambda$5319 line, and the bottom right panel shows the
  $\lambda$5293 line. \label{band}}
\end{figure*}

The Ba abundance was determined from the $\lambda$5854, $\lambda$6142,
and $\lambda$6494 lines shown in Figure~\ref{band}. Though available,
the $\lambda$4554 line is on the damping portion of the curve of
growth. The Ba lines in HE~0414$-$0343 are mostly free of molecular C
contamination. We adopted the mean abundance ratio, $\mbox{[Ba/Fe]} =
1.87$.

The La abundance was derived from five clean lines: $\lambda$4740,
$\lambda$4748, $\lambda$4824, $\lambda$6262, and $\lambda$6390. We
adopt the mean abundance, $\mbox{[La/Fe]} = 1.48$. The Ce abundance
was derived from four lines: $\lambda$4739.51, $\lambda$4739.52,
$\lambda$4747, and $\lambda$4882. The $\lambda$4739.51 and
$\lambda$4739.52 lines are heavily blended; however, there is no other
strong feature in the region of these lines, thus we derived their
abundance simultaneously, as it is impossible to determine if either
line yields a different abundance. We adopt $\mbox{[Ce/Fe]} = 1.42$
based on three lines, as we treat the abundances of the Ce $\lambda$4739.51
and Ce $\lambda$4739.52 lines as a single abundance.

The Nd abundance was obtained from three clean lines in the red
portion of the spectrum, shown in Figure~\ref{band}: $\lambda$5310,
$\lambda$5311, and $\lambda$5319. The mean abundance of these
features, which are all in good agreement with each other,
$\mbox{[Nd/Fe]} = 1.63$.

The Sm abundance was derived from $\lambda$4318, $\lambda$4434,
and $\lambda$4519. The $\lambda$4318 line is in a C-rich region, while
the $\lambda$4519 line is blended with C, thus the C abundance was
treated as a free parameter to best fit the observed spectrum in these
syntheses. Despite the blends, we were still able to derive
abundances that are in very good agreement with each other. We also
evaluated $\lambda$4433 and $\lambda$4687, both of which are severely
blended with C (and also Fe in the case of $\lambda$4687) and
determined upper limits for both features, that are very close to
the final value.  We obtain an average value of the abundance derived
from the three measured lines: $\mbox{[Sm/Fe]} = 1.42$.

\begin{figure*}[!H]
\includegraphics[width=0.85\textwidth]{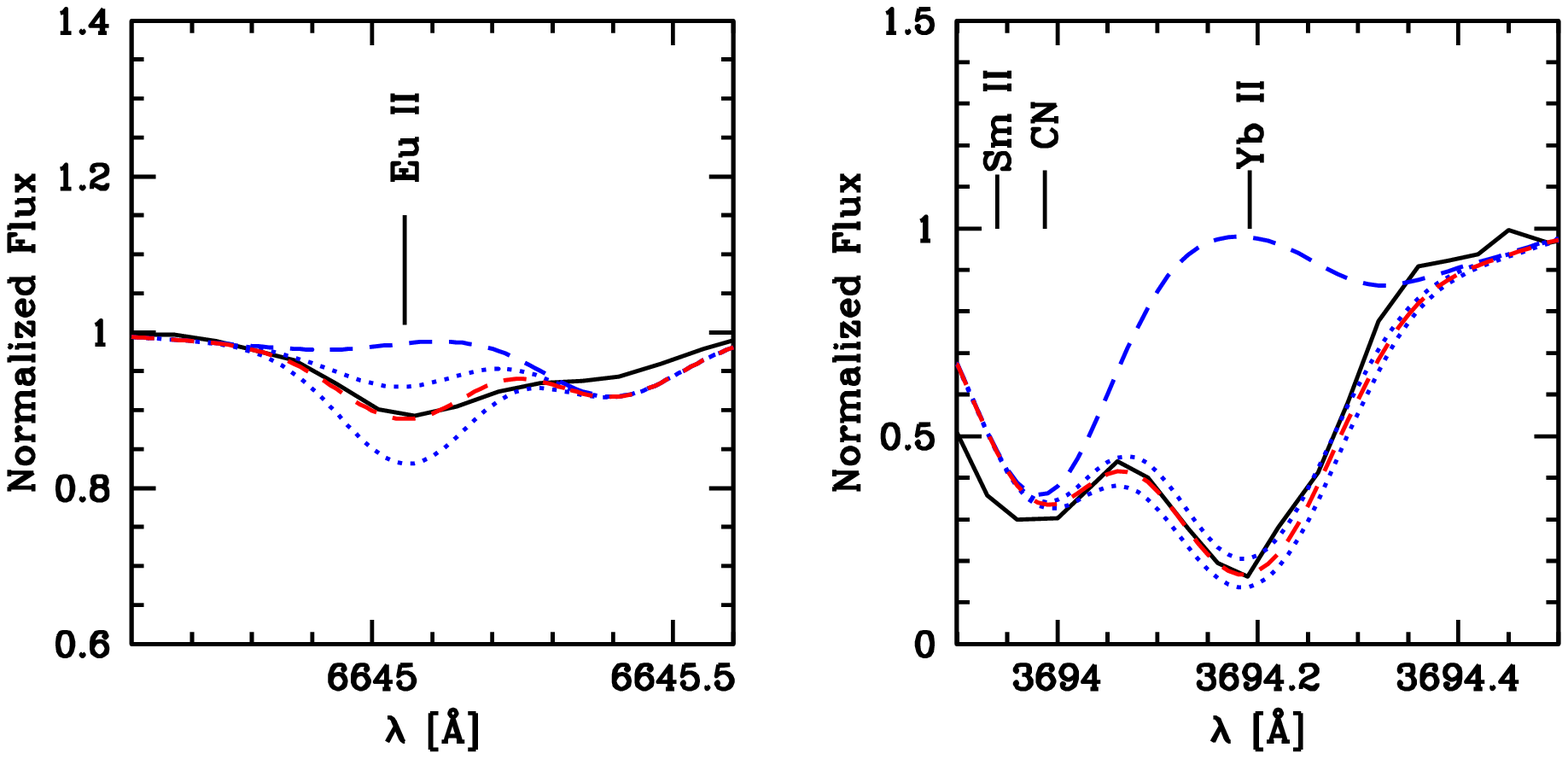}\\
\includegraphics[width=0.85\textwidth]{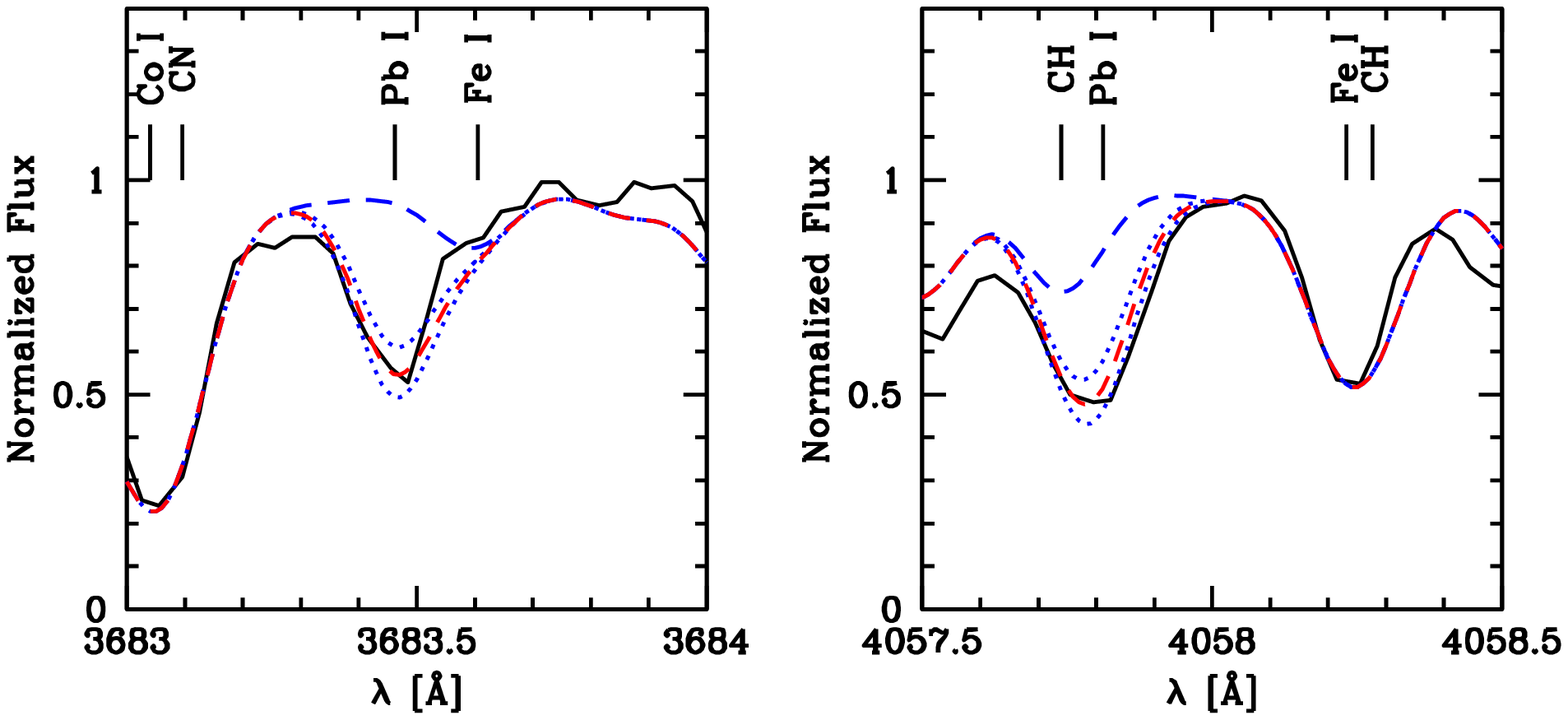}
\caption{ \label{euby} Top: Eu feature at $\lambda$6645\,{\AA} (left
  panel) in the HE~0414$-$0343 spectrum (solid black line), along with
  the best-fit abundance (red dashed line), Eu abundances changed a
  factor of two above and below the best fit synthetic abundance (blue
  dotted line) of $\mbox{[Eu/Fe]} = 1.23$, and a synthetic spectrum
  for which no Eu is present (blue dashed line).  The Yb feature
  located at 3694\,{\AA} (right panel) is plotted in the same way as
  Eu and yields a [Yb/Fe] ratio of 1.43.  Bottom: Pb features at
  3683\,{\AA} (left panel) and 4057\,{\AA} (right panel) in the
  HE~0414$-$0343 spectrum (solid black line), along with the best fit
  abundance (red dashed line), Pb abundances changed a factor of two
  above and below the best fit synthetic abundance (blue dotted line),
  and a synthetic spectrum for which no Pb is present (blue dashed
  line). The $\lambda$3683 and $\lambda$4057 lines yield [Pb/Fe]
  abundance ratios of 2.53 and 2.60, respectively.  }
\end{figure*}

The Eu abundance was determined solely from the $\lambda$6645 line,
yielding $\mbox{[Eu/Fe]} = 1.23$. As seen in the left panel of
Figure~\ref{euby}, this line is blended with CN. Typically, the
$\lambda$4129 line is used; however, the region surrounding that line
is heavily blended with molecular CH features, as well as other
neutron-capture elements, preventing any abundance measurements.

The Dy abundance was determined from the $\lambda$4449 feature, which
is heavily blended with molecular C.  We had to increase the C
abundance to match the features in this region.  We therefore present
the Dy abundance of $\mbox{[Dy/Fe]} = 1.59$ with a larger uncertainty
of 0.3\,dex. However, we do include it in our analysis for diagnostic
purposes with regard to the nature of the nucleosynthetic origins of
HE~0414$-$0343. The magnitude of its abundance is similar to other
abundances of neutron-capture elements that were easier to determine.

The Er abundance was derived from the $\lambda3682$ line yielding
$\mbox{[Er/Fe]} = 1.56$. Given the low S/N and CH molecular features
in the region, this abundance has a large uncertainty of 0.30\,dex;
however, similar to Dy, the magnitude of the derived abundance for Er
is consistent with other neutron-capture elements in the star.

The Yb abundance was determined from the $\lambda$3694 line in the
blue portion of the spectrum, as shown in the right panel of
Figure~\ref{euby}.  It resides on the red side of a blended feature
which includes neutron-capture and Fe-peak elements, and has
isotopic splitting.  We derive $\mbox{[Yb/Fe]} = 1.43$.

Finally, we obtained the Pb abundance from the $\lambda$3683 and
$\lambda$4057 features. We considered isotopic splitting by adopting
the solar isotopic Pb ratios. We derived $\mbox{[Pb/Fe]} = 2.45$ from
the $\lambda$3683 line. The $\lambda$4057 Pb feature is blended with
CH. The linelist available for the CH features in that region is not
complete, which leaves many lines unidentified. To reduce blending
effects, we adjusted the wavelengths of some of the features in the
linelist to reflect those listed in Moore's Solar Atlas
\citep{mooreatlas}. We also adjusted the C abundance to match that of
a nearby CH feature at 4058.2\,{\AA} and we modified the oscillator
strengths of some nearby CH features to better reflect the observed
spectrum, with no impact on our Pb measurement. We derived a mean
value $\mbox{[Pb/Fe]} = 2.53$.  While each of the two lines
individually has a large uncertainty due to many unknown (likely CH)
blends and low S/N in the case of the $\lambda$3683 line, their
agreement is encouraging. Given the strength of both lines, as seen in
Figure~\ref{euby}, it is obvious that the Pb abundance is significant in
this star. Since Pb is a neutral neutron-capture species, we use the
NLTE-corrected value for our analysis and interpretation. We note here
that non-LTE effects of neutral Pb in metal-poor stars are strong
\citep{mashonkina_pb} which become larger for cooler stars and lower
metallicities. The Pb\,I correction for HE~0414$-$0343 is
$\Delta_{\rm{NLTE}}=0.56$\,dex for the $\lambda$4057 line, which would
increase our Pb abundance to [Pb/Fe]$\sim3.09$.

\section{Classification of stars with overabundances in neutron-capture elements 
associated with the s-process}\label{interpretation}

The abundances of s- and r-process stars reflect different
nucleosynthetic processes that have either occurred before the star's
formation (in the case of the r-process) or in a companion star during
stellar evolution (in the case of the s-process). Hence, criteria for
identifying stars with these enrichment patterns have been suggested
(e.g., \citealt{bc05}) and classes of objects have been established
based on certain abundance ratios, i.e., CEMP-s, CEMP-r/s and CEMP-r
stars. The aim is to better understand the nature of these objects
and, more generally, metal-poor stars enriched in neutron-capture
elements. HE~0414$-$0343 shows signs of an s-process enrichment so 
we aim to classify it, given its abundance pattern to learn about its
origins.

One classification method is to compare the stellar abundance patterns
with the scaled Solar System s- and r-process patterns. While this has
worked extremely well for strongly r-process enhanced stars (owing to
the universality of the r-process pattern, e.g., \citealt{sneden2000},
\citealt{he1523}), the s-process is more complex because metallicity
strongly affects the s-process abundance pattern
  \citep[e.g.,][]{gallino1998}. Ba and Eu abundances have been used
as a proxy for distinguishing between s- and r-process element
contributions. According to the \citet{bc05} definitions, s-process
enhanced metal-poor stars are classified by $\mbox{[Ba/Fe]} > 1.0$ and
$\mbox{[Ba/Eu]} > 0.5$, while r/s-enhanced stars fall within the range
of $0.0 < \mbox{[Ba/Eu]} < 0.5$ and also having $\mbox{[Ba/Fe]} >
1.0$. Accordingly, HE~0414$-$0343 would be classified as a CEMP-r/s
star. As described further below, we find that the [Ba/Eu] ratio does
not well correlate with the nature of the abundance patterns of many
of these stars found in the literature, however.

\subsection{Assessing neutron-capture abundances ratios}\label{ass}

\begin{figure}[!t]
\centering
\includegraphics[width=0.3\textwidth]{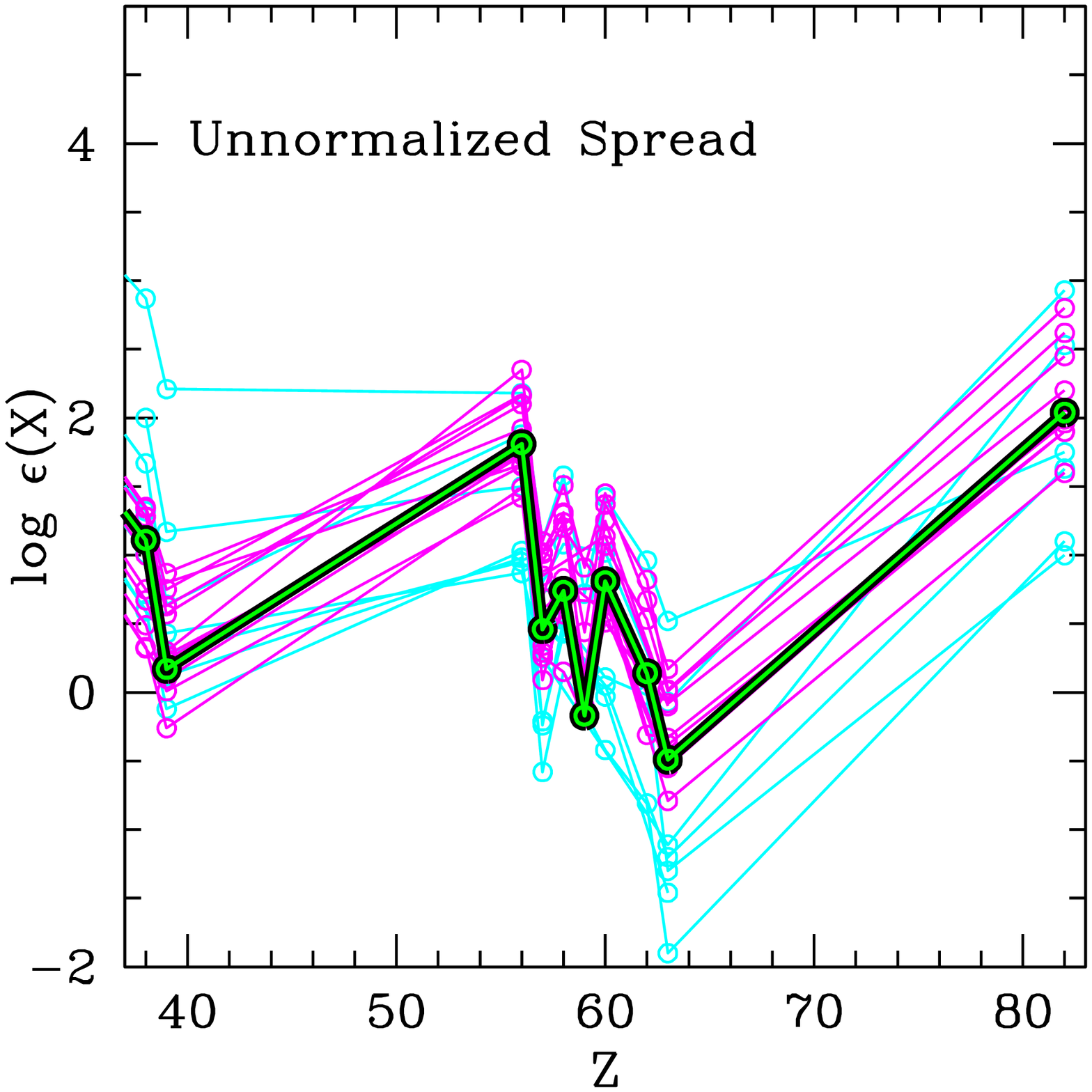}\\
\includegraphics[width=0.3\textwidth]{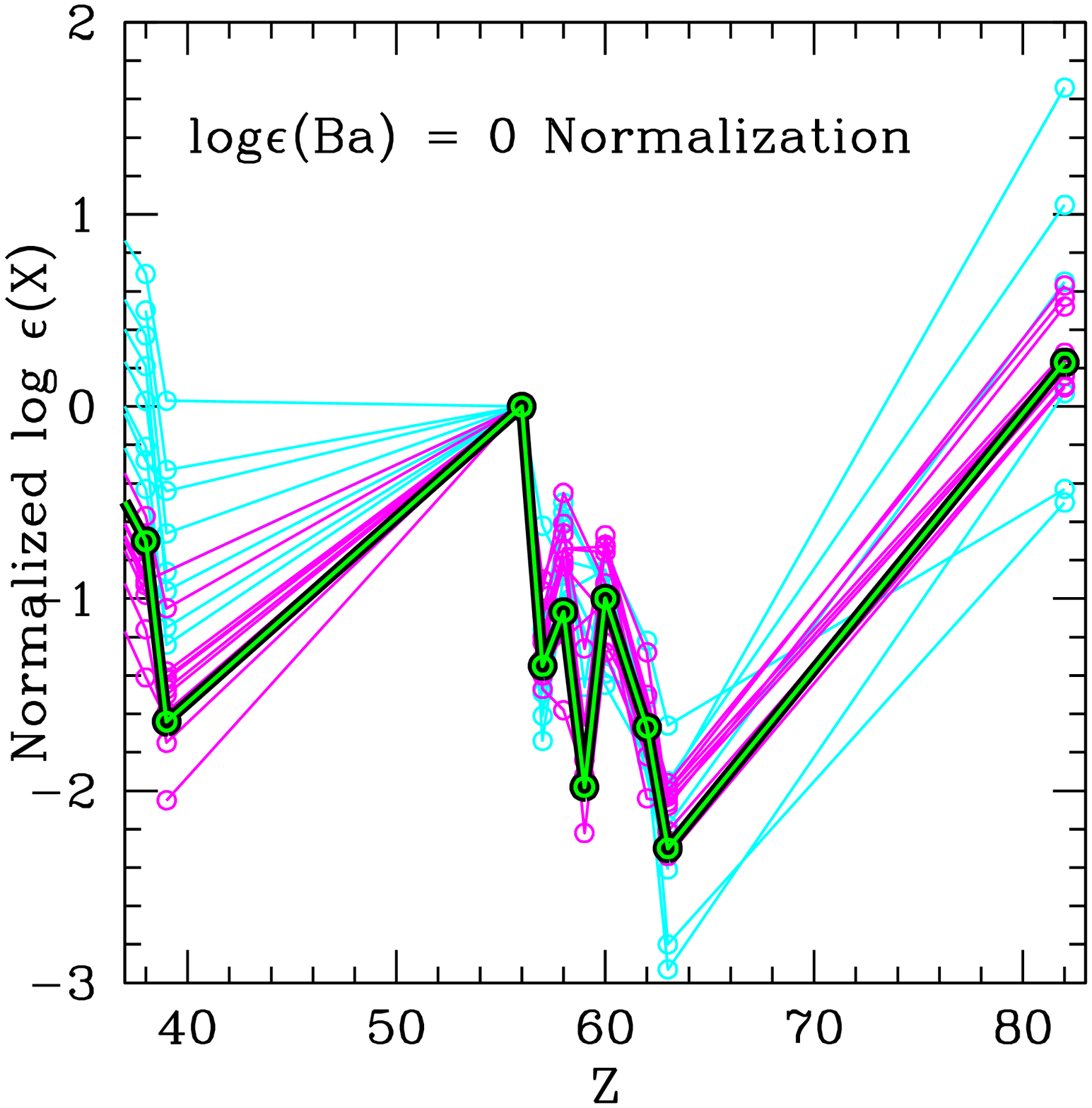}\\
\includegraphics[width=0.3\textwidth]{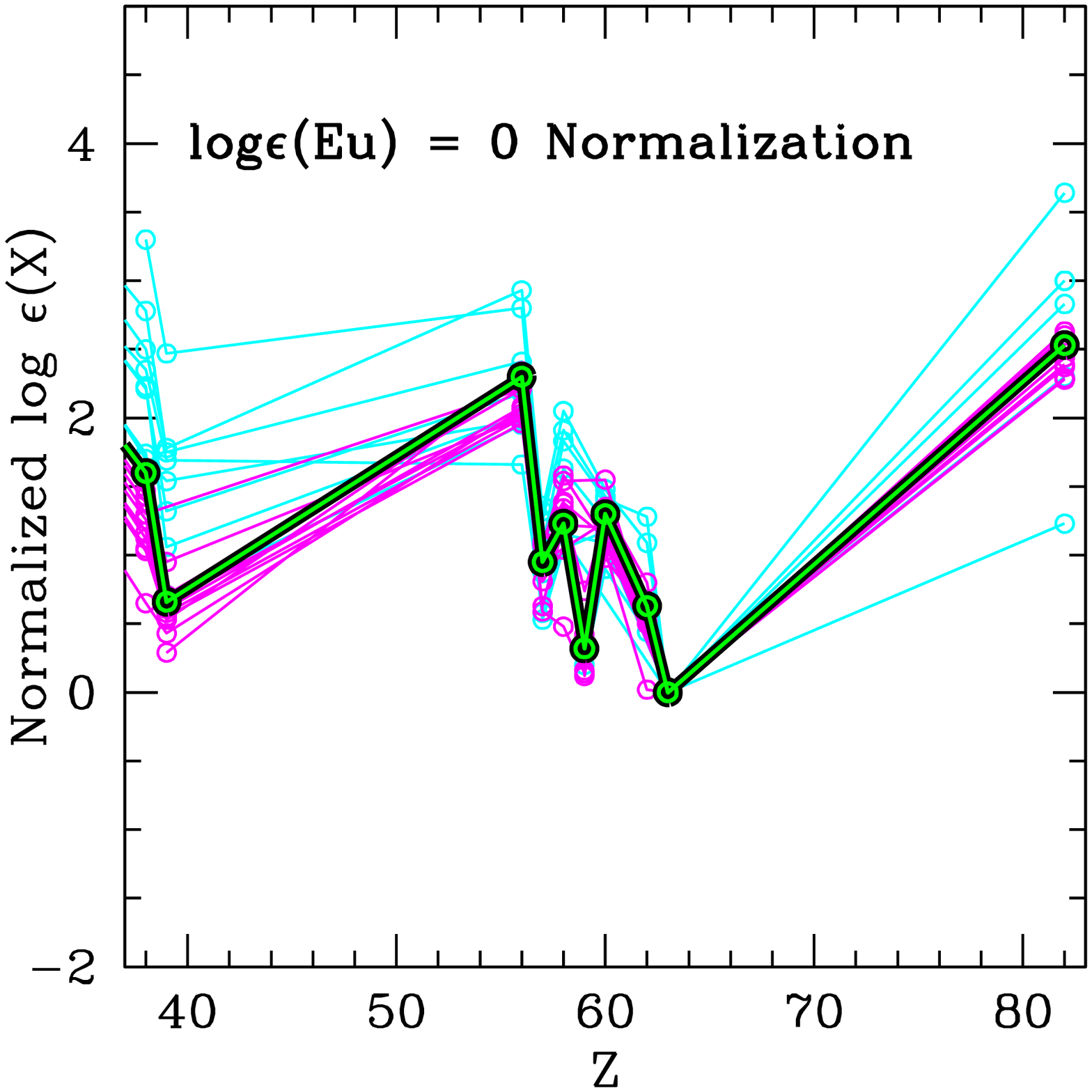}
\caption{$\log\epsilon$(X) abundances plotted against atomic number,
  Z, (top), $\log\epsilon$(X) values that have been normalized to
  $\log\epsilon$(Ba) = 0.0 (middle), and $\log\epsilon(\rm{Eu}) = 0.0$
  (bottom) of selected neutron-capture elements. These normalization
  values were chosen because a majority of the Ba and Eu abundances,
  respectively, were close to those values (but the values are
  arbitrary).  
Color coding according to the traditional ``s''
  and ``r/s'' classification. Cyan corresponds to CEMP-s stars,
  magenta to CEMP-r/s stars, and green to HE~0414$-$0343.  
}

\label{atno}
\end{figure}

We use the CEMP-s and CEMP-r/s stars from Table~6 of
\citet{placco2013}, along with the two new stars presented in that
paper, HE~1405$-$0822 \citep{he1405} as well as HE~0414$-$0343 to
investigate the nature of their abundance patterns, in particular to
learn about the origin of CEMP-r/s stars. In the top left panel of
Figure~\ref{atno}, we plot their $\log$$\epsilon$(X) abundances of the
(un-normalized) neutron-capture elements versus atomic number, Z.  We
are here switching to the $\log$$\epsilon$ abundance notation to
investigate an absolute abundance scale rather than that normalized to
the Sun.  Only stars with $-2.8 < \mbox{[Fe/H]} < -2.3$ are included
in an attempt to remove gross metallicity effects expected to play a
role in s-process nucleosynthesis. This reduced the sample to 11
CEMP-s and 14 CEMP-r/s stars. 

CEMP-r/s stars have higher $\log\epsilon$(heavy neutron-capture
element) abundances than the CEMP-s stars. The overall ranges in the
$\log\epsilon$(Ba) and $\log \epsilon$(Pb) abundances covered within
this sample are very large at $\sim2.0$\,dex, and the range in
$\log\epsilon$(Y) and $\log\epsilon$(Eu) even larger at
$\sim2.5$\,dex. Interestingly, we find that the stars that make up our
sample cover these large abundance ranges rather evenly, suggesting a
continuum.

\begin{figure}[!t]
\centering
\includegraphics[width=0.5\textwidth]{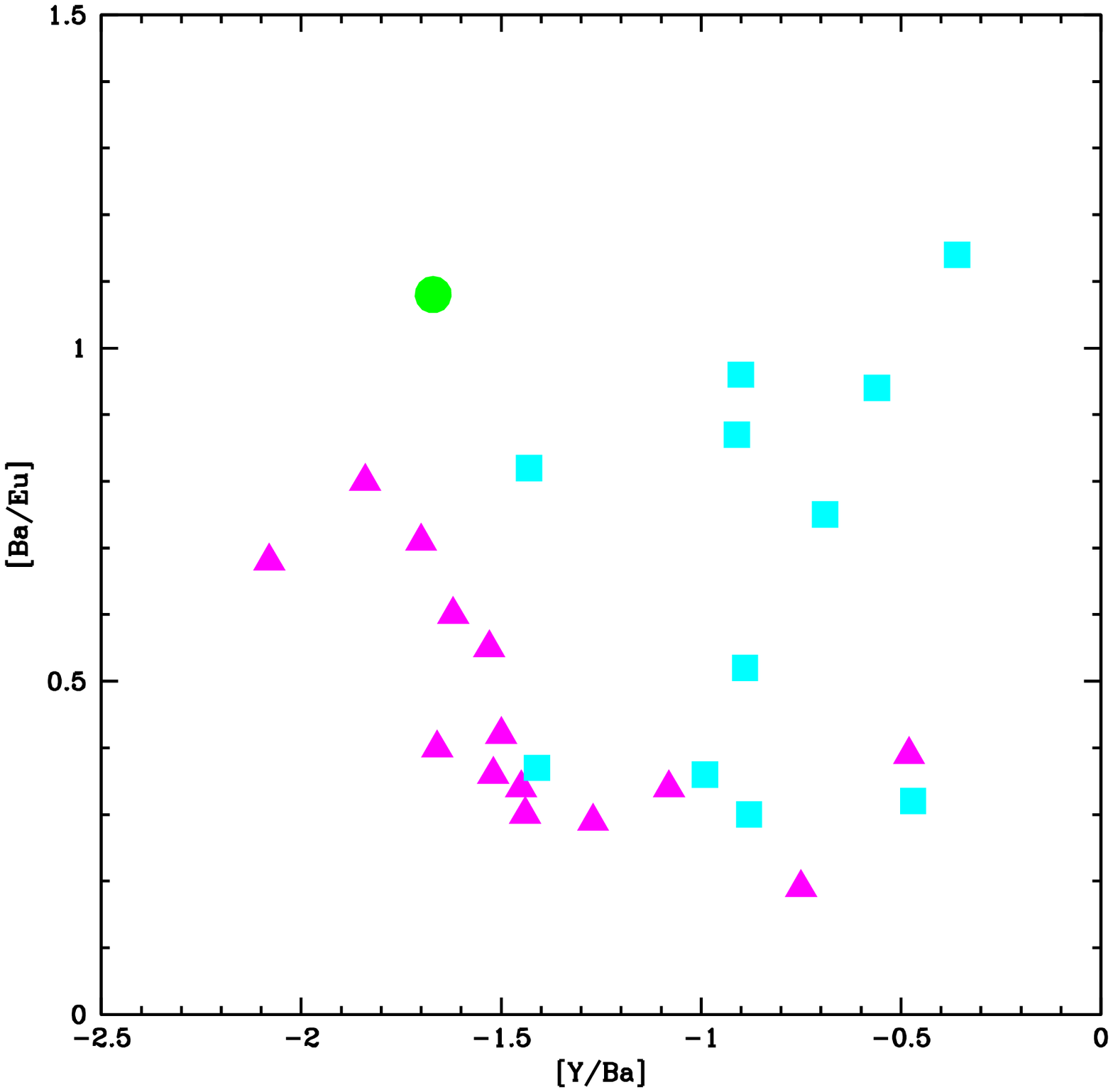}
\caption{[Ba/Eu] abundance ratio for the entire sample plotted against
  [Y/Ba]. It depicts the traditional ([Ba/Eu]) against our proposed
  ([Y/Ba]) classification of CEMP-s and CEMP-r/s stars. The
  distinction between CEMP-s and CEMP-r/s stars has so far been made
  at $\mbox{[Ba/Eu]} =0.5$. As can be seen, it does not well separate
  the stars according to the corresponding CEMP-s and CEMP-r/s
  classes. [Y/Ba] provides a better way to group and classify
  the stars. Cyan squares correspond to stars traditionally
    defined as CEMP-s, magenta triangles to stars defined as
    CEMP-r/s}. The green filled circle corresponds to HE~0414$-$0343.
\label{atno3}
\end{figure}

To better assess the origin and range of the observed neutron-capture
patterns, we then normalize the abundances to
  $\log\epsilon$(Ba) = 0 (middle panel of Figure~\ref{atno}). We use
Ba for normalization because it is considered to be produced mainly in
the s-process that enriched the Solar System
\citep{arlandini1999}. The CEMP-s stars have an average [Y/Ba] =
$(\log \epsilon({\rm Y})_{\star}-\log \epsilon({\rm Y})_{\odot})-
(\log \epsilon({\rm Ba})_{\star}-\log \epsilon({\rm Ba})_{\odot}) =
(\log \epsilon({\rm Y})_{\star}-\log \epsilon({\rm Ba})_{\star}) -
0.03 = \log(\epsilon({\rm Y})/\epsilon({\rm Ba}))$ ratio of
$\sim-0.5$\, but with a spread of $\sim1$\,dex for Y abundances, when
considering the Ba-normalized values. The CEMP-r/s stars have a lower
average $\log(\epsilon({\rm Y})/\epsilon({\rm Ba}))$ ratio of $\sim
-1$, but a similar-sized spread of 1\,dex for the Y abundances. By
comparison, the Solar System ratio is $\sim0.0$, which falls within
the CEMP-s regime. For those stars with measured Pb abundances -- a
difficult task in CEMP stars -- the $\log(\epsilon({\rm
  Pb})/\epsilon({\rm Ba}))$ ratio varies greatly, $> 2$\,dex, in the
CEMP-s stars. This apparent spread in Pb is even larger in the
Ba-normalized abundance patterns than in the absolute overall
abundance patterns. Moreover, the $\log(\epsilon({\rm
  Pb})/\epsilon({\rm Ba}))$ ratio is often negative. In the CEMP-r/s
stars, the $\log(\epsilon({\rm Pb})/\epsilon({\rm Ba}))$ratio is $\sim
0$ for most stars, while the Pb spread is only $\sim 0.75$
dex. Finally, we normalize the abundance patterns to
$\log\epsilon(\rm{Eu})$ = 0, see bottom left panel of
Figure~\ref{atno}. We find that the $\log(\epsilon({\rm
  Pb})/\epsilon({\rm Eu}))$ ratios for the CEMP-r/s stars are
generally lower than those of CEMP-s stars. Also, there is a smaller
spread among the $\log\epsilon$(Pb) abundances. Overall, the behavior
is similar to the Ba-normalized case.

Considering the $\log(\epsilon({\rm Y})/\epsilon({\rm Ba}))$ ratio,
there is a large spread but there is a fairly smooth transition
between CEMP-s and CEMP-r/s stars with no sharp dividing line. This
transition can also be seen in Figure 1 of
\citet{masseroncemp2010}. Together with the $\log(\epsilon({\rm
  Y})/\epsilon({\rm Ba}))$ ratios, the $\log(\epsilon({\rm
  Pb})/\epsilon({\rm Ba}))$ and $\log(\epsilon({\rm Pb})/\epsilon({\rm
  Eu}))$ ratios suggest that there is a transition and no clear,
distinct separation into subgroups. Our star, HE~0414$-$0343, also
demonstrates the continuum nature of the transition between CEMP-s and
CEMP-r/s stars, as its abundance pattern is located between the more
extreme CEMP-s and the CEMP-r/s stars, as can be seen in
Figure~\ref{atno} (green line in all panels).

From this exercise we conclude that considering only observed [Ba/Eu]   
abundance ratios to classify stars into the CEMP-s and CEMP-r/s stars
classes is not sufficient. Indeed, as can be seen in
Figure~\ref{atno3}, using $\mbox{[Ba/Eu]} =0.5$ as the criterion to
distinguish between CEMP-s ($\mbox{[Ba/Eu]} >0.5$) and CEMP-r/s stars
($0<\mbox{[Ba/Eu]} <0.5$) results in less than $\sim50$\% of the cases
being in correctly classified, compared to results based on a detailed
abundances analysis assessment. Therefore, the usefulness of
  the [Ba/Eu] ratio in classifying CEMP-s and CEMP-r/s into two
  distinct subcategories appears to be limited.

Moreover, the labeling of ``CEMP-s'' and ``CEMP-r/s'' suggests that
there is a different underlying nucleosynthetic origins for these
stars, and specifically, that there is an r-process component involved
for r/s stars. But none has yet been found (e.g., \citealt{jonsell06})
and one has to question the existence of such an r-process component.
A more comprehensive assessment and especially a physically-motivated
explanation is thus needed to better understand metal-poor stars showing
neutron-capture element enhancements associated with the s-process. 

The s-process builds up in AGB stars over each thermal pulse after the
third dredge up events begin. The timing of the mass transfer
  of the enriched material in relation to the number of thermal pulses
  experienced by the donor AGB star will help determine what the
  observed abundances will be. The evolutionary status of the observed
  CEMP stars today (e.g., dwarf, giant) also plays an important role
  \citep{stancliffe08,placco14}.

As Figure~\ref{atno3} further shows, the [Y/Ba] ratio seems to provide
a better way to group and classify these stars, although large
  spreads in other neutron-capture elements may be found in each
  class. Such classification shall be introduced in
Section~\ref{origins}. Both Y and Ba are predominantly made in the
s-process. Their relative contributions might thus shed light on the
s-process and the build up of the s-process peaks in AGB stars over
each thermal pulse after the third dredge up events begin. 

To further investigate this apparent continuum of s-process
enhancements in our sample, we now compare the individual stellar
abundance patterns in detail with predictions of the s-process in
models of AGB stars.

\section{Comparison with model AGB s-process yields}\label{placcomodel}

The s-process is thought to occur in thermally-pulsating AGB stars of
$\simeq$0.8 to 8 M$_{\odot}$ \citep{busso1999,karakas14}. The neutrons
that fuel the s-process are primarily produced via the
$^{13}$C($\alpha$,n)$^{16}$O reaction that occurs as a result of
partial CN cycling \citep{abia2001,smithlambert1990}. The
neutron-capture occurs in the He-shell and the newly created s-process
elements are brought to the surface as a result of third dredge-up
mixing episodes. Detailed calculations by \citet{gallino1998},
\citet{bisterzo09}, \citet{cristallo11} and \citet{lugaro}, among
others, have been performed to better understand the s-process. To
reproduce observed s-process abundances in metal-poor companion stars,
stellar models have been created that e.g., vary the mass and the
metallicity of the AGB star, the size of the $^{13}$C pocket, and take
into account dilution effects.

We present a new version of the AGB model of a 1.3\,M$_{\odot}$ star
with [Fe/H] = $-2.5$ previously calculated for \citet{placco2013}. The
new model was calculated with the the Mount Stromlo Stellar
Evolutionary code \citep{karakas2010}, using the same input physics as
before, except with updated low-temperature molecular opacities.  The
  model star was evolved from the zero age main sequence to the
  AGB. The model star underwent 21 thermal pulses during the AGB
  phase, most of which included third dredge up events, resulting in a
  total dredge up of 0.049\,M$_{\odot}$ from the core of the star to
  its surface. Using the \citet{vassil} prescription for mass loss, we
  determined that 0.5\,M$_{\odot}$ is lost during the AGB stage. The
  resultant abundances at each thermal pulse were calculated in the
  same manner as in \citet{lugaro}. The final abundance is reached at
  thermal pulse number 19, thereafter the surface abundances do not
  change.

\citet{placco2013} found two new stars showing signs of s-process
nucleosynthesis and compared their abundances to the yields of an
earlier version of this AGB model, along with the abundances of
several stars classified as CEMP-s and CEMP-r/s stars in the
literature. For a comparison, they also used a range of other AGB star
models spanning 0.9 M$_{\odot} \leq$ M $\leq$ 6\,M$_{\odot}$ at
$\mbox{[Fe/H]} = -2.2$, some of which include pre-enrichment from the
r-process. These models are described in detail in
\citet{karakas2010a} and \citet{lugaro}. \citet{placco2013} also
considered details of the mass transfer event across a binary system,
which necessarily results in dilution of the s-process material once
it is mixed onto the observed star's outer atmosphere.

To seek a physical motivation for the variety of s-process enriched
stars, we set out to investigate the physics of the s-process in AGB
stars and whether the resulting yields can explain the
abundances of HE~0414$-$0343 and the sample of literature stars.

\subsection{HE~0414$-$0343 Abundance Pattern Analysis}\label{ana}

We first extend the Placco et al. analysis to HE~0414$-$0343 and
then later also to their sample (see Section~\ref{litsampanal}). To
account for dilution of the s-process material in the receiver stars'
outer atmosphere, we considered two different options: 5\% and 50\%
cases, where the latter imitates a mass transfer event when the
recipient star is a red giant having 50\% of its mass in the
convection zone and the 5\% case represents a less-evolved star. Since
we do not know how long ago the mass transfer event took place, we
consider both options for HE~0414$-$0343 and each of the sample stars.

The analysis specific to HE~0414$-$0343 then consisted of two steps:
a) comparison of its abundance pattern to the same set of AGB models
as in \citet{placco2013}, specifically, the 1.3\,M$_{\odot}$
  described above and the set of $\mbox{[Fe/H]} = -2.2$ models from
  \citet{lugaro}. For the
  second step b) we extract an r-process pattern from the overall
  observed abundance pattern, and compare the ``decontaminated''
  abundance pattern to the thermal pulse abundance distributions of
  all models.

Regarding step a) and following \citet{placco2013}, we
  minimized the differences between the surface abundance distribution
  after each thermal pulse and the observed abundances for
  HE~0414$-$0343 to find the best match. In particular, we select the
best match based on the smallest overall residual (see e.g.,
Figure~\ref{resdil1}) over different neutron-capture element mass
ranges.  We distinguish five cases. Case i) the full observed
neutron-capture abundance pattern (``full residual''), case ii) just
abundances near the first peak of the s-process (Sr, Zr, Y; ``first
peak residual''), case iii) just abundances near the second peak of
the s-process (Ba, Sm), case iv) just for abundances of the heaviest
neutron-capture elements, e.g., Eu and above (``Eu peak residual''),
and case v) only for elements with $Z > 40$, which excludes the first
peak abundances.  Breaking up the matching procedure into these
element groups helps to disentangle the build up of s-process elements
at the surface of the AGB star. Over time, heavier and heavier
s-process elements are created and dredged-up in the AGB star with
each thermal pulse. The relative contributions to elements in groups
ii) to iv) should be reflected in the residual of the respective
group.

The results of this step are best understood in the context of the
entire sample that we analyzed in an analogous way, and which is described
in Section~\ref{litsampanal}. Hence, in Section~\ref{litsampanal}, we
discuss the analysis results of the entire sample (including
HE~0414$-$0343), together with our conclusion regarding the nature and
origin of s-enriched CEMP stars.

But we note here already that, similar to what was found in
\citet{placco2013}, one of the thermal pulse abundance distributions
of the updated low-metallicity AGB model provides the best match to
the overall abundance pattern of HE~0414$-$0343. However, r-process
elements such as Eu and Dy around the second peak are still
overabundant compared to this best model match. This kind of
discrepancy was also found for the CEMP-r/s stars analyzed in
\citet{placco2013}, suggesting HE~0414$-$343 to be in the same
category. To first order this confirms why CEMP-r/s stars are not
classified as CEMP-s -- they have a curious relative overabundance of
heavy neutron-capture elements compared to lighter ones. This could
indicate a contribution of r-process elements.

Regarding step b), to investigate this heavy-element discrepancy as
well as the nature of the CEMP-r/s abundance pattern and especially
the alleged ``r'' component of ``CEMP-r/s'' stars, we ``extract'' an
arbitrary amount of r-process material (but following the r-process
pattern) from the abundance pattern of HE~0414$-$0343. If the binary
system formed in an r-process enriched gas cloud, extracting an
r-process signature should leave a cleaner s-process signature since
the s-process elements present were only later created during the AGB
phase of the more massive companion.

For the extraction, we decrease the log$\epsilon$~(Eu) abundance in
HE~0414$-$0343 by 0.5 and 1.0\,dex. We use the abundance pattern of
the r-process star CS~22892$-$052 \citep{sneden03} to calculate the
ratios of the neutron-capture elements relative to Eu in order to
extract the putative r-process signature from these stars. These two
``decontaminated'' abundance patterns of HE~0414$-$0343 are then
compared with the models, just like in step a). Again, the best
overall match still did not reproduce the two decontaminated abundance
patterns well. Although the heavy neutron-capture elements above Eu
are better matched than before, the first and second peaks of the
s-process are now very poorly reproduced. Interestingly, using our
$\mbox{[Fe/H]} =-2.5$ AGB model and also a model with a pre-enrichment
of 0.4\,dex of r-process material both yielded similar, equally badly
overall matches, with deviations ranging from 0.5 to 0.9\,dex.

The important conclusion here is that our analysis indicates
  that there is no discernible r-process component present in the
  star.  This is in contrast to the \citet{bc05} criteria which would
  classify HE~0414$-$0343 as a CEMP-r/s given its [Ba/Eu] ratio, and
  implying the existence of an r-process contribution. Our results
  thus render the ``-r/s'' classification unsatisfactory for
  explaining the origins of this star, and potential other stars with
  similar abundance patterns. A different classification seems
  necessary, as well as additional study of these stars in order to
  correctly identify their origin.

\subsection{Literature Sample Analysis} \label{litsampanal}

We extend our detailed model comparison, step a), to the sample of
literature stars used by (\citealt{placco2013}, their Table~6)
including their two new stars as well as HE~1405$-$0822
\citep{he1405}. However, we used just our updated low-metallicity AGB
star model to obtain best matches to the five cases of the different
atomic mass regions. We then compare the residuals of the each of the
five cases to the measured [Y/Ba] ratios of every star in the sample.

\begin{figure*}[!thb]
\includegraphics[width=0.99\textwidth]{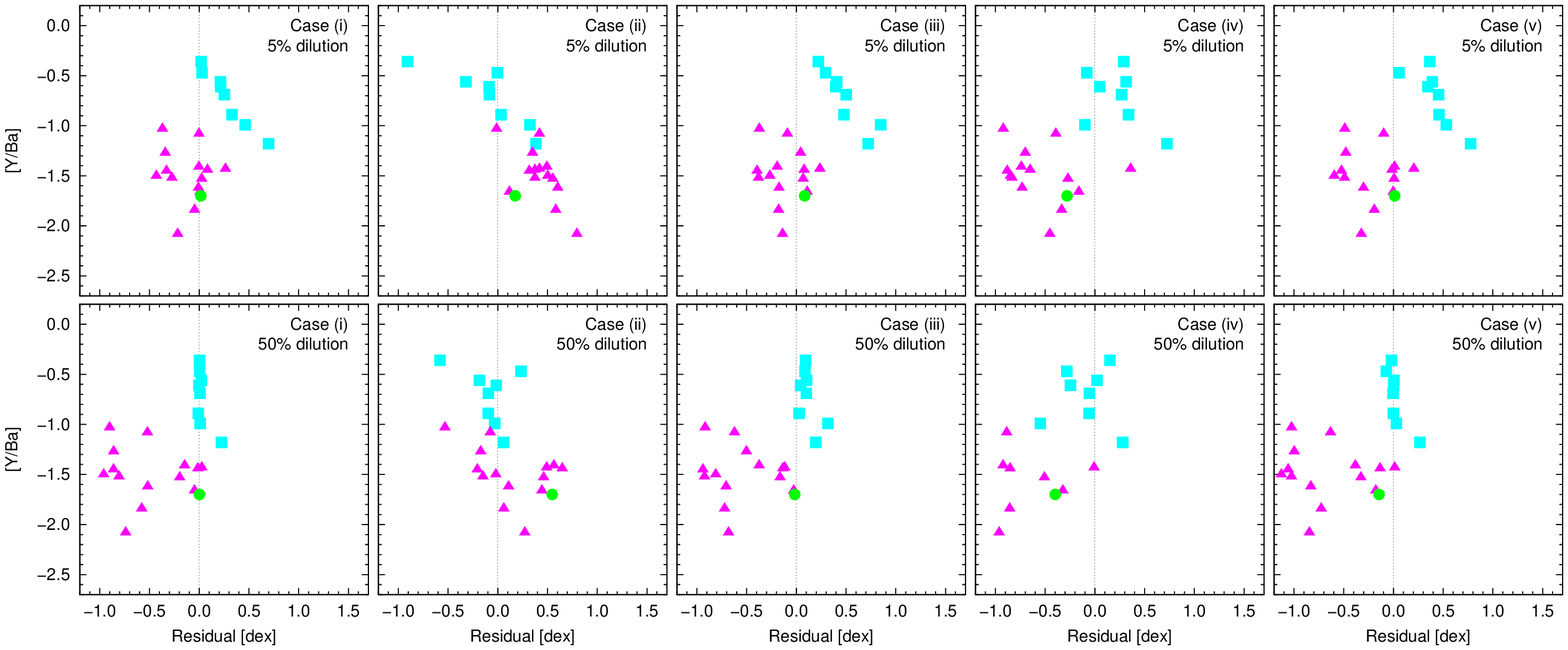}
\caption{\label{resdil1} [Y/Ba] ratio plotted against the five
    cases of residuals obtained from comparing the s-process
  nucleosynthesis results of a $\mbox{[Fe/H]} =-2.5$ AGB star with the
  abundances of star traditionally defined as CEMP-s stars (cyan
  squares), CEMP-r/s stars (magenta triangles), and HE~0414$-$0343
  (green circle). The top row of panels depicts the 5\% dilution case,
  the bottom panels the 50\% dilution case. See text for discussion.}
\end{figure*}

Despite the fact that the best matches are not always
  satisfactory, there is still information in the amount and direction
  of any discrepancy between the thermal pulse abundances and the
  observed abundance patterns. In Figure~\ref{resdil1}, we show cases
  i)-v) for the 5\% and 50\% dilution scenarios. In general, the
stars classified as CEMP-s stars seem to have larger [Y/Ba] ratios
than the CEMP-r/s stars, though there is overlap from $-1$ to
$-1.25$. We already noted this behavior in Section~\ref{ass}.  But
here we additionally find that for the CEMP-s stars, there is an
anticorrelation between the full residual and the [Y/Ba] ratio in the
5\% dilution case, and no correlation in the 50\% case. In the 5\%
case, strong anticorrelations are apparent between the second peak and
Eu peak residuals with the [Y/Ba] ratio. In the 50\% case, the first
and second peak are relatively well matched, while there is a $\sim1$
dex spread in the Eu peak residual. The CEMP-r/s stars show no
correlation for the full residual across all peaks with [Y/Ba] in the
5\% case and a weak correlation in the 50\% case. In the 5\% case,
there is a weak anticorrelation in the first peak residual with the
[Y/Ba] ratio whereas in the Eu peak residual, there is perhaps a
weaker correlation with [Y/Ba]. The 50\% case essentially indicates no
correlations between the residual of any peak and the [Y/Ba]
ratio. Additional results of this analysis will be given in
Section~\ref{tpns}.

\begin{figure*}[!thp]                                                           
\begin{center}
\vspace{4cm}
\includegraphics[width=0.9\textwidth]{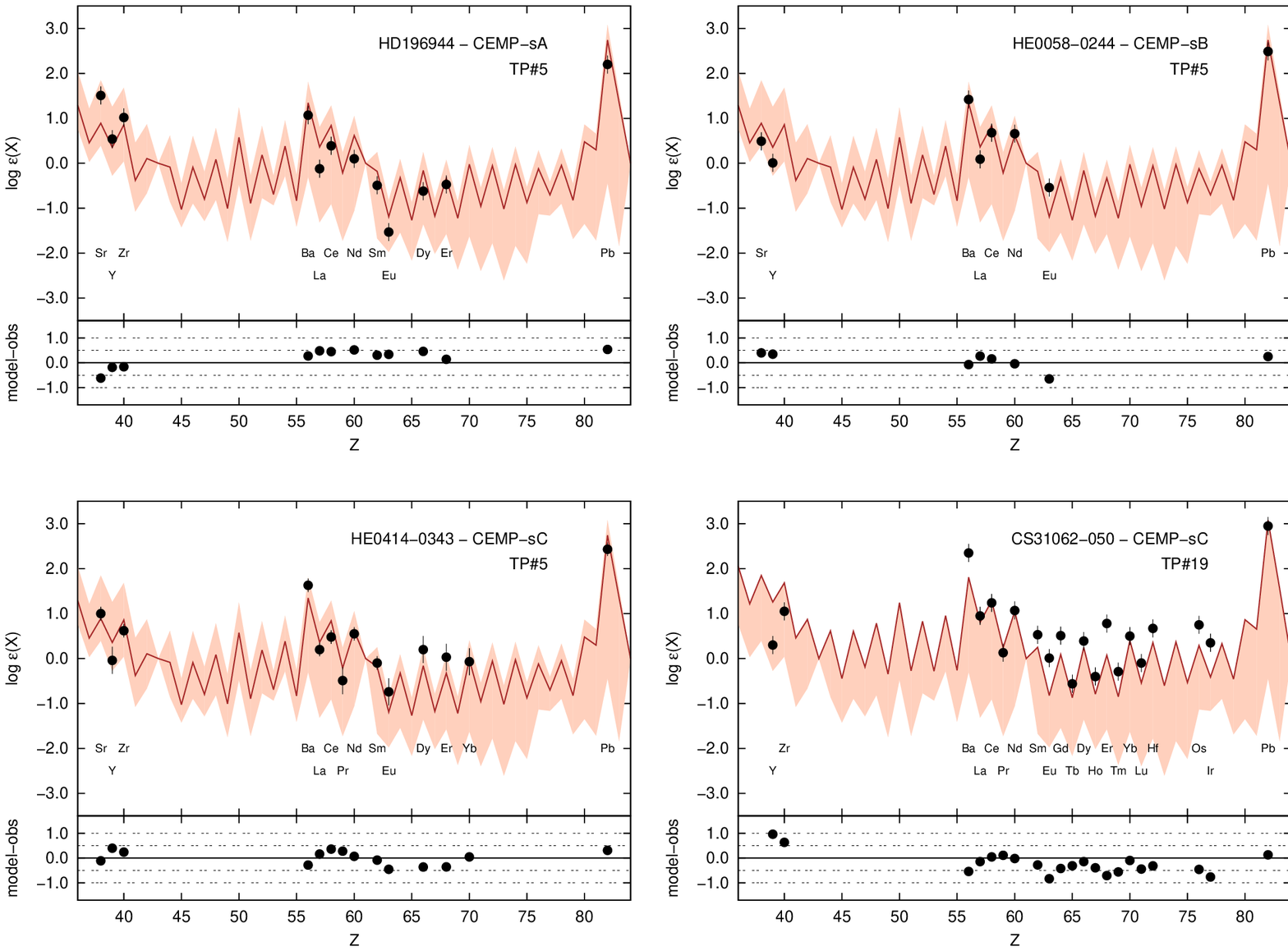}
\caption{\label{ncap} $\log \epsilon$(X) values plotted against atomic
  number for four stars, including our program star,
  HE~0414$-$0343. In the upper panels, the black dots represent the
  stellar abundances, with bars showing  uncertainties. The red
  line belongs to the best thermal pulse abundance distribution of our
  low-metallicity AGB model (for case v), and the peach shaded
  region corresponds to the range of abundance yields accumulating
  over each thermal pulse until thermal pulse 19. The bottom-most part
  of the peach section corresponds to the initial abundances at the
  start of the AGB phase in the model and the top-most part
  corresponds to the final abundance yields. In the bottom panels,
  the black dots represent the difference between the best
  thermal pulse distribution and the observed abundances.  In the
  upper right corner of each plot, we show the star name, its
  classification, and which thermal pulse number corresponds to the
  best match. We have included a star from each classification as well
  as HE~0414$-$0343. The error bars included are the standard errors
  (see Table~3). }
\end{center} 
\end{figure*}

Given the complicated nature of the full abundance pattern
residual, we adopt the results of case v which are presented in
Figures~\ref{resdil1} and \ref{ncap}. We made this choice because
it is generally difficult to interpret the first-peak neutron-capture
elements in the sample stars due to the many possible nucleosynthetic
pathways (e.g., a poorly-understood light element primary process -
\citealt{lepp}) that led to their creation. In the appendix,
we present the abundance patterns of all sample stars together with
their respective best matches from our thermal pulse abundance
distribution comparisons.

\section{The classification and origin of CEMP-sA,sB,sC stars}\label{origins}

Given the smooth transition between stars traditionally
  distinctly defined as CEMP-s and CEMP-r/s stars, i.e., as seen in
Figure~\ref{atno}, we explore a new classification scheme for our
sample which does not assume any r-process contribution. The overall
goal is to find a scheme that accounts for the different
  patterns as well as the smooth transition regarding the magnitude of the
s-process abundance enhancement.  We designate the new classes
CEMP-sA, CEMP-sB, and CEMP-sC. CEMP-sA stars have the ``most
traditional'' s-process abundance pattern and the least negative
[Y/Ba] ratios spanning the range $-0.9 < \mbox{[Y/Ba]} < -0.3$,
whereas CEMP-sC stars have an abundance pattern that deviates the most from
the usual s-process abundance pattern, while still maintaining an
s-process enhancement and, correspondingly, have the largest negative
[Y/Ba] ratios, with $\mbox{[Y/Ba]} < -1.5$. CEMP-sB stars fall in the
middle, with $-1.5 < \mbox{[Y/Ba]} < -0.9$. Table~5 lists our
definitions and classifications for the literature sample we have
employed in this study.

To facilitate future classifications of undiscovered s-process stars, we have
provide a web-based
tool\footnote{\href{http://www.nd.edu/~vplacco/sprocess-classes.html}{http://www.nd.edu/$\sim$vplacco/sprocess-classes.html}}  
that automatically classifies user-input stars based on their abundance
patterns. It determines the best match of a given thermal pulse abundance
distribution of our low-metallicity AGB model and provides plots of other stars
that have been classified the same way, such that a comparison of the entire
neutron-capture element pattern is made available.

\begin{deluxetable}{lcc}
\tabletypesize{\scriptsize}
  \tablecolumns{3} 
  \tablewidth{0pc} 
  \tablecaption{CEMP-sA,-sB,-sC classification scheme}
  \tablehead{\colhead{}& \colhead{Definition} &\colhead{Abbreviation}}
\label{class}
\startdata
   & $\mbox{[Ba/Fe]} > 1.0$ and $-0.9 < \mbox{[Y/Ba]} < -0.3$ & CEMP-sA \\
   & $\mbox{[Ba/Fe]} > 1.0$ and $-1.5 < \mbox{[Y/Ba]} < -0.9$ & CEMP-sB\\
   & $\mbox{[Ba/Fe]} > 1.0$ and        $\mbox{[Y/Ba]} < -1.5$ & CEMP-sC \\
   & $\mbox{[Ba/Fe]} < 0.0$ & CEMP-no\tablenotemark{a}\\\\\hline

&&\\
\textbf{Star} & \textbf{[Y/Ba]} & \textbf{Classification} \\\\\hline
CS~22898$-$027  &   $-$1.52 & CEMP-sC   \\
CS~22942$-$019  &   $-$0.36 & CEMP-sA	  \\
CS~22947$-$187  &   $-$0.99 & CEMP-sB	  \\
CS~22948$-$027  &   $-$1.08 & CEMP-sB	  \\
CS~22964$-$161  &   $-$0.89 & CEMP-sA	  \\
CS~29497$-$030  &   $-$1.27 & CEMP-sB	  \\
CS~29497$-$034  &   $-$0.75 & CEMP-sA	  \\
CS~29526$-$110  &   $-$0.48 & CEMP-sA	  \\
CS~31062$-$012  &   $-$1.41 & CEMP-sB	  \\
CS~31062$-$050  &   $-$2.08 & CEMP-sC	  \\
HD~196944       &   $-$0.56 & CEMP-sA	  \\
HE~0024$-$2523  &   $-$0.47 & CEMP-sA	  \\
HE~0058$-$0244  &   $-$1.44 & CEMP-sB	  \\
HE~0202$-$2204  &   $-$0.91 & CEMP-sB	  \\
HE~0338$-$3945  &   $-$1.50 & CEMP-sB    \\                   
HE~0414$-$0343  &   $-$1.70 & CEMP-sC	  \\
HE~1031$-$0020  &   $-$0.88 & CEMP-sA	  \\
HE~1105+0027    &   $-$1.62 & CEMP-sC	  \\
HE~1135+0139    &   $-$0.69 & CEMP-sA	  \\
HE~1405$-$0822  &   $-$1.66 & CEMP-sC	  \\
HE~1509$-$0806  &   $-$0.90 & CEMP-sA     \\           
HE~2138$-$3336  &   $-$1.43 & CEMP-sB	  \\
HE~2148$-$1247  &   $-$1.45 & CEMP-sB	  \\
HE~2258$-$6358  &   $-$1.53 & CEMP-sC	  \\
LP~62$-$544     &   $-$1.84 & CEMP-sC	 
\enddata
\tablecomments{All stars have $\mbox{[C/Fe]} > 0.7$.} 
\tablenotetext{a}{Carbon-enhanced metal-poor stars with normal
  neutron-capture element abundances (CEMP-no; \citealt{bc05}) stars
  are not part of the classification CEMP-sA-C scheme. We include it
  here for completeness.}
\end{deluxetable}

Following the new classification it is important to investigate if
there is an underlying physical mechanism that could explain the
origin of the continuum of s-process patterns. Three plausible
evolutionary scenarios for the smooth transition from CEMP-sA to
CEMP-sB to CEMP-sC are discussed below.

\subsection{Thermal Pulse Number Stratification}\label{tpns}

The abundance distribution of our low-metallicity AGB model's best
matched thermal pulse is a potential predictor of the CEMP
classification. The abundance patterns of the CEMP-sA stars are best
matched by the corresponding nucleosynthesis yields of thermal pulse
no. 5 in the 5\% dilution case. Residuals are shown in
Figure~\ref{resdil1}. The abundances of the first s-process peak are
consistently under-predicted in the AGB model by $\sim 0.2$\,dex. The
second s-process peak is over-predicted in the model by $\sim
0.3$\,dex and the Eu region and third s-process peak are
over-predicted by $\sim 0.2$\,dex, even when a +0.5\,dex NLTE
correction is applied to the Pb abundance. The CEMP-sC stars show the
largest deviations between the observations and best-matched model
yields. In that case, the first s-process peak is over-predicted by
$\sim 0.5$\,dex, the second peak is under-predicted by $\sim
0.1$\,dex, and the third peak is even more under-predicted by $\sim
0.4$\,dex. The third peak is especially under-predicted when the NLTE
correction is applied to Pb. Given the large abundances of second and
third peak s-process elements, later thermal pulses, which have
increased heavy neutron-capture element abundances, more closely match
these stars. The CEMP-sB stars, predictably, fall in between the
CEMP-sA and CEMP-sC stars.

The 50\% dilution case is similar in that it reveals that the CEMP-sA
stars and the CEMP-sC stars form distinct groups. This is especially
apparent in the first s-process peak residuals. The first peak is
under-predicted by $\sim 0.2$\,dex by best-matched model yields in the
CEMP-sA stars and over-predicted by $\sim 0.3$\,dex in the CEMP-sC
stars. The second peak is well matched in the CEMP-sA stars with a
slight overprediction of only $\sim 0.1$\,dex, but under-predicted in
the CEMP-sC stars by $\sim 0.4$\,dex. The third peak is underpredicted
in all stars but one, with the average underprediction for the CEMP-sA
stars being $\sim 0.3$\,dex and $\sim 0.8$\,dex for the CEMP-sC
stars. The best matched thermal pulse numbers are larger in the
CEMP-sA stars in this dilution case (7-19), which is expected since
each thermal pulse represents more s-process material.  The CEMP-sC
stars are all best matched at thermal pulse 19, i.e., the
  final abundance of the model. As with the 5\% dilution case, the
CEMP-sB stars fall in the middle of these two extremes. While it is
difficult to directly map the best matched thermal pulse number to the
classification, it does show some correlation.

We then compared the best matched thermal pulse number with the [Y/Ba]
ratio. The CEMP-s and CEMP-r/s stars form
distinct groups. We further investigated the best matched thermal
pulse number as a potential key to the physical origin of the [Y/Ba]
continuum. We accomplished this in two stages. First, we attempt to
identify the relationship between the CEMP s-process
sub-classifications (both traditional and new) and the best matched thermal
pulse numbers.  Second, we examine how well the abundance patterns are
fitted by the best matched thermal pulse numbers to seek a physical
explanation for the observed abundance distribution.

In the 5\% dilution case, there is a large spread of $\sim 1$\,dex in
the [Y/Ba] ratio after 5 thermal pulses. A large number of stars are
also best matched after 19 thermal pulses.  All the CEMP-s stars are
best matched at 5 thermal pulses, while the CEMP-r/s stars span a
larger range of best matched thermal pulse numbers from 5-19, with a
majority of the stars being best matched at thermal pulse number
19. Before thermal pulse 5, the AGB star is not (yet) a suitable donor
star. The first third dredge up event happens at that thermal pulse
number in the model and then the increasing pulses allow for the AGB
star to build up s-process material on its surface which later gets
transferred onto the surface of the observed metal-poor star.

In the 50\% dilution case, we see this same degeneracy of best matched
thermal pulse numbers, again at 19 thermal pulses, but not at the low
number end. The abundances of CEMP-r/s stars are best matched over a
tight range of thermal pulse numbers (18-19), while the CEMP-s stars'
abundances span a range from 5-19 thermal pulses. The degeneracy at 19
thermal pulses is likely due to the fact that the model's surface
abundances do not change after the model star has undergone 19 thermal
pulses and further thermal pulses do not alter the AGB star surface
abundances anymore.

Given the apparent connection between the traditional classifications
and the best matched thermal pulse number, we investigate how the new
designations corresponded to the best matched thermal pulse
number. Indeed, the thermal pulse number and the classification of
CEMP-sA, -sB, and -sC are correlated such that low numbers ($\sim 5$)
tend to correspond with CEMP-sA stars and high numbers of thermal
pulses ($\sim 19$) correspond with CEMP-sC stars, while the CEMP-sB
star abundances lead to thermal pulse numbers between 5 and
19. We note, however, that many of the best matches have
  residuals of more than $\pm0.5$\,dex for some elements. In some of
  these cases, it is apparent that the residual analysis breaks down
  since the model element abundance are not large enough, even at the
  highest thermal pulse numbers, to reproduce the observations.

Given these results, we consider the thermal pulse number as a proxy
for the processes of the mass transfer event. Specifically, it could
represent the timing of the mass transfer. Lower thermal pulse number
indicate an earlier transfer from the thermally pulsing AGB companion
star, soon after the onset of its s-process nucleosynthesis. Thus, the
s-process pattern would not have built up nearly as much of the
heavier neutron-capture elements as a mass transfer that occurred at a
later time, i.e. one with a higher thermal pulse number.

It could also represent the physical distance between the stars in the
binary system. Two stars with close physical separation will undergo
an earlier mass transfer, corresponding to a lower thermal pulse
number. If the thermal pulse number alone is what causes the observed
spread in the [Y/Ba] ratios, then only a small dispersion in the
[Y/Ba] ratio would be expected for a given thermal pulse number.
Nevertheless, the degeneracy at both low and high thermal pulse
numbers suggests these to be only a partial explanation of the CEMP-sA
to CEMP-sC transition.

\subsection{Neutron sources in early AGB stars}

While the thermal pulse number seemed to be a promising explanation
for the traditional classification scheme, it falls short with the
[Y/Ba] ratio diagnostic. That is not to say that the thermal pulse
number holds no information. The model we are using describes a
1.3\,M$_{\odot}$ AGB star which we can compare to more massive AGB
star models at $\mbox{[Fe/H]} = -2.2$ using the predictions from
\citet{lugaro}. \citet{abate15} recently performed a similar analysis
using binary population synthesis models and a range of AGB masses
from models of $\mbox{[Fe/H]} = -2.2$.

Models of higher-mass than 1.3\,$M_{\odot}$ but less than
  3\,M$_{\odot}$ result in an s-process pattern characterized by higher
  second and third peak s-process element abundances. This occurs
  because models of less than $\approx 3$\,M$_{\odot}$ experience more
  thermal pulses and deeper third dredge-up than the 1.3\,M$_{\odot}$
  case. For models above 3\,M$_{\odot}$, the
  $^{22}$Ne$(\alpha,n)^{25}$Mg reaction starts to dominate and the
  s-process distribution changes such that there is more first
  s-process peak elements than second or third peak (that is, [Y/Ba]
  $>0$) \citep{lugaro,fishlock14}.  We also note that variations in
  the size of the $^{13}$C pocket can change the ratio of the
  s-process elements from light ``ls'' to heavy ``hs''.  The ratio [ls/Fe]
  is an average of elements at the first s-peak, e.g., Sr, Y, Zr,
  whereas [hs/Fe] is an average of elements at the second peak, e.g.,
  Ba, La and Ce \citep[see e.g.,][]{masseroncemp2010,lugaro}.  The
  ratios of [hs/ls] and [Pb/hs] increase with increasing $^{13}$C
  pocket size, indicating that a larger number of neutrons relative to
  Fe-seed nuclei produces more Ba, La, Ce and Pb, relative to elements
  at the first peak \citep{bisterzo11,lugaro,fishlock14}. 

The CEMP-sA stars present better residuals when using our low-metallicity,
low-mass AGB model, especially with regard to the first peak elements.
However, there are systematic discrepancies between the abundance
patterns of the CEMP-sC stars (and to a lesser degree, of the CEMP-sB
stars) and the low-mass model. While some variations in the AGB
companion mass may provide a better fit for some of the CEMP stars, 
it is not the only solution to the transition from
CEMP-sA to CEMP-sC. This is because the [Y/Ba] ratio varies from 
$\approx -0.7$ to positive values as the stellar mass is increased.
\citet{lugaro} discuss this point in more detail, noting that
AGB models can only produce [ls/hs] $>-1$ as a consequence of the 
way in which the s-process operates \citep[see also][]{busso01}. 
The increase of the [Y/Ba] (or [ls/hs]) ratio with increasing stellar 
mass is a direct consequence of the activation of the 
 $^{22}$Ne$(\alpha,n)^{25}$Mg neutron source. Other suggestions for the
origin of the CEMP-sC stars may be found in physics not currently included
in 1-D AGB models such as proton-ingestion episodes near the tip of the AGB
phase that result in high neutron densities for short times 
\citep[such as in post-AGB stars,][]{herwig11}.

\begin{figure}[!th]
\begin{center}
\includegraphics[width=0.5\textwidth]{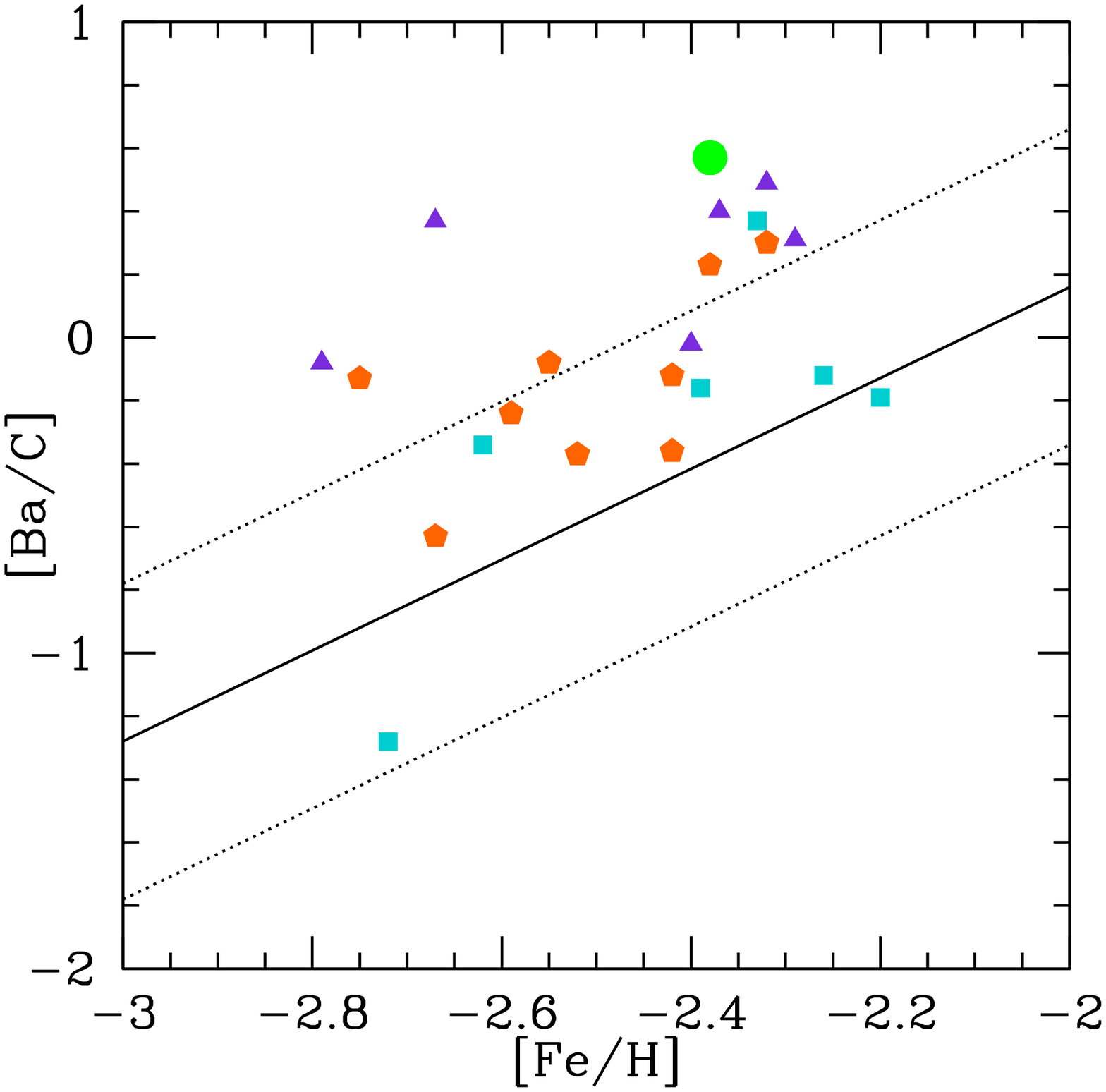}
\caption{[Ba/C] abundance ratio plotted as a function of [Fe/H] of
  HE~0414$-$0343 and the entire literature sample. The green circle
  corresponds to HE~0414$-$0343, the turquoise squares represent the
  CEMP-sA stars, the orange pentagons correspond to the CEMP-sB stars,
  and the purple triangles correspond to the CEMP-sC stars. The solid
  black line corresponds to the relation that would be expected
  between [Ba/C] and [Fe/H] if the main neutron source for the
  s-process is C, empirically adopted from Figure 5 of
  \citet{masseroncemp2010}. The black dotted lines represent the area
  on the plot where the majority of the CEMP-s stars resided in Figure
  5 of \citet{masseroncemp2010} and correspond to the shaded region in
  the upper panel of their Figure 7. \label{bac}}
\end{center}
\end{figure}

At the lowest metallicities the efficiency of the s-process is strongly 
dependent on the number of available Fe seed nuclei. One way to confirm this 
observationally is to examine the [Ba/C] ratio as a function of metallicity. 
The right panel of Figure 5 in \citet{masseroncemp2010} explores this and they
find a tight correlation between [Ba/C] and [Fe/H] in their CEMP-s stars. 
In Figure~\ref{bac}, we plot the same axes, along with the
empirical trend from Figure 5 of \citet{masseroncemp2010} in the black
solid line, with the black dotted lines representing the area in which
most of the CEMP-s stars are plotted. We find that our CEMP-sA stars
mostly fall within this same region even over the narrow range of
metallicity that was chosen. 

That the CEMP-sA stars fall within the same region of
  Figure~\ref{bac} is an indicator that the source of carbon and
  s-process elements is similar. It also indicates that the neutron
  source operating in the AGB companion star was similar.  In Figure 7
  of \citet{masseroncemp2010}, they do not find a correlation between
  [Ba/C] and [Fe/H] for their CEMP-rs stars. Similarly, the CEMP-sC
  stars in our Figure~\ref{bac} do not show the tight correlation
  between [Ba/C] and [Fe/H] that the CEMP-sA stars of our sample and
  the CEMP-s stars from \citet{masseroncemp2010} do. This is an
  indicator that a different neutron source operates in the AGB
  companions of the CEMP-sC stars. AGB models cannot explain the
  alleged high r-process contribution in the CEMP-sC stars nor the
  very low [Y/Ba] $<-1$ \citep[e.g.,][]{lugaro}.  This indicates that
  AGB stars (covering a range of masses) are not responsible for the
  CEMP-sC abundance patterns \citep[see detailed discussion
    by][]{cohen03}.

We can rule out the $^{22}$Ne($\alpha$,n)$^{25}$Mg reaction
  operating in high-mass AGB star models but what about this neutron
  source in rotating massive stars?  Yields of s-process elements from
  rotating massive stars indicate that elements at the first s-process
  peak are predominantly overproduced, with little barium or lead
  synthesized \citep{pignatari08,frischknecht12}. This suggests that
  the s-process in massive stars also cannot be responsible for the
  abundance distribution of CEMP-sC stars, unless the AGB companions
  also (somehow) experience considerable r-process nucleosynthesis
  much beyond what is currently predicted.

\subsection{Initial neutron-capture element abundances of CEMP-sA,sB,sC stars}

In plots of [Ba/Fe] or [Sr/Fe] versus [Fe/H] of halo field stars
(e.g., bottom panels of Figure~18 in \citealt{frebelnorris10}), there
is a huge spread of nearly 3\,dex in regular halo stars with
$\mbox{[Fe/H]} < -2.0$ (and no strong overabundance in carbon) that do
not show any particular enhancement in neutron-capture elements, i.e.,
stars with [Ba/Fe] or $\mbox{[Sr/Fe]} < 1$. Similarly, in stars with
enhanced s-process abundances, i.e., stars with [Ba/Fe] or
$\mbox{[Sr/Fe]} > 1$, we also observe a $ > 2$\,dex spread (e.g., top
panels of Figure~18 in \citealt{frebelnorris10}). Given that the
spread in neutron-capture element abundances are roughly of the same
magnitude in both regular metal-poor stars and s-process stars, one
proposal is that the differences between CEMP-sA, CEMP-sB, and CEMP-sC
stars are caused by variations in the level of pre-enrichment in
neutron-capture elements in the birth gas clouds by progenitor
generations (due to the chemical evolution and analogous to the
alleged r-process contribution discussed earlier) in addition to the
s-process material received from the AGB companion at later times.

S-process rich stars have only been found at metallicities of
$\mbox{[Fe/H]} > -3$ (albeit with one exception, i.e.,
\citealt{sivarani2006}) and the rise of s-process enrichment through
stellar winds from AGB stars has been placed at $\mbox{[Fe/H]}
\sim-2.6$ \citep{simmerer2004}. This
suggests that various chemical enrichment processes were already
operating at $\mbox{[Fe/H]} \sim -2.5$, including different kinds of
neutron-capture processes occurring in supernovae and the more massive
AGB stars. Altogether, in a yet to be understood way, chemical
evolution produced stars with huge spreads in neutron-capture
abundances whereas their light element ($Z\le30$) abundance ratios
(e.g., [Ca/Fe]) are nearly identical (e.g., \citealt{cayrel2004,
  frebelnorris10}).

We examine the [Sr/Fe] and [Ba/Fe] abundance ratios of our sample
stars to test if they could have partially originated from large
variations of the neutron-capture elemental abundances in the stars'
birth clouds. Specifically, we attempt to map them to the [Y/Ba] ratio
to determine if there is a connection between the initial birth
abundances of the CEMP-sA, CEMP-sB, and CEMP-sC stars and the values
that are now observed after the mass transfer event.  As in the case
of our r-process pattern extraction, we now attempt the reverse. We
adjusted the [Sr/Fe] and [Ba/Fe] abundance ratios by subtracting an
amount to represent the mass transfer s-process material. The adjusted
abundances should then reflect the initial amount of neutron-capture
elements of the receiver stars prior to the mass transfer of
neutron-capture- and carbon-rich matter. 

However, the adjusted abundance patterns of the receiver stars do not
map well from the CEMP-sA, -sB, and -sC stars to the
neutron-capture/carbon-normal metal-poor stars. Given this mismatch
between the CEMP-sA, CEMP-sB, and CEMP-sC stars and the
neutron-capture- and C-normal metal-poor stars, the idea that an
initial spread in the abundances in the gas cloud would cause the
observed continuum between CEMP-sA to CEMP-sC will likely take a more
sophisticated approach. This is in part due to uncertainties in which
s-process pattern to subtract, as it is likely different for each
star. More high-resolution spectra of CEMP stars need to be obtained
and analyzed to better investigate this scenario with a much larger
sample.

\section{Caveats and Considerations}\label{caveats}

In this work we have primarily used one low-metallicity AGB model to
match to all stellar abundance patterns.  The main reason we
  used the 1.3\,M$_{\odot}$ model so extensively is because it has a
  metallicity closest to the observed star, at $\mbox{[Fe/H]} = -2.5$.
  Nevertheless, we have used the $\mbox{[Fe/H]} = -2.2$ models to
  guide us in our interpretations regarding the behavior of general
  stellar and nucleosynthesis properties. These models, while at
  slightly higher metallicity, cover an extensive range of mass 0.9 to
  6\,$M_{\odot}$ \citep{lugaro,karakas2010}.

We have made the assumption that the s-process pattern is built up
in the same way for all stellar masses for a given metallicity
although as a function of thermal pulses. Out of necessity, the
biggest assumption is perhaps that the companion mass is the same for
all our stars, but we discuss this further below. Despite this
limitation, these assumptions are justified because we made a
reasonable metallicity cut to select sample stars have similar
metallicities from $-2.7$ to $-2.3$. This ensures that our stars span
over the stellar model metallicity of $\mbox{[Fe/H]} =-2.5$. Thermal
pulse abundances in this model do give the best match to a CEMP-sA star
compared to those of other models.

A more general issue that all AGB modelling faces is the formation and
size of the $^{13}$C pocket, from which the neutrons necessary for the
s-process originate. The size of the pocket can be somewhat
constrained by comparisons to observations of AGB stars and their progeny: 
post-AGB stars and planetary nebulae \citep[see discussion in][]{karakas14}.
In metal-poor post-AGB stars, spreads in the size of the pocket in the models of a
factor of 3-6 are needed to account for the observation data
\citep{axel07b, desmedt2012}. Furthermore, in CEMP stars the spread
needed to match the observational data can be upwards of a factor of
10 or higher \citep{bisterzo11,lugaro}. Central to the issue of 
how large the  $^{13}$C pocket is that we do not understand their
formation process in AGB stars. Of particular interest to this work is
that increasing the size of the $^{13}$C pocket will decrease
the [Y/Ba] ratio. For example, from the models in \citet{lugaro}
we see that a 2$M_{\odot}$ model of [Fe/H] $=-2.2$ will move from
a positive [Y/Ba] of 0.39 with no pocket to [Y/Ba] = $-0.57$ with
the model with the largest $^{13}$C pocket.
As more low-metallicity AGB models with [Fe/H] $\approx -3$ become available 
we will further analyze how different $^{13}$C pocket assumptions would affect
the [Y/Ba] ratio.

For future studies it would be helpful to have a better understanding
about the details of the mass transfer so that the dilution of the
transferred s-process material can be more accurately described
\citep[e.g.,][]{abate15}.  

To summarize, in order to further investigate CEMP s-process star
abundances, more AGB star models are needed that cover a larger range
of both initial mass and metallicity, especially for metallicities
below [Fe/H] = $-2.5$.  More observations of newly discovered
s-process stars are also needed in order to create a larger
statistical sample. These models and observed abundances can then be
used to better probe AGB star mass and metallicity effects on the
abundance patterns for a full physical explanation of the
observations.

\section{Summary and Conclusions}\label{conclusions}

We have presented the red giant HE~0414$-$0343, a CEMP star with
$\mbox{[Fe/H]} = -2.24$. Following a detailed abundance analysis, we
find the star to possess a strong enhancement in the s-process
elements. Based on its [Ba/Eu] and [Ba/Fe] ratios, this star falls
under the traditional category of ``CEMP-r/s'' stars.  We find
HE~0414$-$0343 to most likely be in a binary system since its radial
velocity has been varying over the course of our observations from
2004 to 2011.

To better understand HE~0414$-$0343 and the nature of ``CEMP-r/s''
stars, we also analyzed the abundance patterns of a sample of
literature ``CEMP-s'' and ``CEMP-r/s'' stars. The CEMP-s and CEMP-r/s
categories as defined by \citet{bc05} based on the [Ba/Eu] ratio have
been fundamental in recognizing the diverse nature of CEMP star
abundances. We find, however, that this traditional way of using a
cutoff in the [Ba/Eu] ratio to classify these stars corresponds only
poorly to ``CEMP-s'' and ``CEMP-r/s'' stars once the availability of
their detailed abundances allows a more encompassing assessment of
their abundances signatures.  By instead utilizing the [Y/Ba] ratio to
characterize CEMP stars with neutron-capture element enhancements
associated with the s-process, we can gain a better understanding of the
origin of the s-process diversity. Investigation of the [Y/Ba]
abundance ratio in our sample shows that there is a continuum between
the ``CEMP-s'' and ``CEMP-r/s'' stars, rather than a distinct cut off
separating the two groups of objects with different origins.

We suggested a new classification scheme for s-process stars, CEMP-sA,
CEMP-sB, and CEMP-sC, based on the different levels of [Y/Ba] ratio
values, rather than a different physical mechanism for their origin.
We assign each sample star to one of the three new groups. The
traditional ``CEMP-s'' would loosely correspond to CEMP-sA class,
``CEMP-r/s'' to CEMP-sC, and those stars who fill in the continuum
between are CEMP-sB.  The advantage of using the elements Y and Ba is
that they are both easily measurable in CEMP stars. Moreover, using
[Y/Ba] provides an observable to confirm or refute our hypothesis that
there is only one underlying physical mechanism that causes the large
range of s-process abundance variations in these stars.

We compared the abundance patterns to AGB nucleosynthesis models and
found that certain thermal pulse abundance distributions of the new
low-metallicity 1.3\,M$_{\odot}$ model produced the best match to all
stellar abundance patterns in our sample. The CEMP-sA stars were best
matched this way. The CEMP-sB and CEMP-sC stars were however
increasingly difficult to match. Their abundance patterns have an
excess in the heavy neutron-capture abundances around Eu and above
compared to lighter elements, such as Sr, and the AGB models.  Upon
investigation, we find that this excess cannot be explained by
assuming these stars to have formed from gas already enriched with
r-process elements.

We considered two cases for diluting the material in the
stellar atmosphere of observed stars after the mass transfer of
s-enriched material from the AGB star. This takes into account the
unknown timing of the mass transfer event during the receiver star's
evolution.  We also explored the build up of the s-process elements in
the low-metallicity AGB star model as a function of its thermal
pulses and compared the abundance yields of each thermal pulse with the
abundance patterns of the sample stars to find the best match. 

Considering elements Ba and heavier, the CEMP-sA stars' patterns can
be reproduced by only few thermal pulses of their AGB star companion,
whereas CEMP-sB and CEMP-sC stars require progressively more thermal
pulses. This can be understood since the relative production of
heavier to lighter neutron-capture elements takes longer, and CEMP-sC
stars require larger amounts of the heaviest elements. This is
consistent with the fact that lower thermal pulse numbers correspond
to an earlier mass transfer event in the evolution of the AGB star
compared to later ones.

In conclusion, the CEMP-sA stars are well-explained by the
1.3\,M$_{\odot}$, [Fe/H] = $-2.5$ model. Following some tests with
higher mass models with a metallicity of [Fe/H] = $-2.2$, we speculate
that the abundance patterns of some of the CEMP-sB and CEMP-sC stars
could better be reproduced by models with masses larger than
1.3\,M$_{\odot}$ because we suspect these models to produce larger
amounts of the heaviest elements. We find the abundance signature of
HE~0414$-$0343 to have arisen from a $>1.3$\,M$_{\odot}$ mass AGB star
in combination with a late-time mass transfer. Correspondingly,
HE~0414$-$0343 is a CEMP-sC star. We speculate on the origin
  of the three classes of CEMP-s stars and suggest that the range of
  abundances are caused by a number of factors from variations in AGB
  mass, the timing of the mass transfer event, or from physics not
  currently included in the 1D calculations such as proton-ingestion
   at the tip of the AGB. Pre-enrichment due to chemical evolution     
  could also play a role.

\acknowledgements 

The Hobby-Eberly Telescope (HET) is a joint project of the University
of Texas at Austin, the Pennsylvania State University, Stanford
University, Ludwig-Maximilians-Universit\"at M\"unchen, and
Georg-August-Universit\"at G\"ottingen. The HET is named in honor of
its principal benefactors, William P. Hobby and Robert E. Eberly.  We
are grateful to the Hobby-Eberly staff for their assistance in
obtaining the data collected for this paper. We appreciate useful
conversations with Christopher Waters and Harriet Dinerstein.
J.K.H. acknowledges support for this work from NSF grants AST-0708245
and AST-0908978. A.F. acknowledges support from NSF grant AST-1255160
and the Silverman (1968) Family Career Development
Professorship. 
J.K.H., A.F. and V.M.P. acknowledge partial support for this work from PHY   
08-22648; Physics Frontier Center/{}Joint Institute for Nuclear Astrophysics
(JINA) and PHY 14-30152; Physics Frontier Center/JINA Center for the Evolution
of the Elements (JINA-CEE), awarded by the US National Science Foundation.
This work was supported in part by NSF
grant AST-1211585 to C.S. A.I.K. was supported through an Australian
Research Council Future Fellowship (FT110100475).

\clearpage
\newpage
\appendix
\section{Results of the comparison of Abundance Patterns of CEMP-sA Sample Stars with the best matched thermal pulse abundance distributions.}
\begin{figure*}[!th]
\begin{center}
\includegraphics[width=0.9\textwidth]{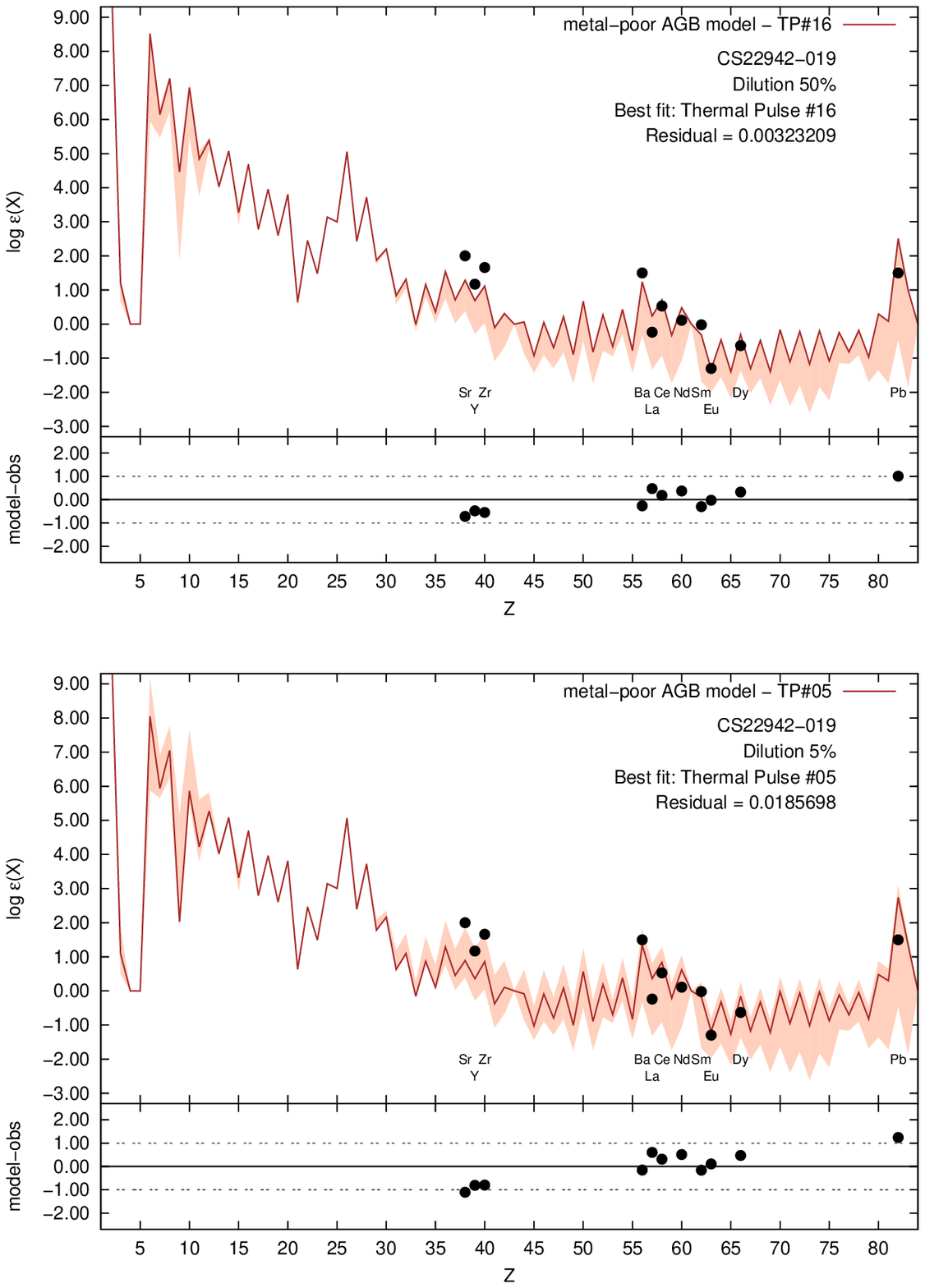}
\caption{Best Fit Model Abundance Comparision of CS~22942$-$019, 50 Percent
Dilution (upper panel), and 5 Percent Dilution (lower panel).}
\end{center}
\end{figure*}

\begin{figure*}[!th]
\begin{center}
\includegraphics[width=0.9\textwidth]{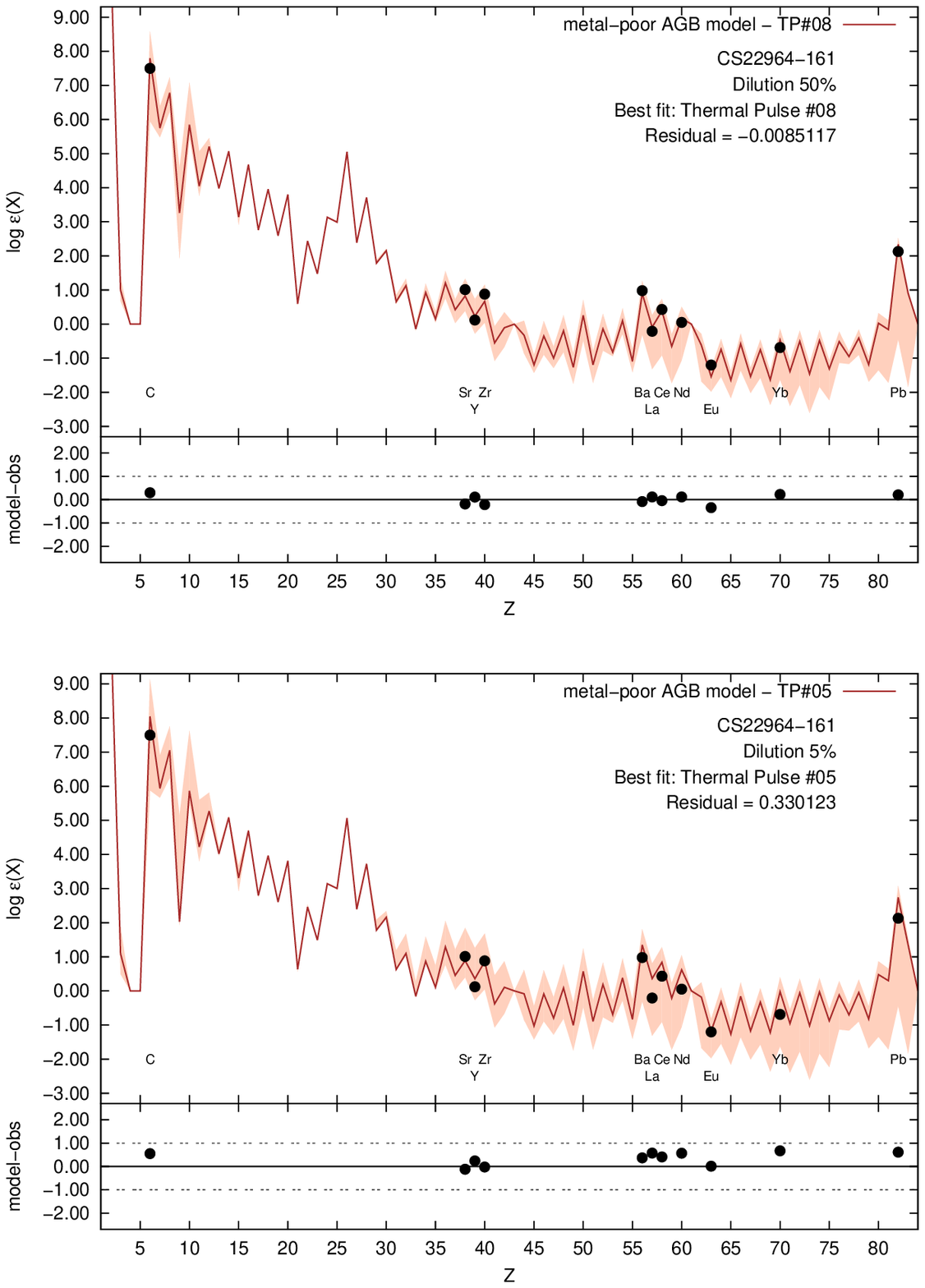}
\caption{Best Fit Model Abundance Comparision of CS~22964$-$161, 50 Percent
Dilution (upper panel), and 5 Percent Dilution (lower panel).}
\end{center}
\end{figure*}
  
\begin{figure*}[!th]
\begin{center}
\includegraphics[width=0.9\textwidth]{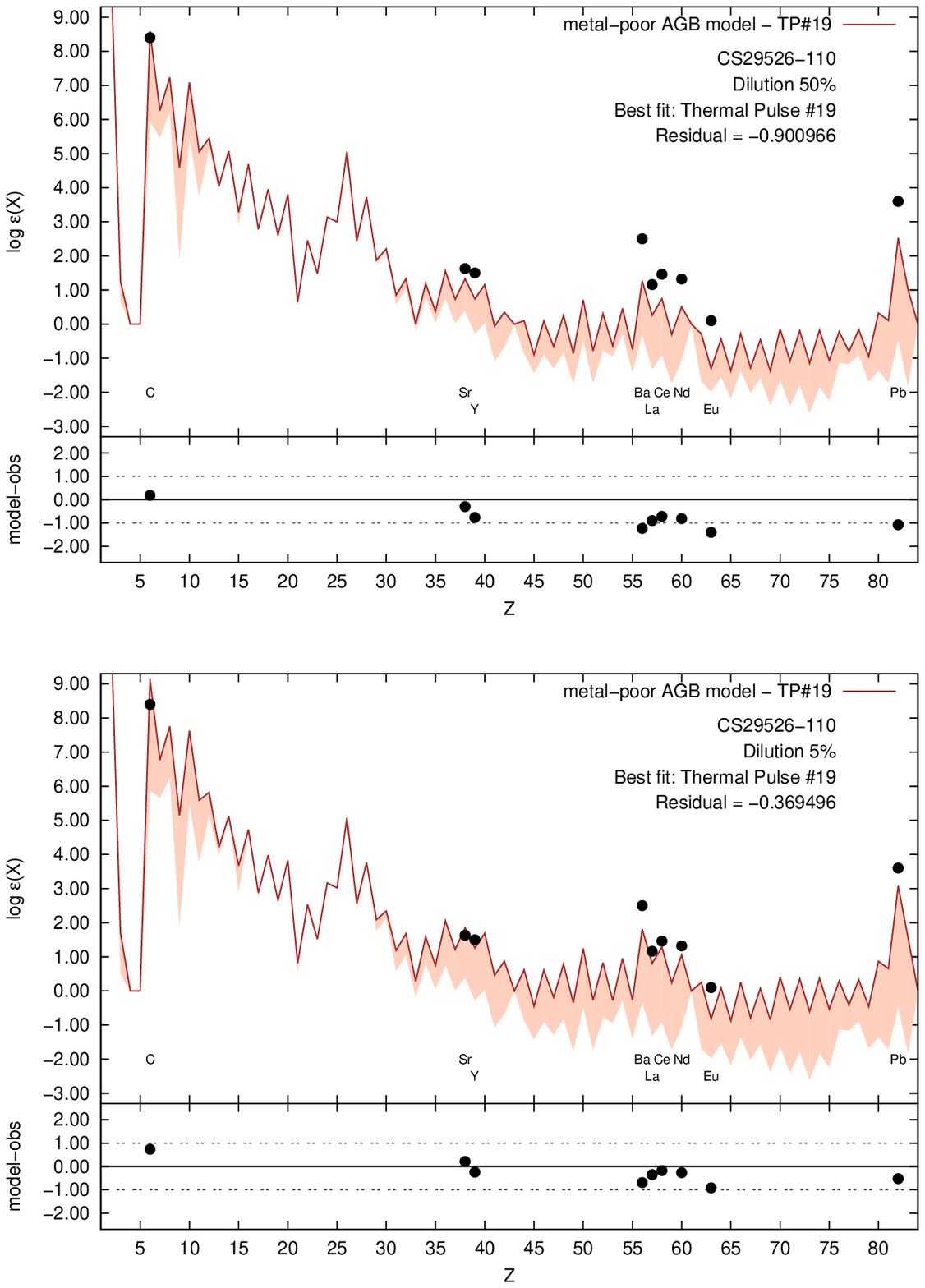}
\caption{Best Fit Model Abundance Comparision of CS~29526$-$110, 50 Percent
Dilution (upper panel), and 5 Percent Dilution (lower panel).}
\end{center}
\end{figure*}
  
\begin{figure*}[!th]
\begin{center}
\includegraphics[width=0.9\textwidth]{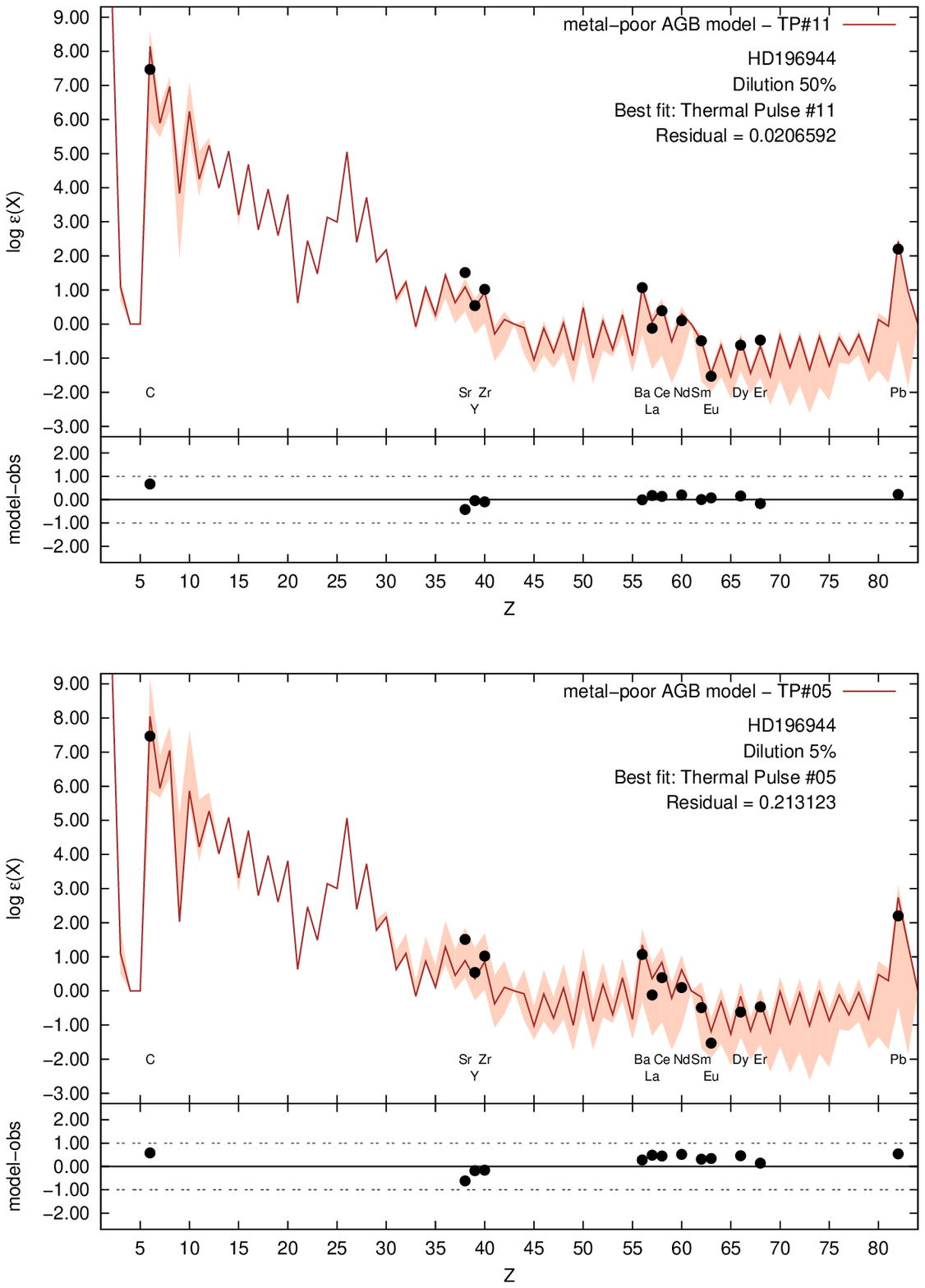}
\caption{Best Fit Model Abundance Comparision of HD~196944, 50 Percent
Dilution (upper panel), and 5 Percent Dilution (lower panel).}
\end{center}
\end{figure*}
  
\begin{figure*}[!th]
\begin{center}
\includegraphics[width=0.9\textwidth]{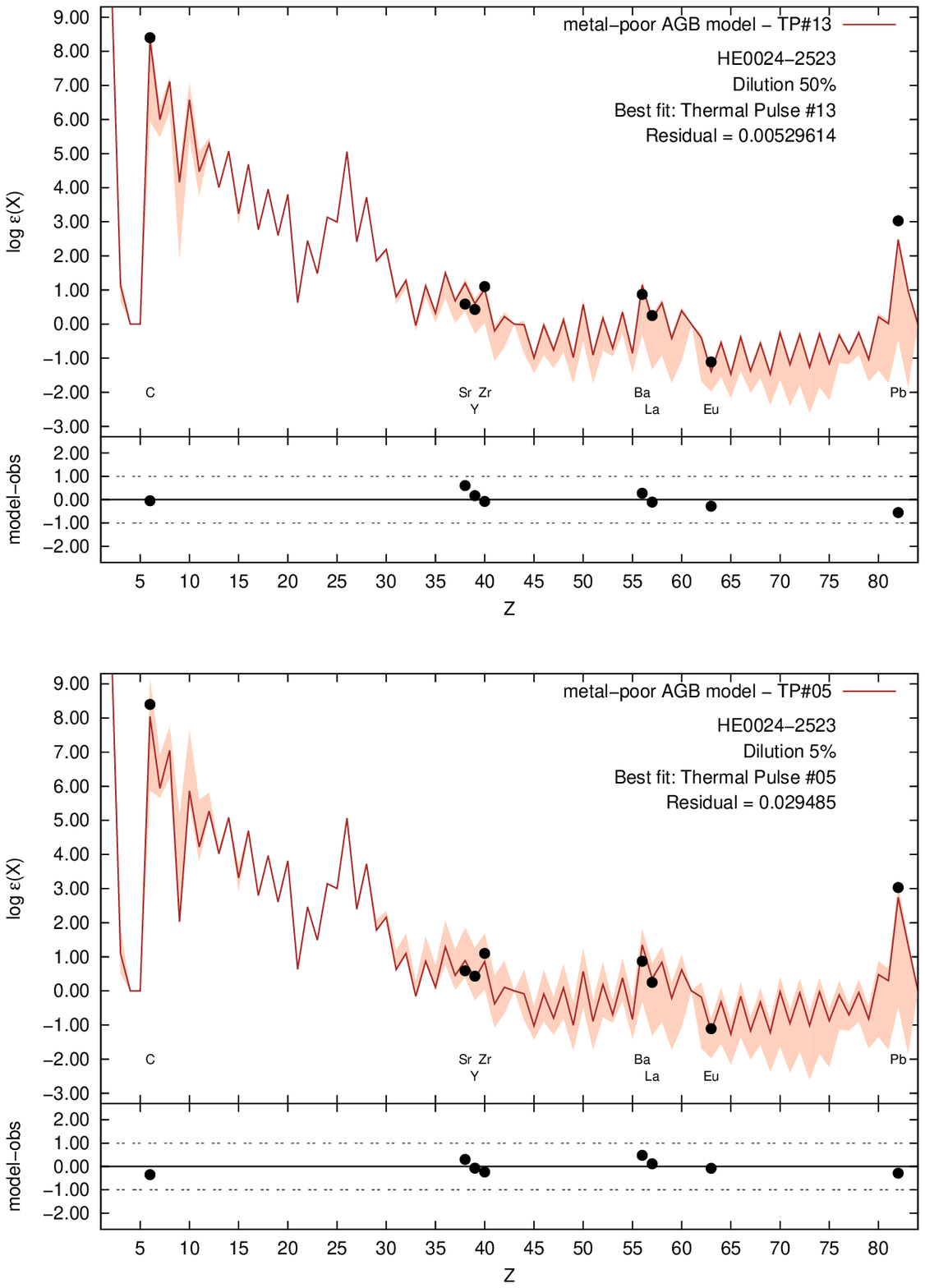}
\caption{Best Fit Model Abundance Comparision of HE~0024$-$2523, 50 Percent
Dilution (upper panel), and 5 Percent Dilution (lower panel).}
\end{center}
\end{figure*}
  
\begin{figure*}[!th]
\begin{center}
\includegraphics[width=0.9\textwidth]{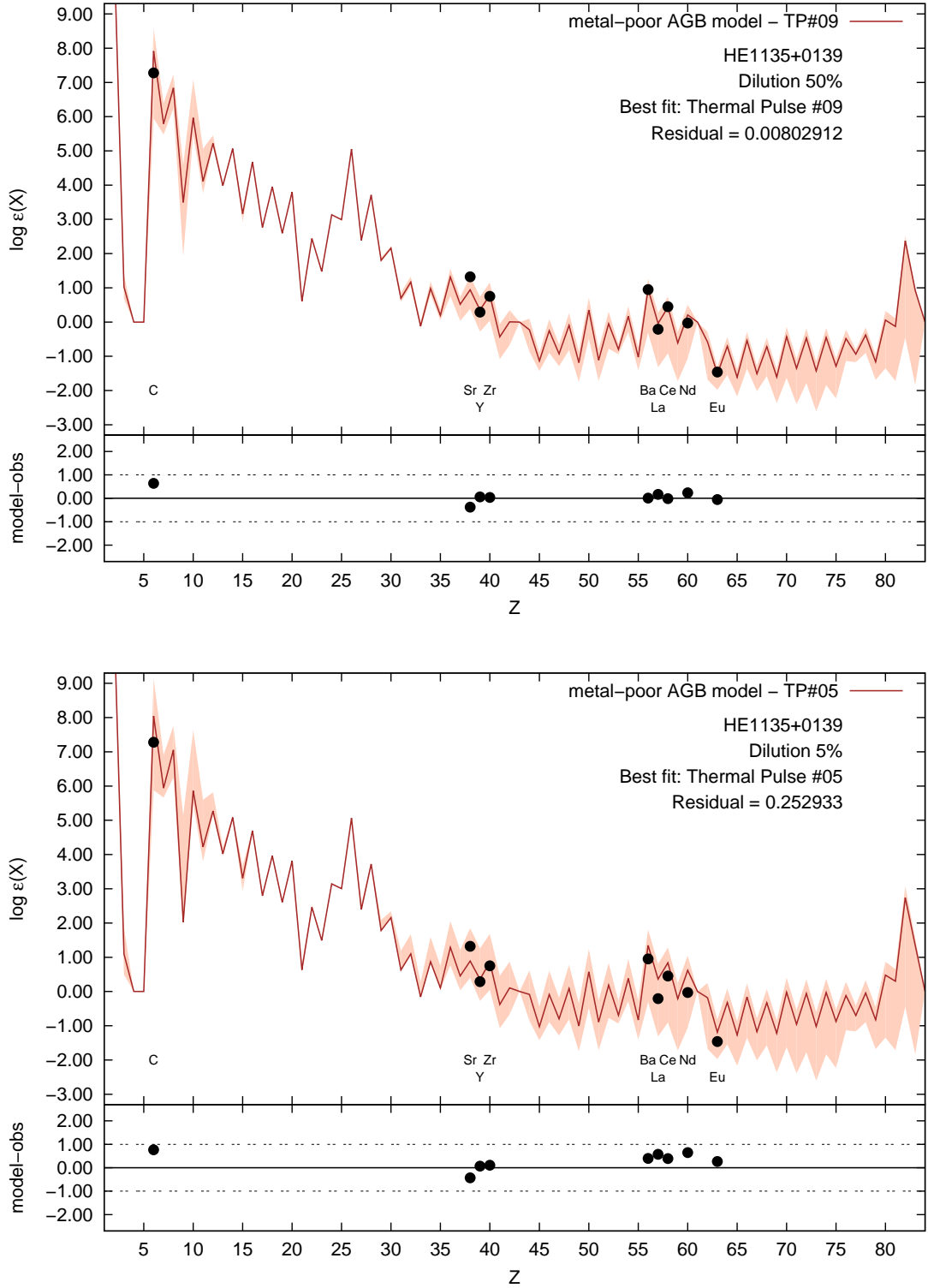}
\caption{Best Fit Model Abundance Comparision of HE~1135$+$0139, 50 Percent
Dilution (upper panel), and 5 Percent Dilution (lower panel).}
\end{center}
\end{figure*}

\clearpage
\newpage
\section{Results of the comparison of Abundance Patterns of CEMP-sB 
Sample Stars with the best matched thermal pulse abundance
distributions.}  \begin{figure*}[!th]
\begin{center}
\includegraphics[width=0.9\textwidth]{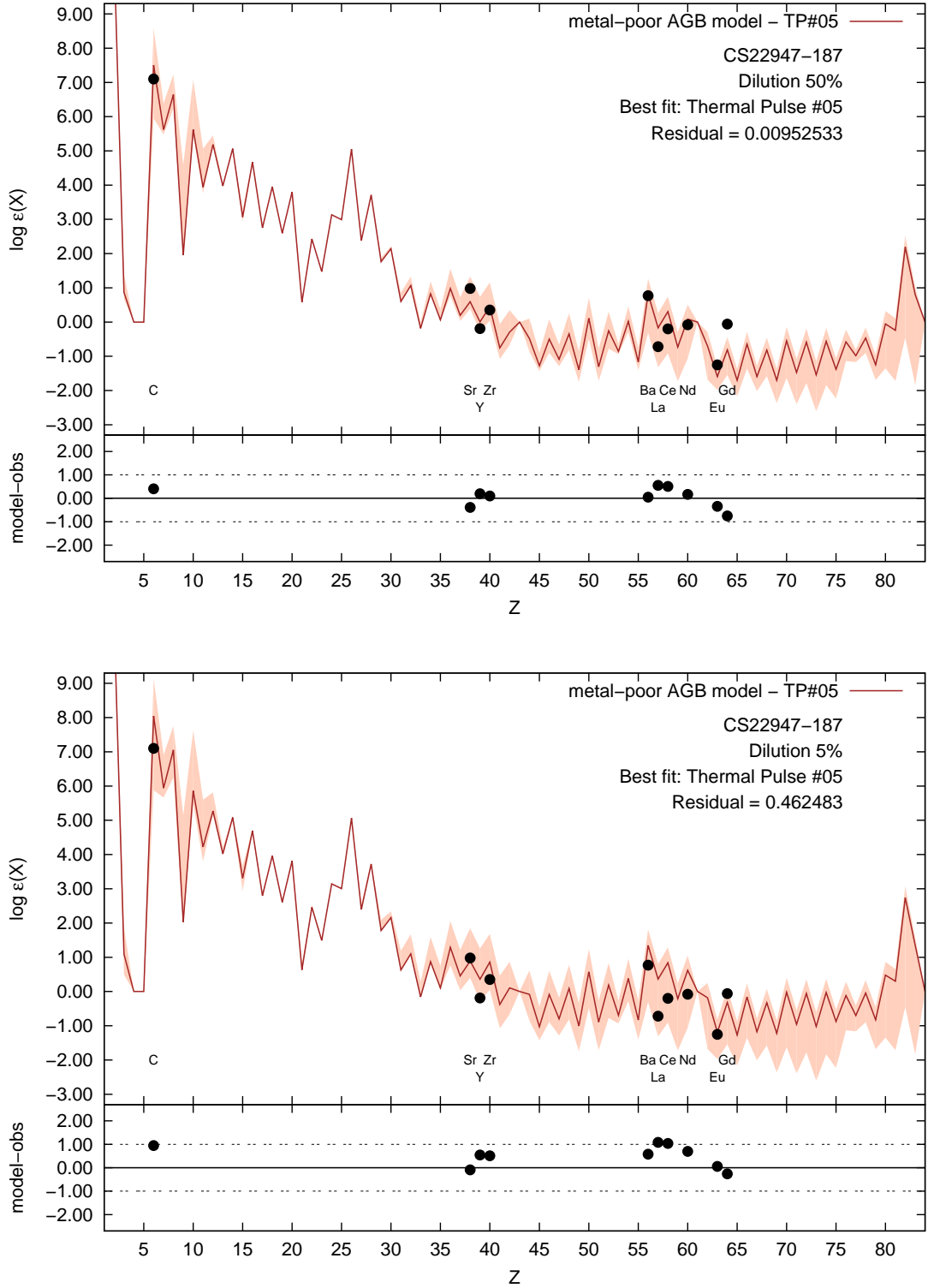}
\caption{Best Fit Model Abundance Comparision of CS~22947$-$187, 50 Percent
Dilution (upper panel), and 5 Percent Dilution (lower panel).}
\end{center}
\end{figure*}
\newpage
  
\begin{figure*}[!th]
\begin{center}
\includegraphics[width=0.9\textwidth]{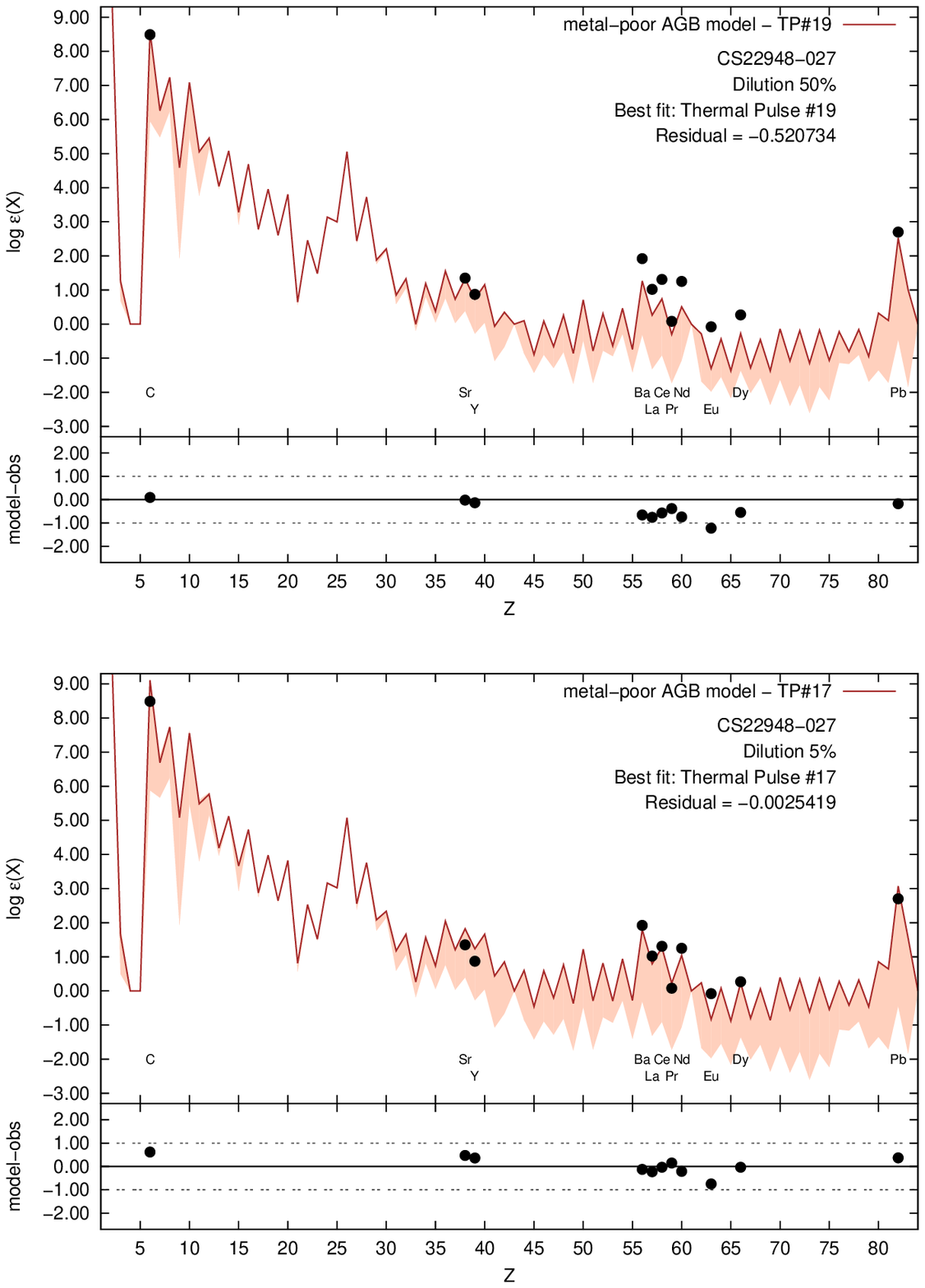}
\caption{Best Fit Model Abundance Comparision of CS~22948$-$027, 50 Percent
Dilution (upper panel), and 5 Percent Dilution (lower panel).}
\end{center}
\end{figure*}
\newpage
  
\begin{figure*}[!th]
\begin{center}
\includegraphics[width=0.9\textwidth]{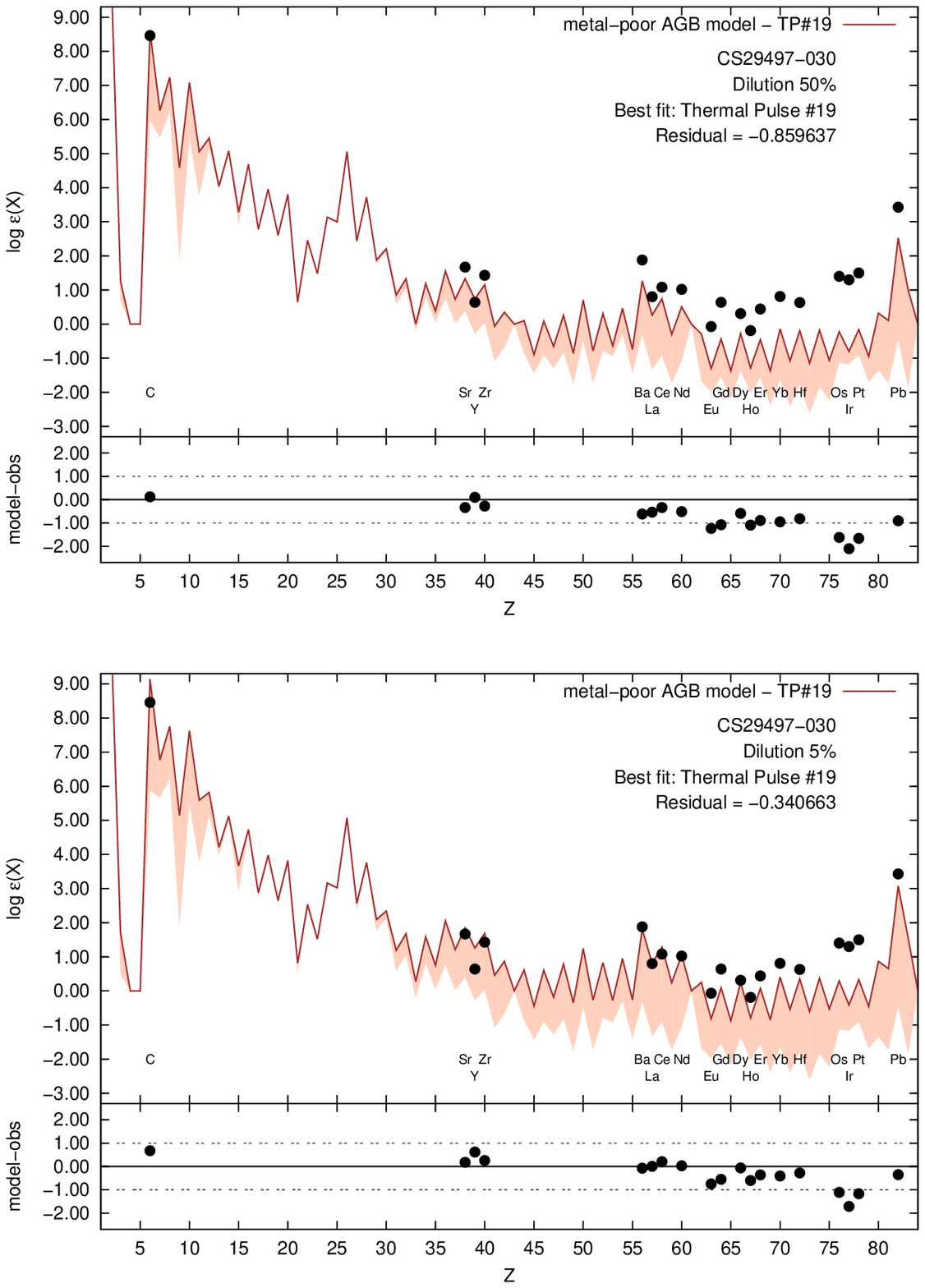}
\caption{Best Fit Model Abundance Comparision of CS~29497$-$030, 50 Percent
Dilution (upper panel), and 5 Percent Dilution (lower panel).}
\end{center}
\end{figure*}
\newpage
  
\begin{figure*}[!th]
\begin{center}
\includegraphics[width=0.9\textwidth]{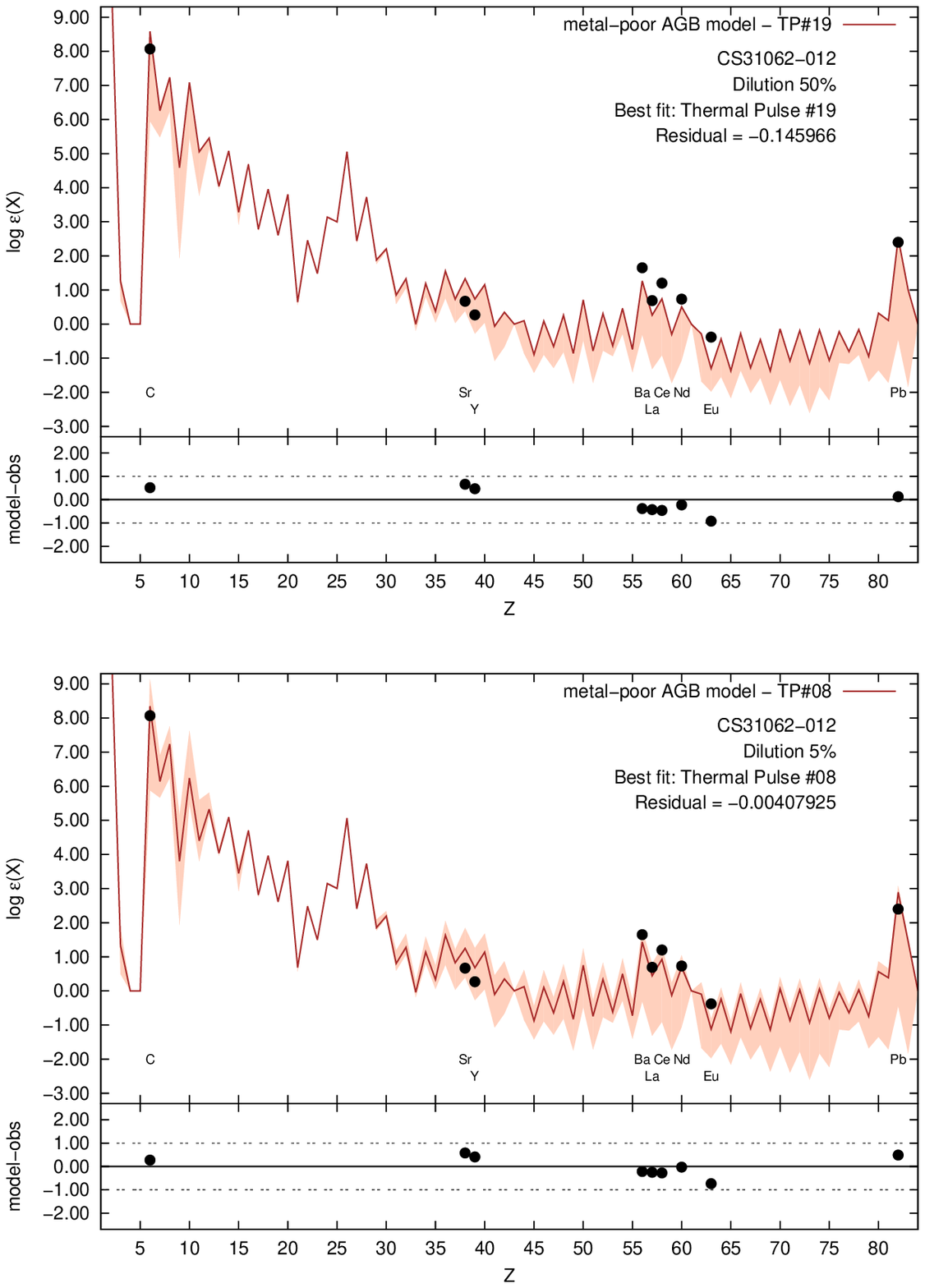}
\caption{Best Fit Model Abundance Comparision of CS~31062$-$012, 50 Percent
Dilution (upper panel), and 5 Percent Dilution (lower panel).}
\end{center}
\end{figure*}
\newpage
  
\begin{figure*}[!th]
\begin{center}
\includegraphics[width=0.9\textwidth]{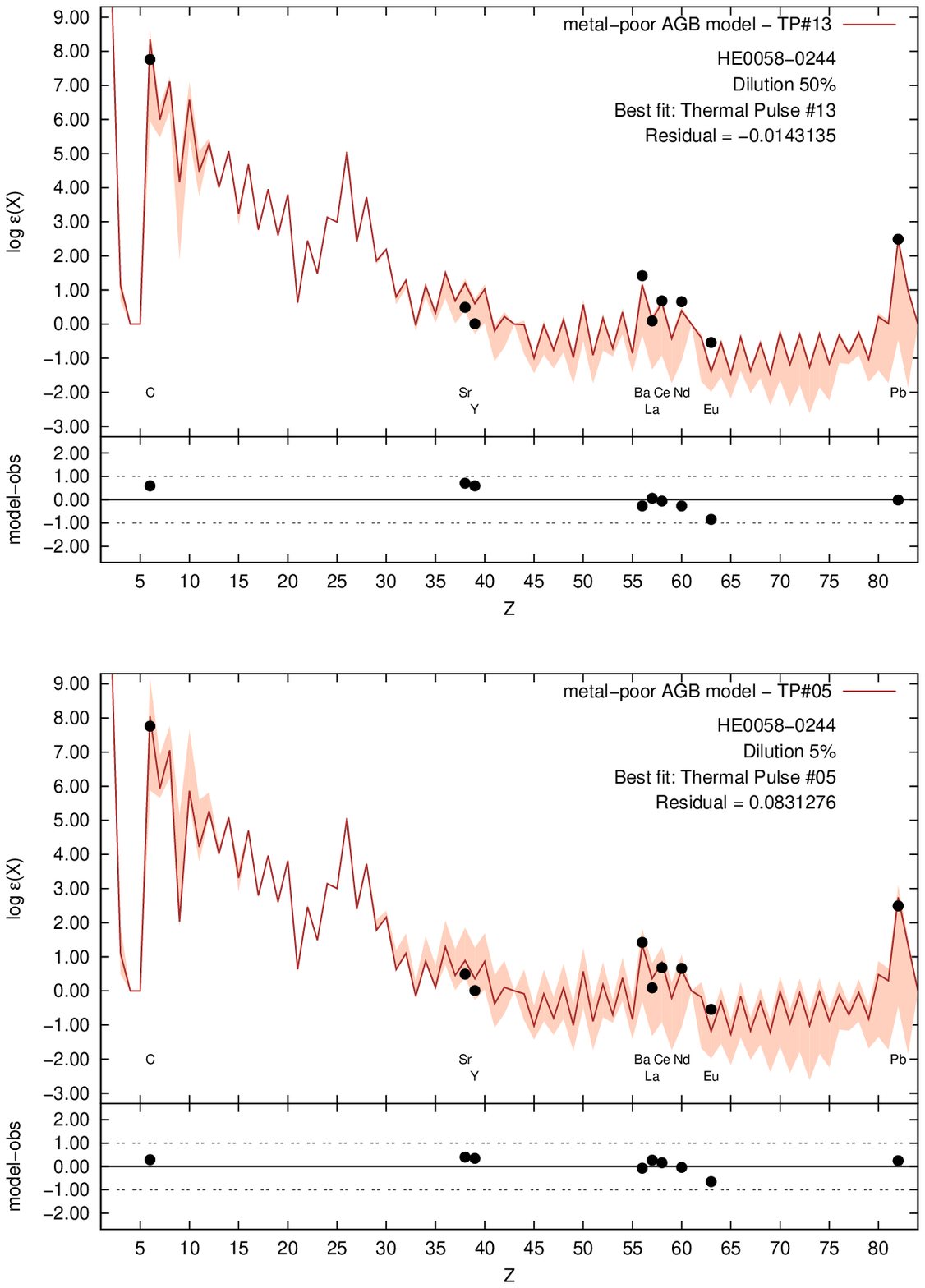}
\caption{Best Fit Model Abundance Comparision of HE~0058$-$0244, 50 Percent
Dilution (upper panel), and 5 Percent Dilution (lower panel).}
\end{center}
\end{figure*}
\newpage
  
\begin{figure*}[!th]
\begin{center}
\includegraphics[width=0.9\textwidth]{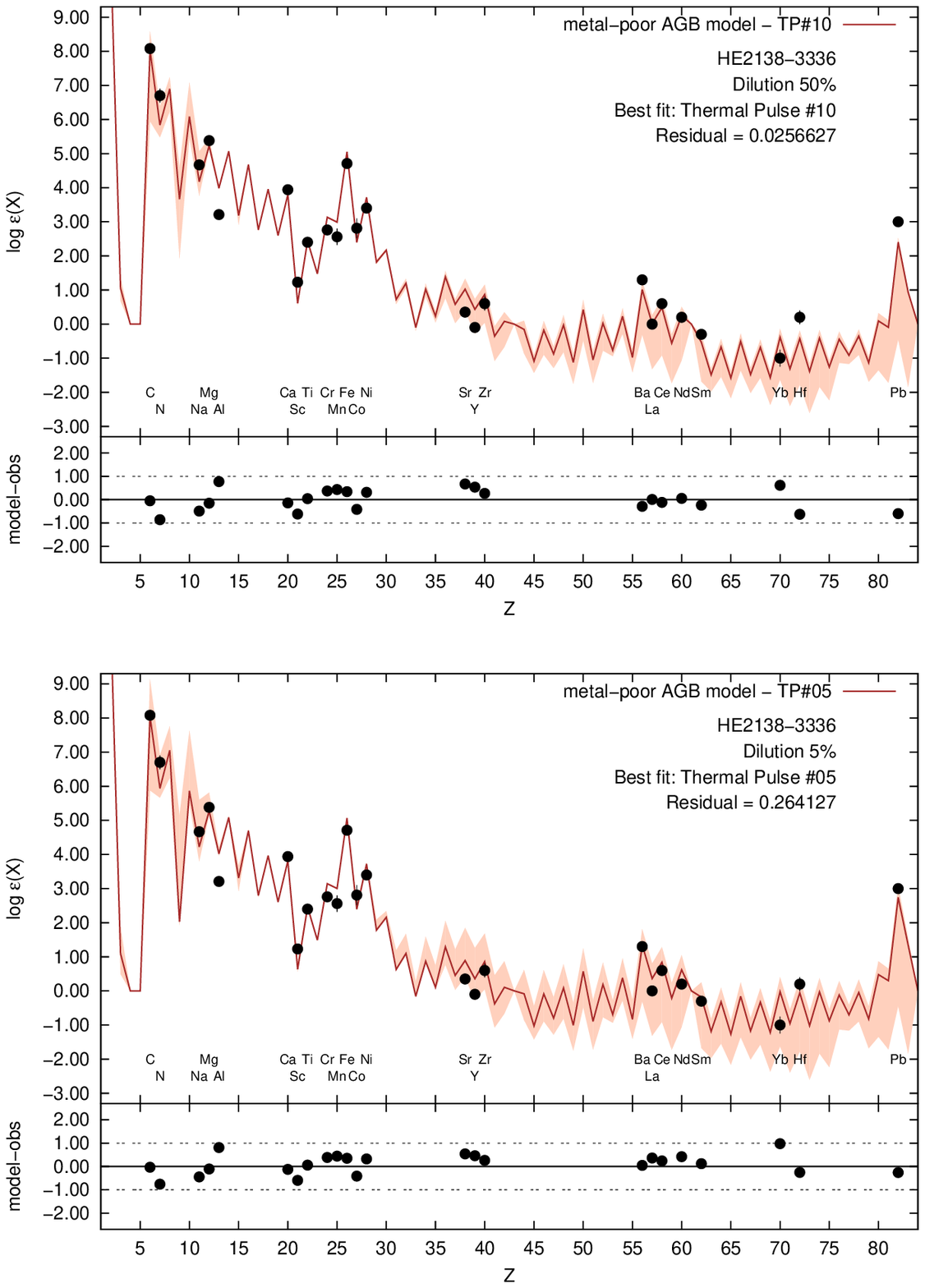}
\caption{Best Fit Model Abundance Comparision of HE~2138$-$3336, 50 Percent
Dilution (upper panel), and 5 Percent Dilution (lower panel).}
\end{center}
\end{figure*}
\newpage
  
\begin{figure*}[!th]
\begin{center}
\includegraphics[width=0.9\textwidth]{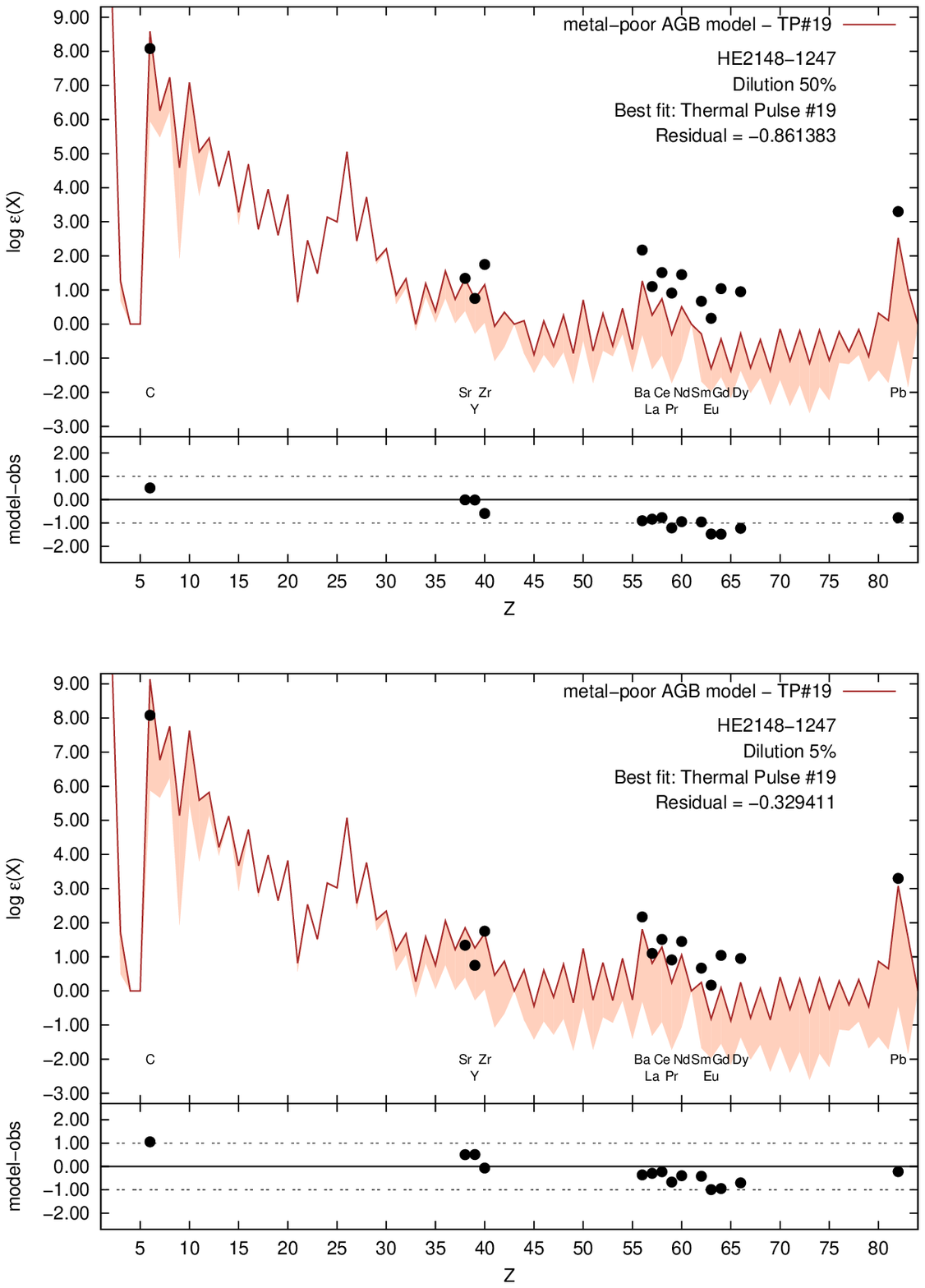}
\caption{Best Fit Model Abundance Comparision of HE~2148$-$1247, 50 Percent
Dilution (upper panel), and 5 Percent Dilution (lower panel).}
\end{center}
\end{figure*}
\newpage
  
\begin{figure*}[!th]
\begin{center}
\includegraphics[width=0.9\textwidth]{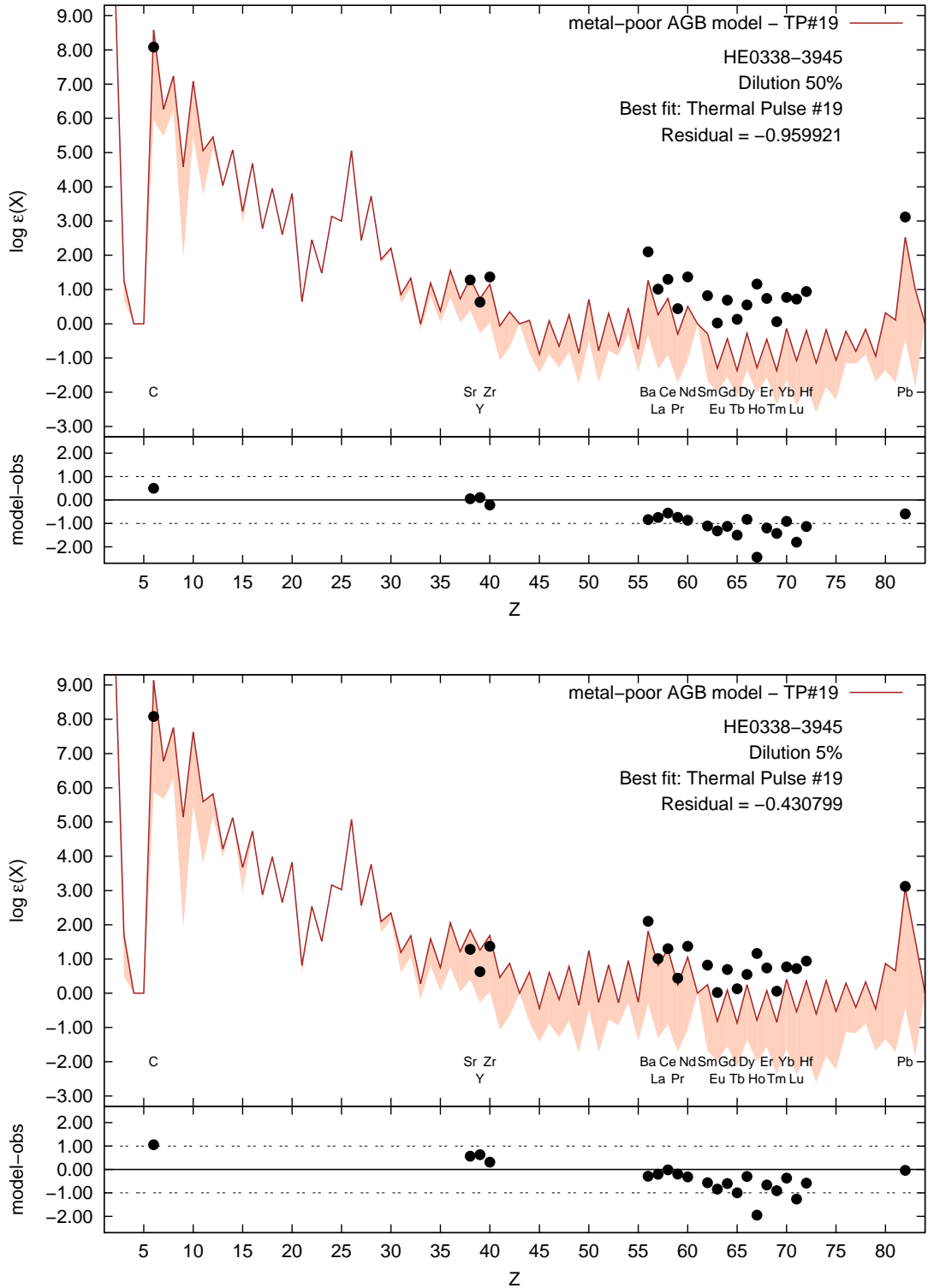}
\caption{Best Fit Model Abundance Comparision of HE~0338$-$3945, 50 Percent
Dilution (upper panel), and 5 Percent Dilution (lower panel).}
\end{center}
\end{figure*}
\newpage

\clearpage
\newpage
\section{Results of the comparison of Abundance Patterns of CEMP-sC                   
Sample Stars with the best matched thermal pulse abundance
distributions.}  \begin{figure*}[!th]
\begin{center}
\includegraphics[width=0.9\textwidth]{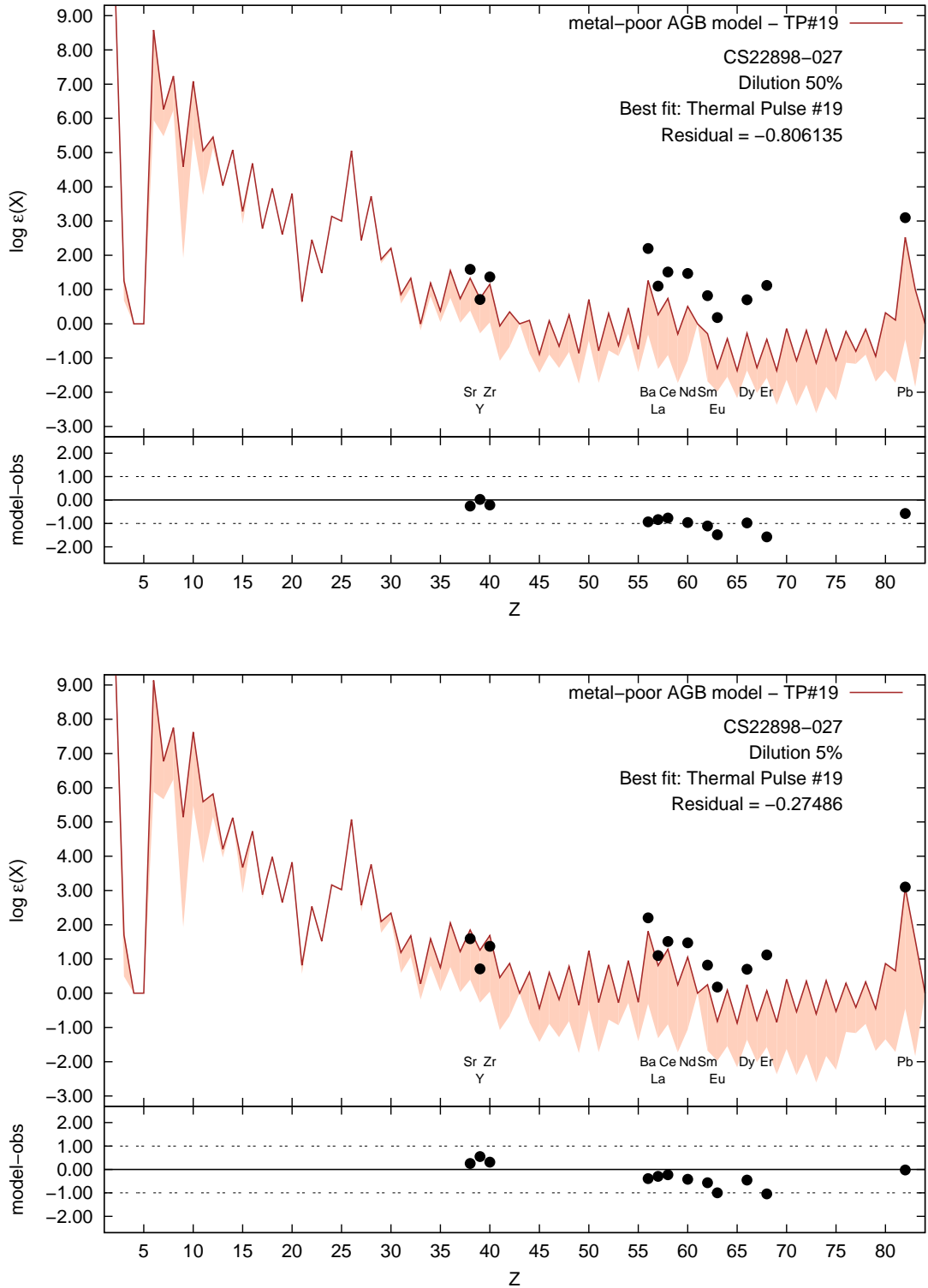}
\caption{Best Fit Model Abundance Comparision of CS~22898$-$027, 50 Percent
Dilution (upper panel), and 5 Percent Dilution (lower panel).}
\end{center}
\end{figure*}
\newpage
  
\begin{figure*}[!th]
\begin{center}
\includegraphics[width=0.9\textwidth]{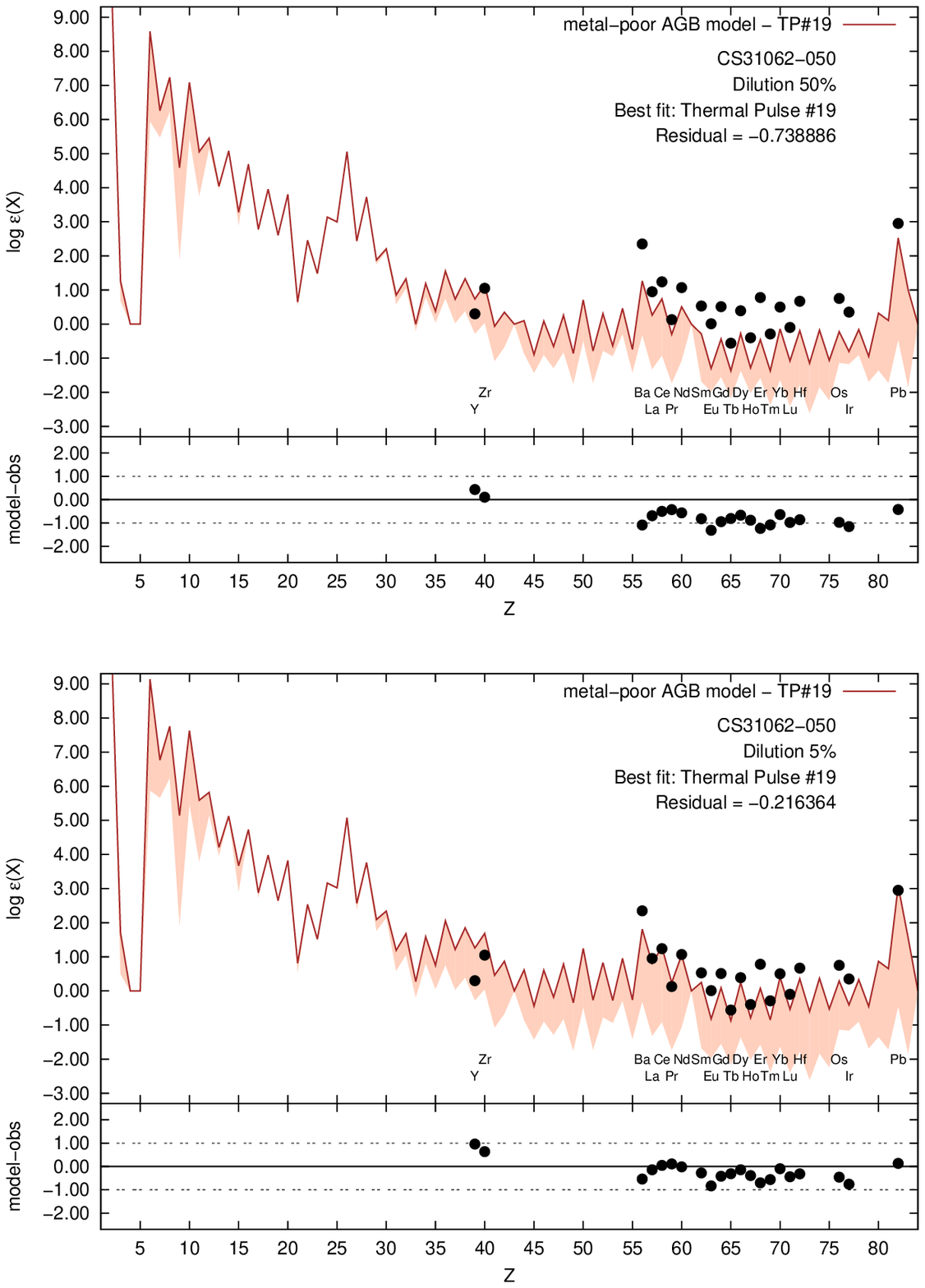}
\caption{Best Fit Model Abundance Comparision of CS~31062$-$050, 50 Percent
Dilution (upper panel), and 5 Percent Dilution (lower panel).}
\end{center}
\end{figure*}
\newpage
  
\begin{figure*}[!th]
\begin{center}
\includegraphics[width=0.9\textwidth]{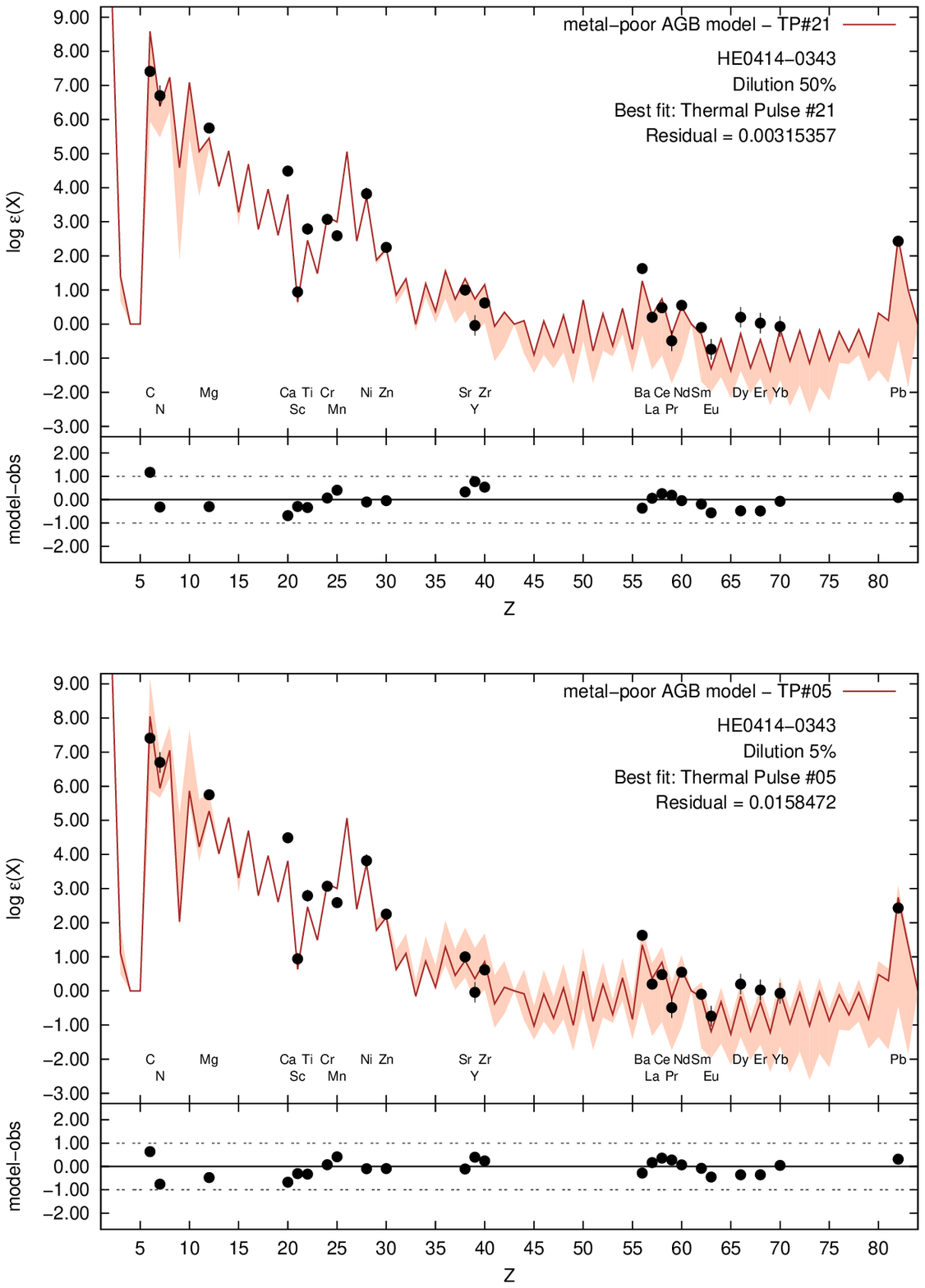}
\caption{Best Fit Model Abundance Comparision of HE~0414$-$0343, 50 Percent
Dilution (upper panel), and 5 Percent Dilution (lower panel).}
\end{center}
\end{figure*}
\newpage
  
\begin{figure*}[!th]
\begin{center}
\includegraphics[width=0.9\textwidth]{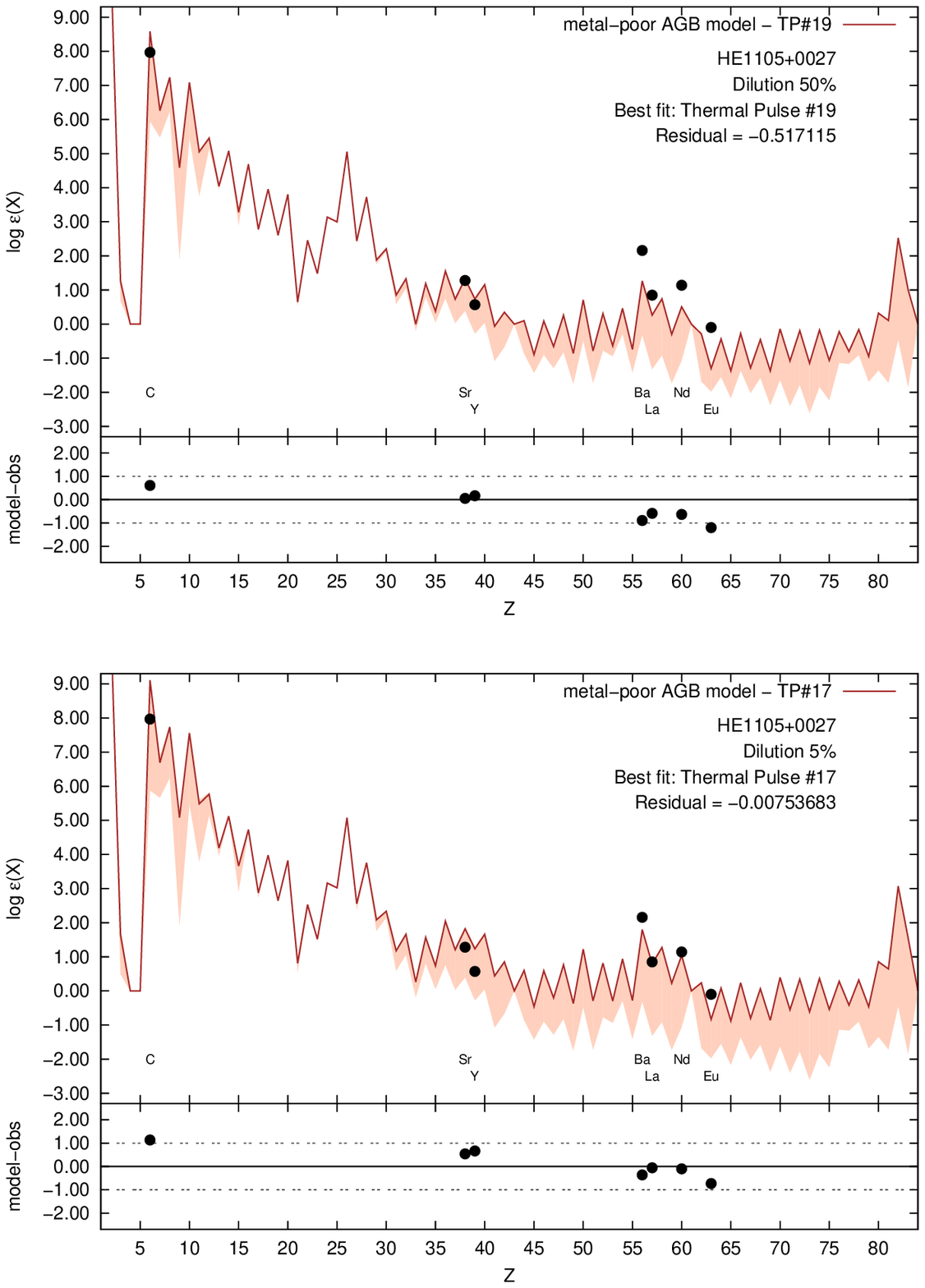}
\caption{Best Fit Model Abundance Comparision of HE~1105$+$0027, 50 Percent
Dilution (upper panel), and 5 Percent Dilution (lower panel).}
\end{center}
\end{figure*}
\newpage
  
\begin{figure*}[!th]
\begin{center}
\includegraphics[width=0.9\textwidth]{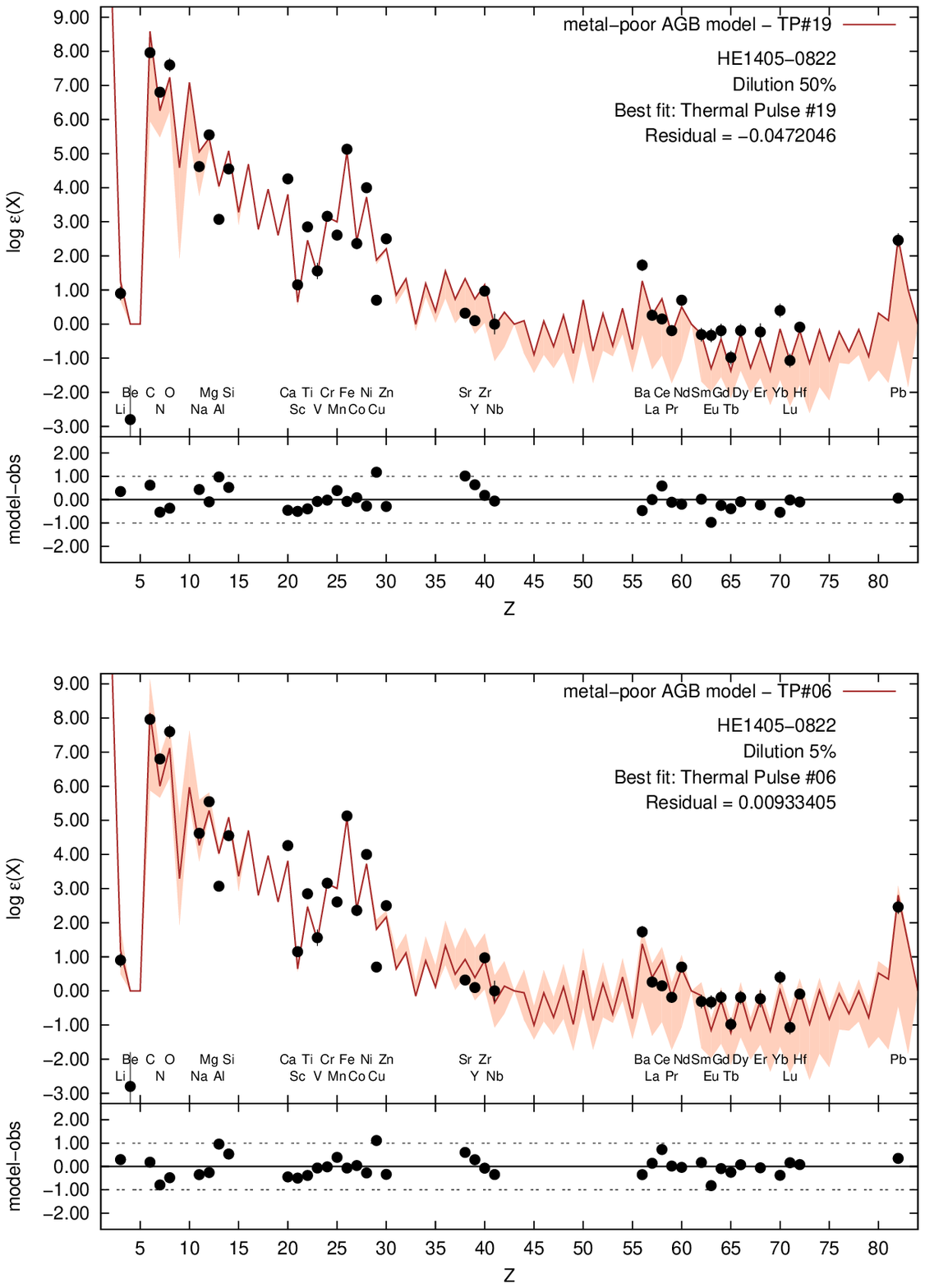}
\caption{Best Fit Model Abundance Comparision of HE~1405$-$0822, 50 Percent
Dilution (upper panel), and 5 Percent Dilution (lower panel).}
\end{center}
\end{figure*}
\newpage
  
\begin{figure*}[!th]
\begin{center}
\includegraphics[width=0.9\textwidth]{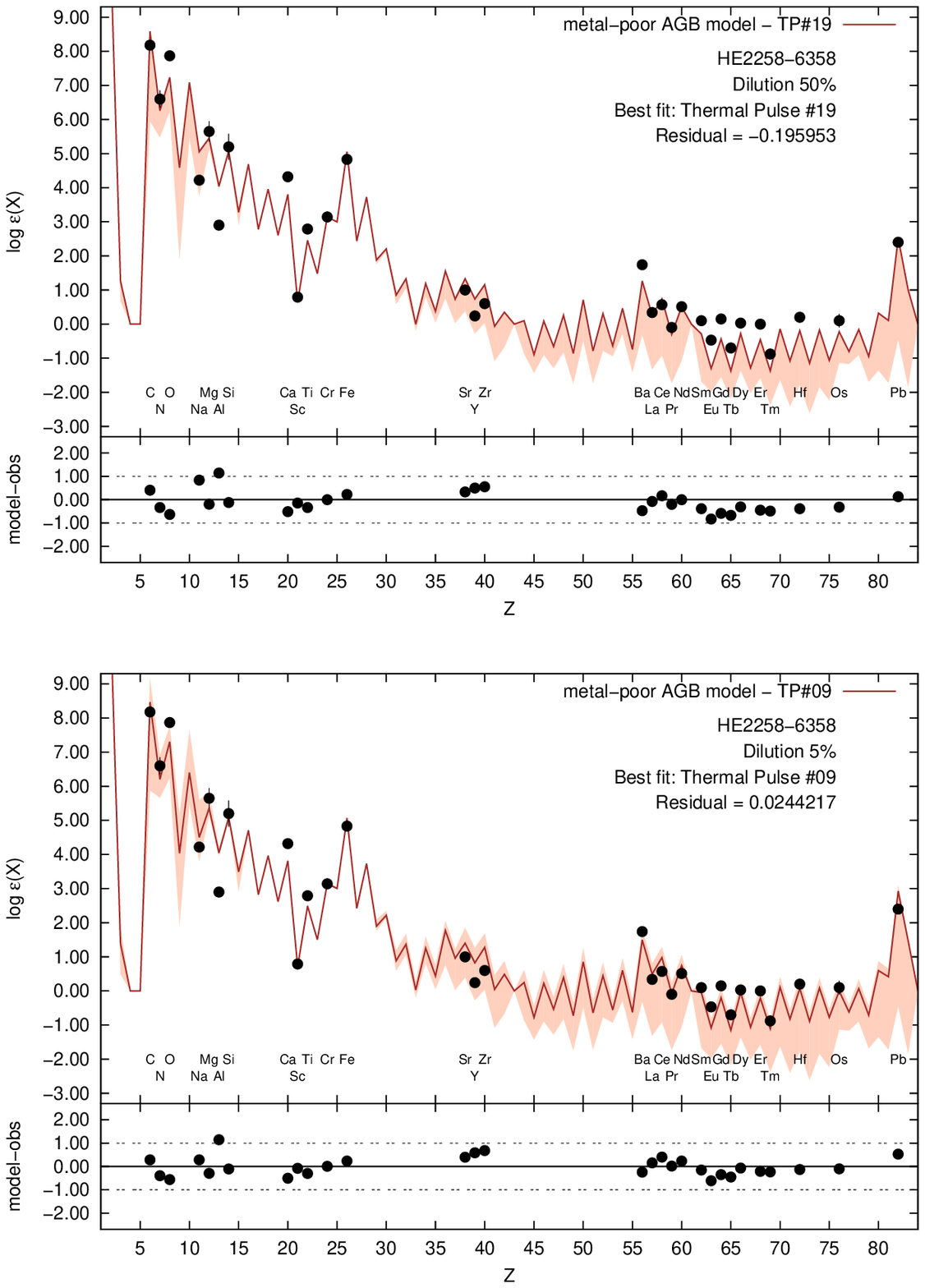}
\caption{Best Fit Model Abundance Comparision of HE~2258$-$6358, 50 Percent
Dilution (upper panel), and 5 Percent Dilution (lower panel).}
\end{center}
\end{figure*}
\newpage

\end{document}